\documentclass[twocolumn]{aastex63}

\usepackage{xspace}
\usepackage{epstopdf}
\usepackage{mathtools}
\usepackage{mathrsfs}
\usepackage{calrsfs}
\usepackage[hang,flushmargin]{footmisc}
\usepackage[autostyle, english = american]{csquotes}
\usepackage{multirow}
\MakeOuterQuote{"}

\newcommand{\rb}{$R_B$\xspace}
\newcommand{\rsoi}{$R_{\rm SOI}$\xspace}

\received{--}
\revised{--}
\accepted{--}

\submitjournal{ApJ}

\shorttitle{Jet Collimation and Acceleration in NGC 315}
\shortauthors{Park et al.}

\graphicspath{{./}{figures/}}

\begin{document}

\title{Jet Collimation and Acceleration in the giant radio galaxy NGC 315}

\correspondingauthor{Jongho Park}
\email{jpark@asiaa.sinica.edu.tw}

\author[0000-0001-6558-9053]{Jongho Park}
\affiliation{Institute of Astronomy and Astrophysics, Academia Sinica, P.O. Box 23-141, Taipei 10617, Taiwan}
\affiliation{Department of Physics and Astronomy, Seoul National University, Gwanak-gu, Seoul 08826, Republic of Korea}

\author[0000-0001-6906-772X]{Kazuhiro Hada}
\affiliation{Mizusawa VLBI Observatory, National Astronomical Observatory of Japan, Osawa, Mitaka, Tokyo 181-8588, Japan}
\affiliation{Department of Astronomical Science, The Graduate University for Advanced Studies (SOKENDAI), 2-21-1 Osawa, Mitaka, Tokyo 181-8588, Japan}

\author[0000-0001-6081-2420]{Masanori Nakamura}
\affiliation{National Institute of Technology, Hachinohe College, 16-1 Uwanotai, Tamonoki, Hachinohe, Aomori 039-1192, Japan}
\affiliation{Institute of Astronomy and Astrophysics, Academia Sinica, P.O. Box 23-141, Taipei 10617, Taiwan}

\author{Keiichi Asada}
\affiliation{Institute of Astronomy and Astrophysics, Academia Sinica, P.O. Box 23-141, Taipei 10617, Taiwan}

\author[0000-0002-4417-1659]{Guangyao Zhao}
\affiliation{Instituto de Astrofísica de Andalucía–CSIC, Glorieta de la Astronomía s/n, E-18008 Granada, Spain}
\affiliation{Korea Astronomy and Space Science Institute, Daedeok-daero 776, Yuseong-gu, Daejeon 34055, Republic of Korea}

\author[0000-0002-2709-7338]{Motoki Kino}
\affiliation{Kogakuin University of Technology \& Engineering, Academic Support Center, 2665-1 Nakano, Hachioji, Tokyo 192-0015, Japan}
\affiliation{National Astronomical Observatory of Japan, 2-21-1 Osawa, Mitaka, Tokyo 181-8588, Japan}

\begin{abstract}

We study the collimation and acceleration of the jets in the nearby giant radio galaxy NGC 315, using multifrequency Very Long Baseline Array observations and archival High Sensitivity Array and Very Large Array data. We find that the jet geometry transitions from a semi-parabolic shape into a conical/hyperbolic shape at a distance of $\approx10^5$ gravitational radii. We constrain the frequency-dependent position of the core, from which we locate the jet base. The jet collimation profile in the parabolic region is in good agreement with the steady axisymmetric force-free electrodynamic solution for the outermost poloidal magnetic field line anchored to the black hole event horizon on the equatorial plane, similar to the nearby radio galaxies M87 and NGC 6251. The velocity field derived from the asymmetry in brightness between the jet and counterjet shows gradual acceleration up to the bulk Lorentz factor of $\Gamma \sim 3$ in the region where the jet collimation occurs, confirming the presence of the jet acceleration and collimation zone. These results suggest that the jets are collimated by the pressure of the surrounding medium and accelerated by converting Poynting flux to kinetic energy flux. We discover limb-brightening of the jet in a limited distance range where the angular resolution of our data is sufficient to resolve the jet transverse structure. This indicates that either the jet has a stratified velocity field of fast-inner and slow-outer layers or the particle acceleration process is more efficient in the outer layer due to the interaction with the surroundings on pc-scales.

\end{abstract}


\section{Introduction} 
\label{sec:intro}

About 10\% of active galactic nuclei (AGNs) in the local Universe releases large amounts of energy in the form of jets \citep[e.g.,][]{Netzer2015, Blandford2019, Hada2019}. AGN jets are often observed to move at relativistic speeds with apparent speeds up to tens of times the speed of light \citep[e.g.,][]{Jorstad2017, Lister2018}, produce rapid time variability at multiple wavelengths \citep[e.g.,][]{PT2014, PT2017}, actively interact with the interstellar and intergalactic medium, and affect the evolution of galaxies and clusters \citep[e.g.,][]{Fabian2012, HC2020}. They are believed to be launched by the accretion of matter and strong magnetic fields in the vicinity of black holes \citep[e.g.,][]{BZ1977, BP1982, Contopoulos1995, NQ2005, McKinney2006, Tchekhovskoy2011, Sadowski2013, Pu2015}.

How AGN jets are collimated and accelerated to relativistic speeds has been a longstanding problem. Many theoretical studies and special/general relativistic magnetohydrodynamic (S/GRMHD) simulations have shown that AGN jets are gradually accelerated by converting the electromagnetic energy into the kinetic energy \citep[e.g.,][]{Li1992, BL1994, VK2004, Komissarov2007, Komissarov2009, Tchekhovskoy2008, Tchekhovskoy2009, Lyubarsky2009}. The MHD jet acceleration occurs more efficiently when the jet and its associated poloidal magnetic fields are being systematically collimated through e.g., the magnetic nozzle effect \citep[e.g.,][]{Camenzind1987, Li1992, BN2006, Komissarov2009, Tchekhovskoy2009, Vlahakis2015}. Jet collimation is thought to be governed by the pressure of an external confining medium \citep{Eichler1993, BL1994, Komissarov2007, Komissarov2009, Lyubarsky2009}, which is presumably non-relativistic gas outflows launched from the accretion disk \citep[e.g.,][]{Sadowski2013, Yuan2015, Nakamura2018}. Therefore, jet acceleration and collimation are believed to occur simultaneously, under the influence of the black hole's gravity, forming a jet acceleration and collimation zone (ACZ) at a distance $\lesssim10^4$--$10^6$ gravitational radii \citep[$R_g$,][]{VK2004, Marscher2008, Meier2012}.

Recent very long baseline interferometry (VLBI) observations have indeed found the existence of the ACZs in nearby radio galaxies and blazars. \cite{AN2012} showed that the jet in M87 is gradually collimated in a semi-parabolic shape inside the Bondi radius, while it freely expands conically outside. The parabolic jet collimation profile appears to be present all the way down to near the jet base \citep{Junor1999, Hada2013, NA2013, Hada2016, Mertens2016, Kim2018, Nakamura2018, Walker2018}. \cite{Asada2014} showed that the M87 jet accelerates to relativistic speeds inside the Bondi radius, followed by gradual deceleration in the outer regions \citep{Biretta1995, Biretta1999, Meyer2013}. Recent VLBI monitoring observations of the M87 jet have suggested that the jet becomes relativistic already at distances $\lesssim1,000\ R_g$ and the velocity field may be stratified \citep{Mertens2016, Park2019a}. Also, \cite{Park2019b} have revealed that the magnitude of Faraday rotation measures in the M87 jet systematically decreases with increasing distance from the black hole. They applied a simple analytical model of hot accretion flows \citep[e.g.,][]{YN2014} and found that the inferred pressure profile of an external confining medium is flat enough to collimate the jet \citep[e.g.,][]{Komissarov2009}. These observations are consistent with the theoretical picture of AGN jet collimation and acceleration.

The detailed view of jet acceleration and collimation processes provided by the extensive observations of M87 has triggered many VLBI observations aiming at finding the ACZs in other radio-loud AGNs. These observations have shown that a jet structural transition is common \citep[e.g.,][]{Tseng2016, Akiyama2018, Hada2018, Nakahara2018, Nakahara2020, Kovalev2020}, while there is a diversity in the jet geometries before and after the transitions \citep[e.g.,][]{Nakahara2020}, in the transition locations \citep[e.g.,][]{Hada2018, Nakahara2018, Nakahara2020}, and there are a few cases with no clear indication of a jet structural transition \citep{Giovannini2018, Nakahara2019}. On the other hand, deriving jet acceleration profiles in the collimation zones has been challenging. This is mainly because the brightness distributions of the jets in nearby radio galaxies are usually smooth, unlike distant blazars which show the jets consisting of several distinct knots, and a robust jet kinematic analysis is allowed only when dense VLBI monitoring data observed with a high-resolution and at a high-cadence are available \citep[see][for a related discussion]{Park2019a}. Jet collimation and acceleration from non-relativistic to relativistic speeds occurring in the same region, to our knowledge, have been observed only in M87, Cygnus A \citep{Boccardi2016b}, and the $\gamma$-ray emitting narrow-line Seyfert 1 galaxy 1H 0323+342 \citep{Hada2018}.

The nearby giant elliptical galaxy NGC 315, at a redshift of 0.01648 \citep{Trager2000} which corresponds to a scale of 0.348 kpc arcsec$^{-1}$ for our adopted cosmology ($H_0 = 67.4\ {\rm km\ s^{-1}}$, $\Omega_m = 0.315$, \citealt{Planck2020}), is a good laboratory for studying jet collimation and acceleration. It hosts a Fanaroff-Riley Class I \citep[FR I,][]{FR1974} radio source whose two-sided jets extend out to several hundred kpc from the galaxy \citep[e.g.,][see also Figure~\ref{fig:images}]{Bridle1976, Laing2006}. Its large black hole mass of $1.6\times10^9\ M_{\odot}$, estimated from the stellar velocity dispersion of $\sigma_v\approx350 {\rm\ km\ s^{-1}}$ \citep{Faber1989, Ene2020} and the latest $M_{\rm BH}$-$\sigma_v$ relation \citep{Sexton2019}, makes it easier to resolve the putative jet collimation and acceleration region of this source.

The jets in NGC 315 have been extensively explored on kpc-scales. \cite{Canvin2005} applied an analytical model which assumes that apparent asymmetries between an approaching and a receding jet in total intensity and linear polarization are caused by relativistic aberration, originally developed by \cite{LB2002}, to deep, high-resolution Very Large Array (VLA) observations at 5 GHz. They constrained the jet viewing angle of NGC 315 to be $\theta\approx38^\circ$ and found that the jets decelerate from $\beta\approx0.9$ to $\beta\approx0.4$ between 8 and 18 kpc from the nucleus, where $\beta$ is the intrinsic jet speed in units of the speed of light. They also found that the jet velocity field is stratified in such a way that the jet edge has a slower speed than the jet on-axis. \cite{LB2014} extended this study by using an improved model and more data\footnote{They also modelled nine other nearby FR I radio galaxies.} and obtained similar results with an updated jet viewing angle of $\theta=49.8^{\circ+0.5}_{-0.2}$. These studies showed that the jets have a region with high synchrotron emissivity (so-called "brightness flaring"), followed by a rapid expansion of the jets with the jet opening angles increasing with distance ("geometrical flaring"). \cite{Laing2006} investigated the spatial distribution of the spectral index of the jets and found a flatter spectrum at the edge than on-axis in the jet deceleration and downstream regions. This transverse spectral index structure might be associated with the electron energy acceleration due to velocity shear developed by the interaction between the jets and the surrounding medium. \cite{Worrall2003, Worrall2007} detected the approaching jet at X-rays out to $\approx10, 30$ arcsec from the nucleus, respecitvely, which also suggests the presence of distributed particle acceleration in the jet.

On pc-scales, \cite{Venturi1993} observed only the north-west jet from multifrequency global-VLBI observations, from which they derived an upper limit on the jet viewing angle of $\theta\lesssim58^\circ$. They also found flat and steep spectral indices for the core and the extended jet, respectively. \cite{Giovannini1994} combined the result of \cite{Venturi1993} with other indirect constraints on the jet speed and viewing angle based on, e.g., the correlation between the core and the total radio power in radio galaxies, and constrained the viewing angle to be in the range of $30^\circ<\theta<41^\circ$. \cite{Cotton1999} performed multi-epoch observations of NGC 315 with the very long baseline array (VLBA) at 5 and 8 GHz with an average interval between observations of about one year. They observed outward moving features in the jet, as well as the receding jet, from which they suggested systematic jet acceleration on pc-scales.

The relatively large jet viewing angle of NGC 315, $\theta\approx50^\circ$ constrained on kpc-scales and $\theta\gtrsim30^\circ$ on pc-scales in previous studies, means that its jets would be less affected by relativistic effects as compared with M87. Thus, investigating the NGC 315 jets helps us to have a unified view of AGN jet collimation and acceleration processes. In this paper, we report the result from our multifrequency VLBA observation of NGC 315, as well as archival high-resolution VLBI data and VLA data analysis, which constrains the jet collimation and acceleration profiles over a wide range of jet distances. We also present the result from our supplementary dense VLBI monitoring observations, which shows a complex jet kinematic structure and constrains the jet viewing angle on pc-scales.

The paper is organized as follows. We describe the observations, archival data we used, and data reduction in Section~\ref{sec:data}. We present the results of our analysis of jet structure, opacity, collimation, and acceleration in Section~\ref{sec:results}. The results are discussed in Section~\ref{sec:discussion}. We summarize our findings and conclude in Section~\ref{sec:summary}.

\section{Observations and data reduction} \label{sec:data}
\subsection{Multifrequency VLBA observation}
\label{sec:vlba}
We observed NGC 315 with the VLBA on 2020 Jan 05 at frequencies of 1.548, 2.284, 4.980, 8.416, 15.256, 22.220, and 43.120 GHz. All ten VLBA stations successfully participated in the observation. The data were recorded in both right and left-hand circular polarizations with two-bit quantization in eight baseband channels (also often called intermediate frequencies; IFs), using the polyphase filterbank (PFB) observing system, at a recording rate of 2 Gbps, yielding a total bandwidth of 256 MHz for each polarization. The total observation time is 12 hours. The on-source time on our target is $\approx40$ minutes at 1.5--8.4 GHz, and $\approx50, 80, 100$ minutes at 15.3, 22.2, and 43.1 GHz, respectively. The weather condition was good and no major technical issue occurred during the observation at all stations. We summarize the basic properties of our data in Table~\ref{tab:data}.

We performed a standard data reduction with the NRAO's Astronomical Image Processing System (AIPS; \citealt{Greisen2003}, see, e.g., Section C in the AIPS cookbook\footnote{\url{http://www.aips.nrao.edu/CookHTML/CookBook.html}}). We updated the Earth Orientation Parameters using more accurate parameters that are available after the data correlation, taken from NASA CDDIS. We corrected the dispersive delays caused by the ionosphere by using the GPS models of the electron content in the ionosphere using the procedure VLBATECR in AIPS. The sampler voltage offsets were corrected by using the autocorrelation spectra. We removed the instrumental delay residuals and phase offsets between IFs by performing a "manual phase-cal" using a scan of bright calibrators such as 0133+476, 3C 84, BL Lac. The bandpass shapes of the cross-power spectra are calibrated for each IF by using calibrators. We corrected a possible offset of the autocorrelation amplitudes from unity caused by the bandpass calibration. A priori amplitude calibration was done by using the antenna gain curves and system temperatures with an atmospheric opacity correction. The antenna parallactic angles were corrected. We performed global fringe fitting \citep{SC1983} for each IF using a solution interval of ten seconds\footnote{This solution interval is shorter than the typical coherence time expected at the frequencies of our interest. The main purpose of using relatively short solution intervals was to capture possible rapid phase variations caused by atmospheric fluctuations on short timescales, especially at low source elevations. We repeated data reduction using longer solution intervals (one minute at $\lesssim8.4$ GHz and 30 seconds at $\gtrsim15.3$ GHz) and found that our results are robust against the solution intervals.}, using a point-source model. We found that the fringe detection rates for the antennas comprising very long baselines, which are HN (Hancock), MK (Mauna Kea), and SC (Saint Croix) stations, at high ($\gtrsim15$ GHz) frequencies are relatively low with this parameter setup. We combined all the IFs and increase the solution interval to 20 seconds in those cases, achieving high ($\gtrsim95$\%) detection rates for all stations and frequencies except at 43 GHz (see Section~\ref{sec:hsa}). The remaining instrumental delay between polarizations at the reference antenna was corrected by using a scan on bright calibrators. The data were averaged over the channels within each IF and in time over ten seconds. We performed CLEAN and phase/amplitude self-calibration iteratively with the Caltech Difmap package \citep{Shepherd1997} and produced source images.

\begin{deluxetable*}{cccccccc}
\tablecaption{Multifrequency Observations and Data of NGC 315\label{tab:data}}
\tablewidth{0pt}
\tablehead{
Proj. Code & Obs. Date & \colhead{Freq.} & \colhead{Beam Size} & \colhead{$I_{\rm p}$} & \colhead{$I_{\rm rms}$} & \colhead{$\theta_{\rm c, modelfit}$}  & \colhead{$\theta_{\rm c, JMFIT}$} \\
&& \colhead{(GHZ)} & \colhead{(mas $\times$ mas, degree)} & \colhead{(Jy/B)} & \colhead{(mJy/B)} & (mas) & (mas) \\
&&&(a)&(b)&(c)&(d)&(e)
}
\startdata
\hline
\multicolumn{8}{c}{VLBA}\\
\hline
\multirow{ 6}{*}{BP243} & \multirow{ 6}{*}{2020 Jan 05} & 1.548 & $9.78\times6.22$, -4.27 & 0.250 & 0.048 & $0.753\pm0.022|_{\rm stat}\pm0.086|_{\rm sys}$ & $0.624\pm0.035|_{\rm stat}\pm0.086|_{\rm sys}$ \\
&& 2.284 & $6.56\times4.35$, -9.15 & 0.277 & 0.136 & $0.645\pm0.040|_{\rm stat}\pm0.031|_{\rm sys}$ & $0.571\pm0.043|_{\rm stat}\pm0.031|_{\rm sys}$ \\
&& 4.980 & $2.85\times1.97$, -5.12 & 0.378 & 0.037 & $0.280\pm0.002|_{\rm stat}\pm0.099|_{\rm sys}$ & $0.140\pm0.007|_{\rm stat}\pm0.099|_{\rm sys}$ \\
&& 8.416 & $1.75\times1.17$, -11.58 & 0.389 & 0.053 & $0.131\pm0.004|_{\rm stat}\pm0.010|_{\rm sys}$ & $0.145\pm0.004|_{\rm stat}\pm0.010|_{\rm sys}$ \\
&& 15.256 & $0.97\times0.68$, -3.40 & 0.285 & 0.068 & $0.104\pm0.004|_{\rm stat}\pm0.000|_{\rm sys}$ &  $0.108\pm0.002|_{\rm stat}\pm0.000|_{\rm sys}$ \\
&& 22.220 & $0.77\times0.53$, -5.66 & 0.264 & 0.073 & $0.095\pm0.003|_{\rm stat}\pm0.000|_{\rm sys}$ &  $0.097\pm0.002|_{\rm stat}\pm0.000|_{\rm sys}$ \\
\hline
\multicolumn{8}{c}{HSA}\\
\hline
BG170B$^1$ & 2008 Feb 03 & 43.212 & $0.36\times0.17$, -13.46 & 0.205 & 0.142 & $0.018\pm0.003|_{\rm stat}\pm0.017|_{\rm sys}$ & $0.042\pm0.002|_{\rm stat}\pm0.017|_{\rm sys}$ \\
\hline
\multicolumn{8}{c}{VLA}\\
\hline
AL538 & 2001 Mar 10 & 1.365 & $6021\times5425$, 89.63 & 0.429 & 0.054 & $-$ & $-$ \\
AC476 & 1996 Nov 02 & 4.860 & $516\times460$, 84.56 & 0.706 & 0.019 & $-$ & $-$ \\
\enddata
\tablecomments{(a) Major axis, minor axis, and position angle of the synthesized beam under the natural weighting of the data. (b) Map peak intensity in units of Jy per beam. (c) Image rms-noise measured in the off-source regions in units of mJy per beam. $I_{\rm rms}$ is relatively high at 2.3 GHz due to (i) the signals in some IFs being blocked by the filters installed in many stations and (ii) severe radio frequency interference in this band. (d) FWHM of core \texttt{modelfit} elliptical Gaussian component along the perpendicular direction to the jet axis. (e) FWHM of core JMFIT elliptical Gaussian component along the perpendicular direction to the jet axis. Statistical and systematic errors in the fitted core sizes in (d) and (e) are noted. $^1$Participating stations: the VLBA (except Mauna Kea station), the Effelsberg 100m station, the Green Bank Telescope, and the phased-up VLA.}
\end{deluxetable*}

\subsection{Archival HSA data at 43 GHz}
\label{sec:hsa}
We found that our VLBA image at 43 GHz is mostly dominated by the compact core, which is not well-suited for investigating the jet collimation profile, the jet-to-counterjet brightness ratio, and so on. The fringe detection rate for this data was at most $\lesssim50, 30, 10\%$ for HN, MK, and SC stations, respectively, even though we tried to combine all IFs and increase the solution interval to a few minutes for the global fringe fitting procedure. This is presumably because the source is fainter, the data is more sensitive to the weather condition, the extended source structure is more easily resolved out, and it suffers more from the antenna pointing offsets than at other frequencies. We instead used an archival High Sensitivity Array (HSA) data observed at 43.212 GHz on 2008 Feb 03 to investigate the jets at a short distance from the black hole. The Effelsberg 100m station, the Green Bank Telescope, and the phased-up VLA\footnote{We found that the phased-up VLA baselines have poor sensitivities than our expectation. We suspect that the phasing efficiency was not very high for this observation for some reasons. The Effelsberg 100m telescope and the Green Bank Telescope baselines show high sensitivities, as expected.} participated in the observation, which enables to achieve a higher sensitivity than our VLBA-only 43 GHz observation. We found that the image rms-noise of the HSA 43 GHz data is lower than the VLBA 43 GHz data by more than a factor of four (see Figure~\ref{fig:map_q2} for comparison of the maps). Therefore, we decided to use this data for our analysis by assuming that there was no significant change in the jet collimation and acceleration properties between the observations (over $\approx12$ years). We calibrated this archival data and imaged as described in Section~\ref{sec:vlba}.

\subsection{Archival VLA data at 1.4 and 4.9 GHz}
We analyzed two historical Very Large Array (VLA) data available in the VLA archive. One is performed on 2001 Mar 10 at 1.365 GHz in the B-configuration and the other on 1996 Nov 02 at 4.860 GHz in the A-configuration. These data were analyzed and presented in previous studies of NGC 315 \citep{Canvin2005, Laing2006, Worrall2007, LB2014}. The data calibration was done in a standard manner with AIPS. We performed imaging and self-calibration in Difmap. 

\subsection{Notes on linear polarization of VLBA/HSA data}
We corrected the instrumental polarization of the VLBA and HSA data by using GPCAL, which is a new pipeline designed for achieving a high calibration accuracy \citep{Park2020}. GPCAL allows (i) to use multiple calibrators simultaneously and (ii) to take into account linear polarization structures of calibrators through so-called "instrumental polarization self-calibration", without being limited by using the conventional way of assuming the linear polarization structures being proportional to the total intensity structures. We applied GPCAL to two calibrators, 3C 84 and BL Lac, for the VLBA data and a single calibrator, 0133+476, for the HSA 43 GHz data, by implementing ten iterations of instrumental polarization self-calibration. However, we could not find any significant linear polarization in the NGC 315 jets at any frequency. This suggests two possibilities; 1. the NGC 315 jets are strongly depolarizaed on pc-scales by complex jet magnetic field structures or a turbulent external Faraday rotating screen \citep[e.g.,][]{Sokoloff1998}, or 2. the NGC 315 jets have very high rotation measures and the polarization signals are cancelled out when we averaged the data over frequency channels \citep[e.g.,][]{Bower2017}. The latter possibility requires calibration of the frequency-dependent instrumental polarization, which is beyond the scope of the present study. We plan to explore it in our future studies.

\section{Analysis and Results}
\label{sec:results}
\begin{figure*}[t!]
\centering
\includegraphics[width = 1.0\textwidth]{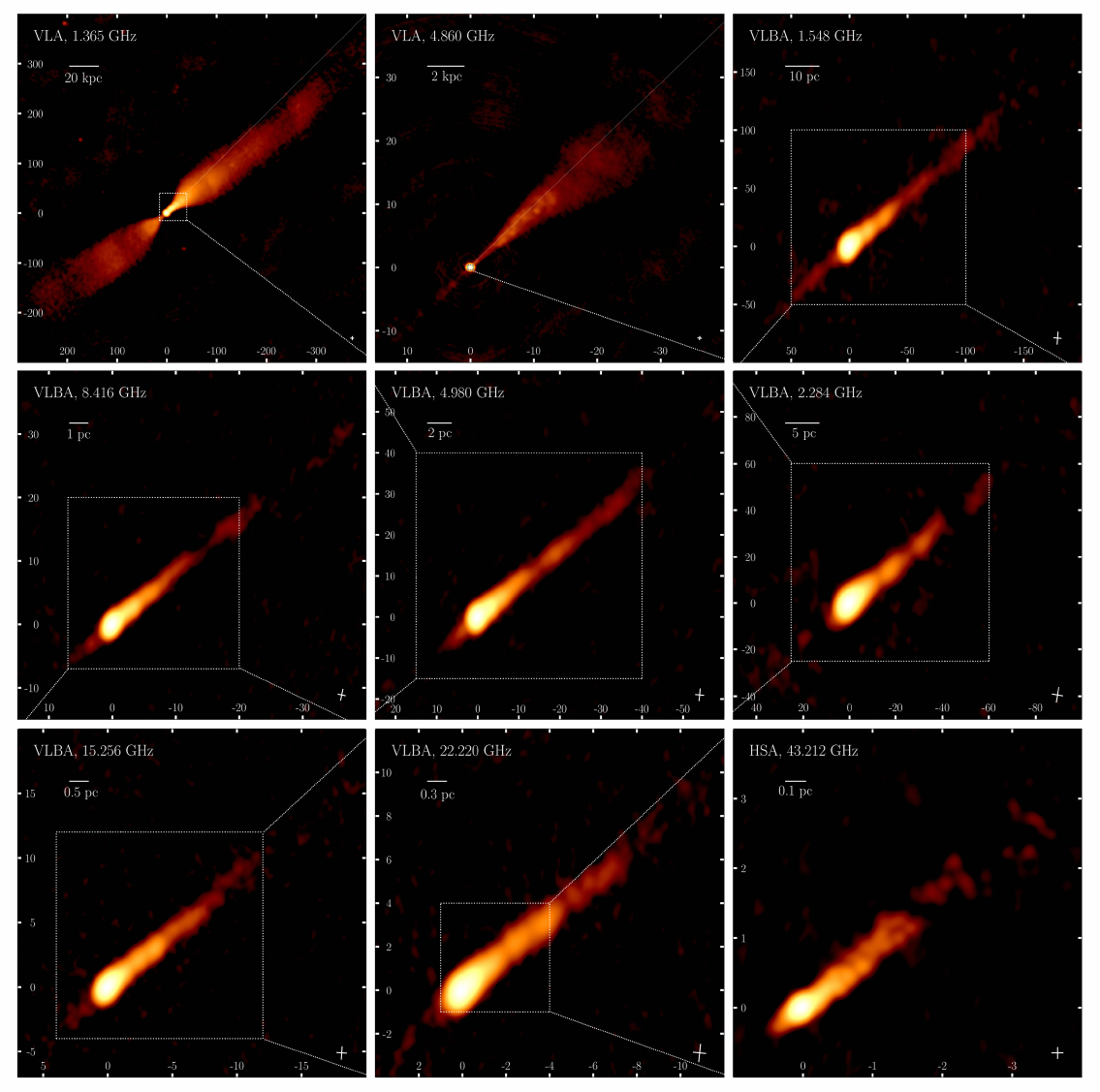}
\caption{Images of the jets of NGC 315 reconstructed from archival VLA data at 1.365 and 4.860 GHz, our VLBA observations at 1.5--22.2 GHz, and archival HSA data at 43.212 GHz. The physical scales corresponding to the white horizontal sticks are noted. The ticks on the axes are in units of arcsec for the VLA images and of mas for the VLBA/HSA images. The white dashed rectangles show the scales of the next maps in the order from left to right for the top and bottom rows and from right to left for the middle row. CLEAN models are restored with the naturally-weighted synthesized beams to produce the images except for the HSA 43 GHz image for which we restored with a circular beam with the minor-axis of the synthesized beam. This is for illustrating an indication of limb-brightening observed only at that frequency. The restoring beam FWHMs are indicated with the white cross in the lower right of each map. On kpc-scales, prominent two-sided jets with a morphology of rapid expansion near the central core and recollimation in the outer region are observed. On pc-scales, the south-east jet is much fainter and less extended than the north-west jet but it is detected at all frequencies. \label{fig:images}}
\end{figure*}

We present the images of NGC 315 in Figure~\ref{fig:images}. The rapidly expanding jet region (the "geometrical flaring" region, \citealt{Canvin2005, LB2014}), followed by the re-collimation of the jet, is shown in both the approaching (the jet in the north west direction) and the receding (in the south east direction) jets in the VLA 1.4 GHz image (we will hereafter use the terminology of the jet and counterjet instead of the approaching and receding jets, respectively). The conically expanding jet and the tenuous counterjet are present in the VLA 4.9 GHz image. These are consistent with the previous studies using the same data\footnote{We note that \citealt{Laing2006} showed the jets extending up to larger distances compared to our images thanks to the high-sensitivity achieved by combining multiple data sets. Our VLA images are enough to constrain the jet collimation profile on kpc-scales, which is the main purpose to re-analyze the VLA data in the present study.} \cite[e.g.,][]{Laing2006}. The overall jet morphology on pc-scales is similar to kpc-scales. The images are dominated by the jet but the counterjet could be imaged at all VLBA/HSA frequencies. Notably, the 43.2 GHz image shows an indication of limb-brightening in the outer part of the jet, which will be discussed later (Section~\ref{sec:limb_brightening}).

\begin{figure}[t!]
\centering
\includegraphics[width = 0.49\textwidth]{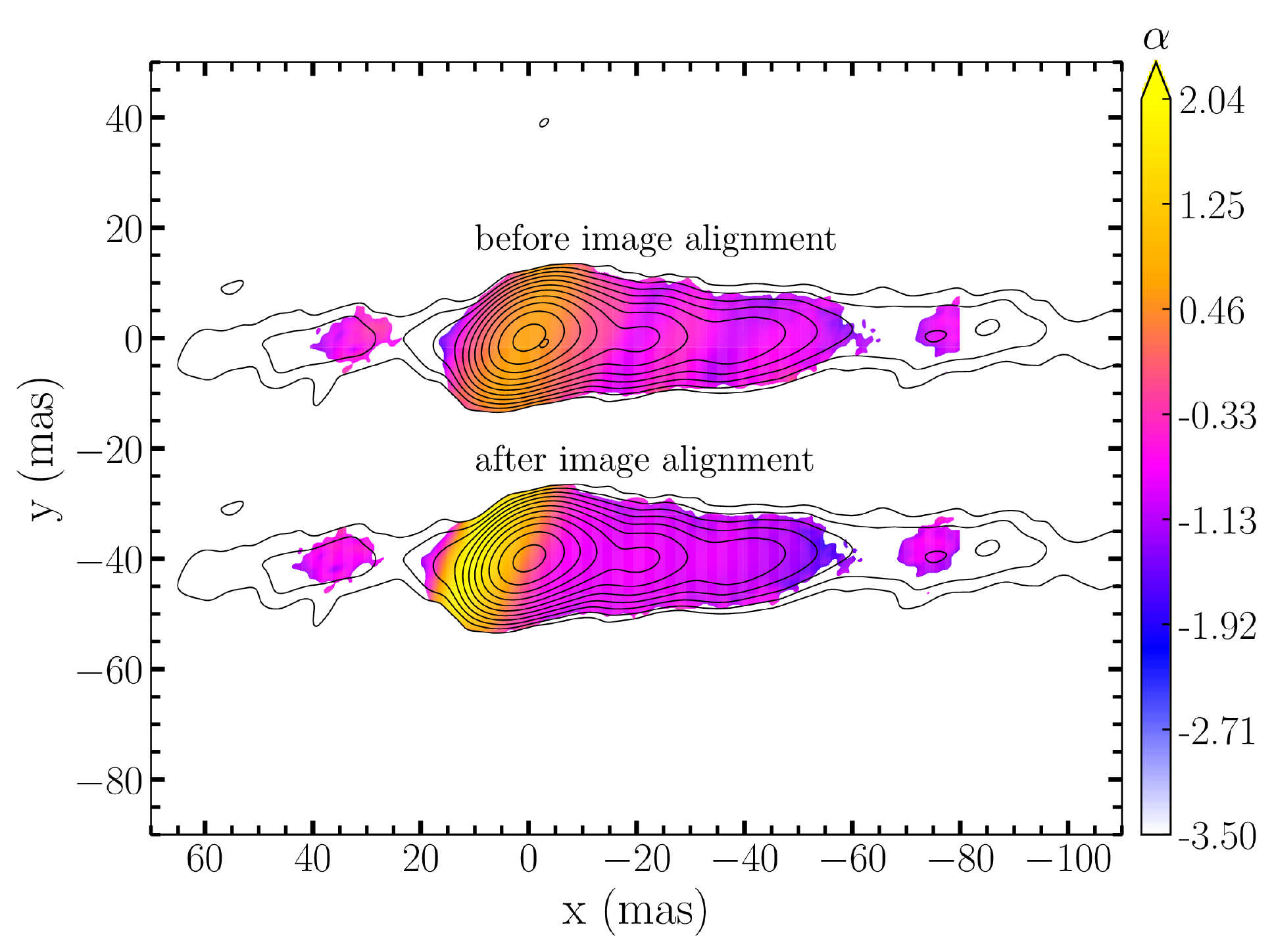}
\caption{Color maps of the distributions of spectral index $\alpha$ (defined as $I_\nu \propto \nu^\alpha$) between 1.5 and 5.0 GHz overlaid on contours of the total intensity at 1.5 GHz before (top) and after (bottom) alignment of the images (Section~\ref{sec:coreshift}). The maps are rotated clockwise by $40^\circ$ and the jet axis is aligned with the horizontal axis. The low-$\alpha$ region upstream of the core (on the counterjet side) and the oscillatory pattern along the jet in the upper map disappear after the image alignment. \label{fig:spix_example}}
\end{figure}

\subsection{Core-shift}
\label{sec:coreshift}
AGN jets are synchrotron emitters and accompany synchrotron self-absorption. Significant absorption usually occurs near the jet base, where the electron density and the magnetic field strength are expected to be high. Thus, the observed base of jet, so-called the "core", is separated from the physical jet base. Since synchrotron self-absorption depends on frequency \citep[e.g.,][]{RL1979}, the observed core-position is also expected to change with frequency. This is known as the "core-shift" effect \citep[e.g.,][]{Konigl1981, Lobanov1998, Hirotani2005} and has been observed in the jets of many blazars \citep[e.g.,][]{Lobanov1998, OG2009, Pushkarev2012, Fromm2013, Hada2018} and radio galaxies \citep[e.g.,][]{Sudou2000, Hada2011, Hada2013, Haga2015}. An accurate measurement of core-shift at multiple frequencies allows us to infer the location of the central engine \citep{Hada2011}, which is crucial to obtain proper "jet distance".

We define the position of "apparent" core as the brightest pixel in the core region in the image. The apparent core may be separated from the physical core, i.e., the $\tau=1$ surface, where $\tau$ is the synchrotron optical depth. This is because of the finite beam size, which can introduce an additional shift of the apparent core position to the extended jet side due to the blending of the physical core and the extended jet \citep[e.g.,][]{Hada2011}. The core-shift effect refers to a shift of the physical core to the jet base at a higher frequency, which occurs due to the physical properties of the source. The separation of the apparent core from the physical core is just due to the limited angular resolution. We will derive the physical core-shift effect below, which is important to constrain the location of the jet base. We constrain the position of the apparent core in Appendix~\ref{appendix:core_identification}, which is necessary to properly convert the apparent jet distance, i.e., the distance between a pixel in the image from the apparent core, to the physical jet distance, i.e., the distance between a pixel from the inferred jet base.

We investigated the core-shift effect in the jets by performing two-dimensional cross-correlation of the optically thin jet emission in the VLBA images at different frequencies \citep{CG2008}. This method is suitable for our analysis because the prominent, extended jet structures could be obtained at all frequencies (Figure~\ref{fig:images}). For each frequency pair, we used the same image size and pixel size. The pixel size of $1/20$ of the minor axis of the synthesized beam at the higher frequency was used \citep{Pushkarev2012, Fromm2013}. We restored the CLEAN models of each frequency pair with the synthesized beam of the low frequency image. We aligned the restored images so that the apparent cores are located at the map origin. We rotated the images clockwise by $40^\circ$ (the jet axis is aligned with the horizontal axis of the map with this rotation) and used the regions separated from the apparent cores by more than the major axis of the convolving beam along the x-axis to avoid the optically thick core in the calculation. We did not use the counterjet because it is much weaker and its length is more different between frequencies as compared with the jet (Figure~\ref{fig:images}). We computed the normalized cross-correlation coefficient between the images as
\begin{equation}
    \rho_{xy} = \frac{\sum_{i=1}^n\sum_{j=1}^n(I_{\nu_1,ij} - \overline{I_{\nu_1}})(I_{\nu_2,ij} - \overline{I_{\nu_2}})}{\sqrt{\sum_{i=1}^n\sum_{j=1}^n(I_{\nu_1,ij} - \overline{I_{\nu_1}})^2\sum_{i=1}^n\sum_{j=1}^n(I_{\nu_2,ij} - \overline{I_{\nu_2}})^2}},
\end{equation}
where $n$ is the number of pixels in each direction, $I_{\nu_1,ij}$ and $I_{\nu_2,ij}$ are the intensities for the maps at frequencies $\nu_1$ and $\nu_2$ at $i$-th and $j$-th pixels along the x and y directions in the rotated maps, respectively, and $\overline{I_{\nu_1}}$ and $\overline{I_{\nu_2}}$ are the mean values of the intensities over the region analyzed. We shifted one of the images along the x and y axes by up to 80--160 pixels and searched for the amount of shift that gives us the maximum correlation coefficients. The maximum coefficients are larger than 0.97 in all the considered frequency pairs. We derived the uncertainties in the core-shifts by investigating the change in a radial spectral index profile in the optically thin jet region when introducing an additional shift to one of the pair images along the jet direction (Appendix~\ref{appendix:2dcc_error}).

We did not include the result of the frequency pairs that show very different lengths of the jet; the maximum correlation coefficients are lower than 0.9 in those cases and they are not considered robust. Also, the jet structure at 43 GHz appears to be quite different from those at other frequencies (Figure~\ref{fig:contours_shift}). This could be due to the long-term evolution of the jet brightness distribution over $\approx12$ years (Table~\ref{tab:data}), which prevents us from deriving a reliable core-shift at 43 GHz\footnote{We tried to obtain the core-shift between 22 and 43 GHz despite the apparent difference in the jet brightness distributions and obtained the maximum cross-correlation coefficient at the shift of the higher frequency map by zero and 0.08 mas along the jet longitudial and transverse directions, respectively. This results in a weird spectral index map and is very different from the trend we consistently see for all the other frequency pairs.}. Thus, we decided to obtain the 43 GHz core position by extrapolating from the core positions constrained at lower frequencies. The core-shifts for different frequency pairs are summarized in Table~\ref{tab:coreshift}.

\begin{deluxetable}{cccc}
\tablecaption{Core-shifts between each pair of frequencies derived from the VLBA data \label{tab:coreshift}}
\tablewidth{0pt}
\tablehead{
\colhead{$\nu_1$} & \colhead{$\nu_2$} & \colhead{$\Delta r$} & \colhead{Angle} \\
 (GHz) & (GHz) & (mas) & ($^\circ$)
}
\startdata
1.548 & 2.284 & 1.090 (0.11) $\pm$ 0.663 & -50.0 \\
1.548 & 4.980 & 3.367 (0.34) $\pm$ 0.991 & -51.7 \\
1.548 & 8.416 & 4.120 (0.42) $\pm$ 0.870 & -48.4 \\
2.284 & 4.980 & 2.081 (0.32) $\pm$ 0.843 & -52.7 \\
2.284 & 8.416 & 2.960 (0.45) $\pm$ 0.581 & -47.8 \\
2.284 & 15.256 & 3.067 (0.47) $\pm$ 0.714 & -46.2 \\
4.980 & 8.416 & 0.296 (0.10) $\pm$ 0.349 & -38.7 \\
4.980 & 15.256 & 0.477 (0.17) $\pm$ 0.340 & -45.9 \\
4.980 & 22.220 & 0.523 (0.18) $\pm$ 0.390 & -44.3 \\
8.416 & 15.256 & 0.238 (0.14) $\pm$ 0.171 & -50.0 \\
8.416 & 22.220 & 0.313 (0.18) $\pm$ 0.247 & -45.2 \\
15.256 & 22.220 & 0.104 (0.11) $\pm$ 0.079 & -50.0 \\
\enddata
\tablecomments{Magnitude and position angle of core-shift measured between each pair of frequencies, derived by performing two-dimensional cross-correlation of the optically thin jet emission in the VLBA images. The values in the parentheses are the ratios of the core-shift magnitudes to the major axes of the restoring beams.}
\end{deluxetable}

We illustrate the impact of image alignment on the spectral index distribution of the jets in Figure~\ref{fig:spix_example}. The spectral index map between 1.5 and 5.0 GHz before the alignment shows an optically thin region in the upstream of the core (on the counterjet side) and an oscillatory pattern in the downstream jet. These features disappear after the alignment, which is consistent with the previous core-shift results on various sources (e.g., \citealt{CG2008}). 

\begin{figure}[t!]
\centering
\includegraphics[width = 0.49\textwidth]{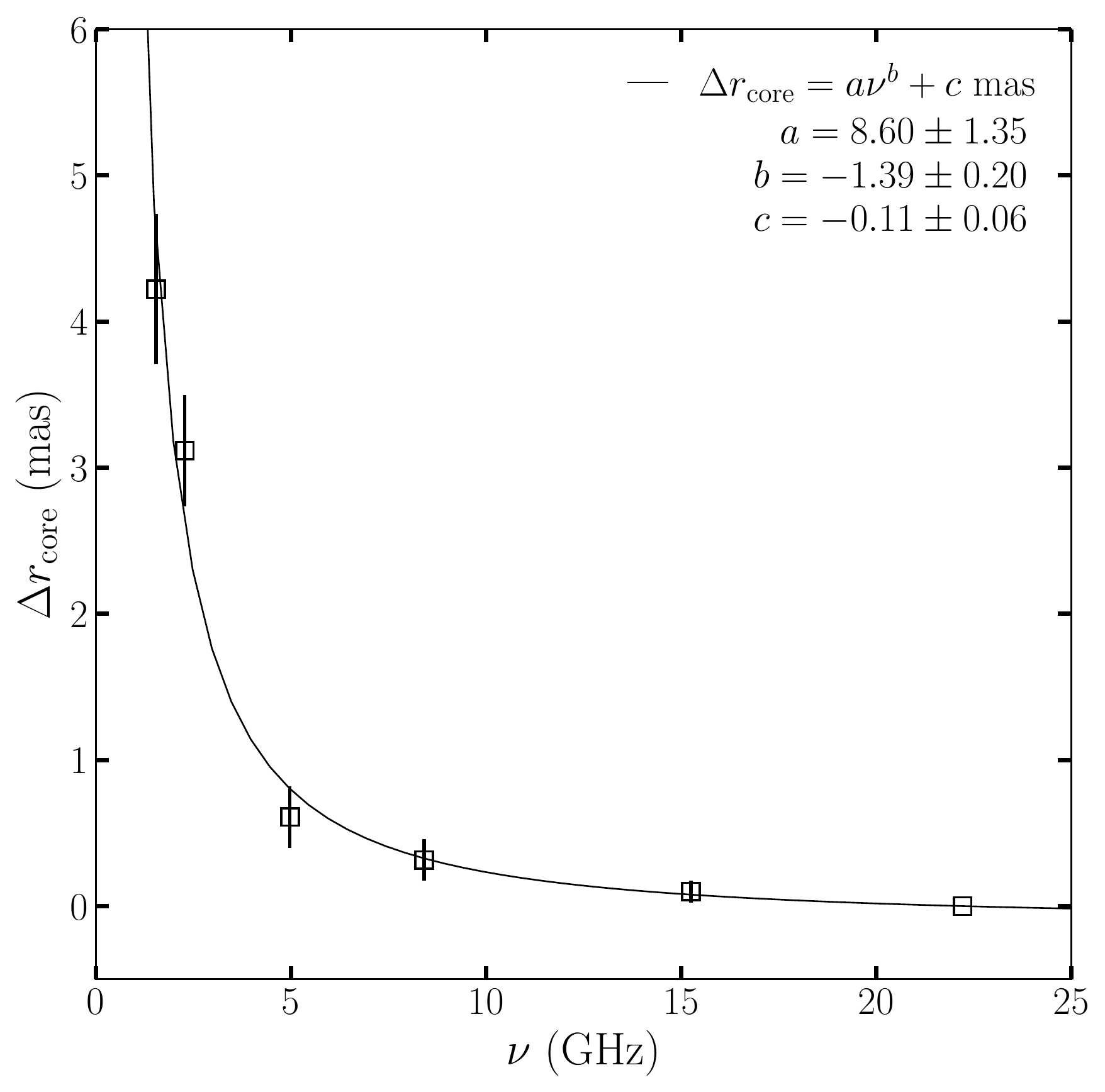}
\caption{Core position as a function of frequency. The physical core positions relative to the 22 GHz core are derived from a least-square fitting to the observed core-shifts between many frequency pairs (Table~\ref{tab:coreshift}). The black solid line shows the best-fit power-law function to the data points and the best-fit parameters of the function are noted on the top right. \label{fig:coreshift}}
\end{figure}

\begin{deluxetable*}{cccccc}
\tablecaption{Best-fit core-positions \label{tab:fitted_coreshift}}
\tablewidth{0pt}
\tablehead{
 & \colhead{1.548 GHz} & \colhead{2.284 GHz} & \colhead{4.980 GHz} & \colhead{8.416 GHz} & \colhead{15.256 GHz}
}
\startdata
$\Delta r_{\rm core}$ & $4.222\pm0.514$ & $3.117\pm0.381$ & $0.610\pm0.211$ & $0.315\pm0.141$ & $0.100\pm0.076$
\enddata
\tablecomments{Best-fit physical core positions relative to the 22 GHz core in units of mas.}
\end{deluxetable*}

Our core-shift estimates are derived after aligning the images at different frequencies convolved with the same beam based on the apparent core positions. Thus, we expect that the separations of the apparent cores from the physical cores for each frequency pair would be similar and the core-shift estimates would represent the relative distances between the physical cores at different frequencies. We constrain the physical core position at each frequency with respect to the core at the reference frequency (assumed to be 22 GHz here) by using the core-shifts for different frequency pairs. We parametrized the relative position between the physical core at a certain frequency and the 22 GHz core and performed a least-square fitting. The best-fit core positions would reproduce the observed core-shifts for all frequency pairs well. We present the best-fit core positions and the uncertainties in Table~\ref{tab:fitted_coreshift}. 

In Figure~\ref{fig:coreshift}, we present the physical core positions relative to the 22 GHz core as a function of frequency. The core position offsets systematically decrease with increasing frequency. We fit a power-law function of $\Delta r_{\rm core} = a\nu^b + c$, where $\Delta r_{\rm core}$ is the core position offset and $\nu$ the frequency, and find the best-fit values of $a=8.60\pm1.35$, $b=-1.39\pm0.20$, and $c=-0.11\pm0.06$. This result allows us to infer the location of the jet base (by taking $\nu \xrightarrow{} \infty$), which is at $\approx0.11$ mas upstream of the 22 GHz physical core.

We note that the observed core-shift is very large up to $\approx4.2$ mas at 1.5 GHz. A similar large core-shift was observed in the nearby radio galaxy NGC 4261 \citep{Haga2015}. We illustrate the large core-shifts in NGC 315 in Figure~\ref{fig:contours_shift}. The images are registered with respect to the inferred jet base position by using the constraints on the apparent core positions (the physical core-shift plus the additional shift of the apparent core from the physical core, see Appendix~\ref{appendix:core_identification}). The jet emission appears to have several knotty (or re-brightened) regions, notably at $\approx3.5$, 5, 25, 44 mas, and their positions are well aligned at different frequencies. However, the apparent core positions change a lot with frequency. We also present the spectral index distributions between adjacent frequencies in colors on top of the contours. Interestingly, the spectral indices become maximum near the location of the inferred jet base, reaching up to $\alpha\approx2.5$ between the lowest frequency pair, which is expected for synchrotron self-absorbed spectrum at low frequencies \citep{RL1979}. This observation is consistent with the expectation that AGN jet emission near the jet base is strongly absorbed by the synchrotron self-absorption process \citep[e.g.,][]{Konigl1981, Lobanov1998, Hirotani2005}.

\begin{figure*}[t!]
\centering
\includegraphics[width = 1.0\textwidth]{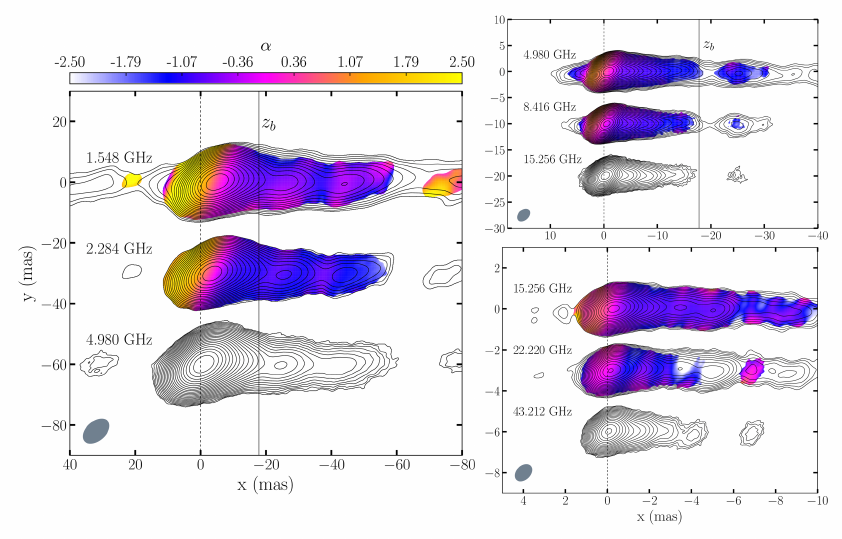}
\caption{Contours show total intensity distributions at different frequencies convolved with the beam at the lowest frequency in each panel, shown on the bottom left. Colors show spectral index distributions between adjacent frequencies. For example, the color on top of the 1.548 GHz contour shows $\alpha$ between 1.548 and 2.284 GHz, convolved with the 1.548 GHz beam. All the maps are rotated clockwise by $40^\circ$ and the jet axis is aligned with the horizontal axis. The contours are shifted along the negative x-axis using the constraints on the apparent core positions with respect to the inferred jet base. Thus, the black dashed vertical line at $x=0$ corresponds to the inferred jet base position. Interestingly, $\alpha$ reaches its maximum near the inferred jet base location between all the frequency pairs, which suggests the presence of severe synchrotron self-absorption near the jet base, as expected for AGN jets \citep[e.g.,][]{Konigl1981}. The black solid vertical line marked as $z_b$ shows the location of the jet collimation break (Figure~\ref{fig:jet_radius}). There are a few re-brightened regions which appear as knots at $\approx-3.5, -5, -25, -44$ mas. The positions of these regions are consistent between the maps after the image registration, except at 43 GHz, which was observed $\approx12$ years earlier than the maps at other frequencies. This implies that there might be a long-term evolution of the jet brightness distribution. \label{fig:contours_shift}}
\end{figure*}

\subsection{Jet collimation profile}
\label{sec:collimation}

We derive jet radius\footnote{The jet radius is assumed to be half the jet width.} as a function of jet distance as follows. We restored the CLEAN model for each frequency with a circular beam having the size of the major axis of the synthesized beam to remove the effect of the restoring beam straightforwardly. We rotated the images clockwise by $40^\circ$, which makes the jet ridge well aligned with the x-axis of the maps (Figure~\ref{fig:contours_shift}). We obtained a transverse intensity profile (along the y-axis of the rotated maps) at each jet distance and fitted a Gaussian function to the profile. We subtracted the restoring beam full width at half maximum (FWHM) from the measured jet FWHM in quadrature to derive the intrinsic jet width. We regarded the jet width as robust only when (i) the amplitude of the fitted Gaussian function exceeds 15 times the off-source image rms-noise and (ii) the measured FWHM is larger than the restoring beam FWHM. We obtained the jet widths at distances separated from the apparent cores by more than the major axis beam sizes to avoid a potential complication originating from the convolution of the brightest core emission; we try to constrain the jet widths in the core regions by employing model fitting on the visibility and image domains (see below).

Although we derive the intrinsic jet width at each distance bin (with a size of the image pixel size), many of those measurements are not independent due to the finite beam size. We binned the jet widths in distance with a bin size of half the major axis beam size. We took the median value of the widths in each bin for a representative jet width and assumed 1/10 of the major axis beam size for an uncertainty of the width\footnote{We derive the jet width only when the peak intensity of the transverse Gaussian profile exceeds 15 times the image rms-noise. Therefore, our assumed errors are larger than the nominal, approximated errors in sizes of model components fitted to VLBI data given by $\sigma_d=d/{\rm SNR}$, where $d$ is the component size and SNR the signal-to-noise ratio \citep[e.g.,][]{Fomalont1999, Lee2008}.}. 

The fact that we have a good constraint on the core-shift of the jet indicates that we could also constrain the jet collimation profile on scales much smaller than the formal resolution limit. This can be achieved by assuming that the size of the core corresponds to the jet width and using the distance of the apparent core from the jet base (Section~\ref{sec:coreshift} and Appendix~\ref{appendix:core_identification}). This approach was taken in previous studies of jet collimation in M87 \citep{NA2013, Hada2013} and it turned out that the jet radii derived from the core sizes are indeed consistent with those from the transverse jet intensity profile analysis \citep{Hada2013, Nakamura2018}.

There are two widely used methods to extract the core size from VLBI data: one is to fit a two-dimensional elliptical Gaussian function in the core region of the image and the other is to fit a Gaussian "component" to the visibilities directly. We take both approaches in this study. We used the task JMFIT in AIPS for the former approach, which was used for a similar analysis in previous studies (e.g., \citealt{Hada2013, Hada2018}). JMFIT provides the intrinsic major and minor axes of the fitted Gaussians after subtracting the convolving beam sizes\footnote{We used the images convolved with the circular Gaussian beams, similar to the transverse intensity profile analysis.}. We used \texttt{modelfit} in Difmap for the second approach and fitted a single elliptical Gaussian component. We found that the offsets of the fitted positions from the apparent core positions are small but not negligible at some frequencies for both JMFIT and \texttt{modelfit} measurments, which were considered to calculate the distance of the core components from the jet base. We calculated the FWHMs along the direction perpendicular to the jet axis and regarded them as the jet widths in the core regions. Each software provides the statistical uncertainties of the parameters for the fits. However, we found that the two measurements at some frequencies deviate from each other more than the formal $1\sigma$ uncertainties, indicating that there might be some systematic uncertainties that could not be captured by the fitting process. In those cases, we estimated the systematic uncertainties which are assumed to be the same for both measurements at each frequency and can make the discrepancy between the measurements equal to the total $1\sigma$ uncertainty. The core widths from two methods and their uncertainties are presented in Table~\ref{tab:data}.

\begin{figure*}[t!]
\centering
\includegraphics[width = 0.85\textwidth]{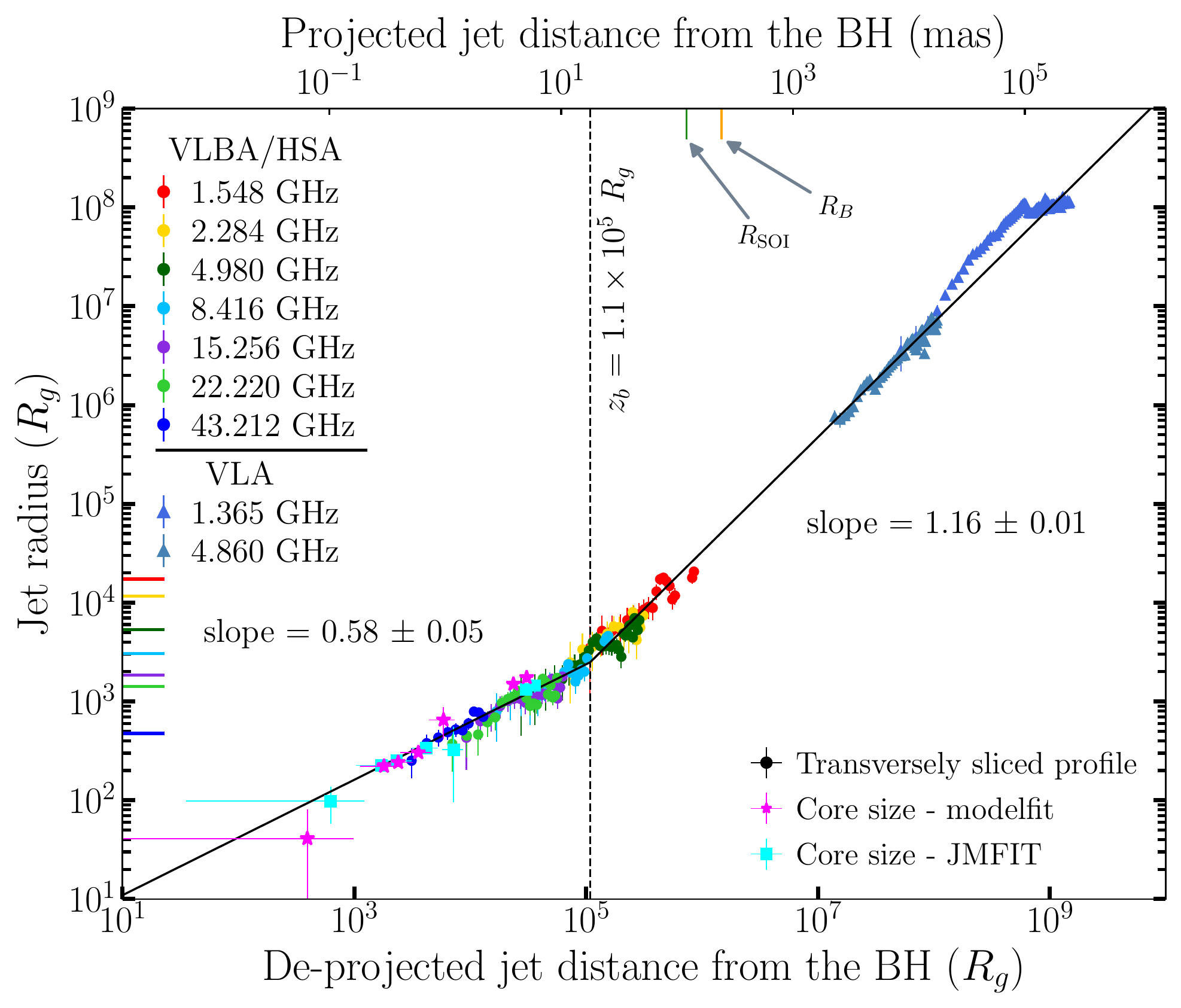}
\caption{Jet radius as a function of de-projected jet distance from the black hole in units of $R_g$. Filled circles and filled triangles are jet radii obtained from the VLBA/HSA and VLA data, respectively, by analyzing transverse jet intensity profiles. Magenta diamonds and cyan squares are jet radii at the cores constrained by \texttt{modelfit} in Difmap on the visibility domain and JMFIT in AIPS on the image domain, respectively. The best-fit broken power-law function is shown with the black solid line. The jet morphology is semi-parabolic (with a slope of $0.58\pm0.05$ in the logarithmic space) and conical/hyperbolic (with a slope of $1.16\pm0.01$) inside and outside of the transition distance of $z_b=(1.1\pm0.2)\times10^5\ R_g$ indicated by the vertical dashed line. The locations of the sphere of influence of the black hole gravity ($R_{\rm SOI}$) and the Bondi radius ($R_{B}$) are indicated by the green and orange solid ticks on the upper y-axis, respectively. The colored ticks on the left y-axis shows the $\rm FWHM/2$ of the synthesized beams along the direction transverse to the jet axis. They are comparable to the maximum observed jet radii except at 43 GHz, which may explain the reason for the observed limb-brightened feature only at that frequency. \label{fig:jet_radius}}
\end{figure*}

In Figure~\ref{fig:jet_radius}, We present the jet radius as a function of de-projected jet distance from the black hole. The jet distance we obtained in our analysis is with respect to the apparent core. We add the distance between the apparent core and the inferred jet base position to derive the physical jet distance from the black hole (Section~\ref{sec:coreshift} and Appendix~\ref{appendix:core_identification}). We used the black hole mass of $M_{\rm BH} = 1.6\times10^9 M_\odot$ (Section~\ref{sec:intro}) to obtain distance in units of $R_g$ and the viewing angle of $\theta = 49.8^\circ$ (Section~\ref{sec:velocity}) for de-projection. We find that the jet radii from multiple frequencies at a similar jet distance are consistent with each other within errors, which implies that we did not underestimate the jet width uncertainties. The jet radii measured from the HSA 43 GHz data are consistent with those from the VLBA data at other frequencies, suggesting that the jet collimation profile is not significantly affected by the possible long-term evolution of the jet brightness distribution (Figure~\ref{fig:contours_shift}). The geometrical flaring of the jet and the jet recollimation are observed at distances of $\approx10^8$--$10^9\ R_g$, as already shown in the previous studies \citep{Canvin2005, LB2014}. We will investigate the origin of the recollimation in more details in a forthcoming paper. Before the jet reaches the flaring region, the jet appears to expand by following the same power-law function of jet shape, i.e., the same slope in the logarithmic space, and this trend continues to the mas-scale. However, the slope becomes flatter at a certain distance around $\approx10^5\ R_g$.

We fitted a broken power-law function to the observed jet radii. The function is given in the form of 
\begin{eqnarray}
R = az + b &, & \quad z<z_b \nonumber \\
R = cz + (a - c)z_b + b &, & \quad z\gtrsim z_b
\end{eqnarray}
in the logarithmic space, where $R$ and $z$ denote the jet radius and distance, respectively. This form guarantees that the two functions are connected at the break location at $z=z_b$. We did not use the VLA 1.4 GHz data for fitting due to the complex evolution of the jet geometry with distance. We considered the uncertainties in both the jet distance (from the core-shift uncertainties) and the jet radius in the fitting; the jet distance uncertainties are important only for the measurements from the core size analysis. We tested fitting with a single power-law function and obtained the reduced chi-square of $\chi^2_{\rm red}\approx9.9$, which is significantly larger than $\chi^2_{\rm red}\approx5.9$ obtained from the broken power-law fitting\footnote{We note that the large $\chi^2_{\rm red}$ is mostly contributed by the VLA 5 GHz data. We obtain $\chi^2_{\rm red}\approx0.95$ when we exclude the VLA data for calculation of $\chi^2_{\rm red}$. This is because the jet is well resolved on kpc-scales and shows oscillations in jet radius with respect to the global broken power-law function (Figure~\ref{fig:jet_radius}). These oscillations have amplitudes much larger than the uncertainties in the jet radii and may be associated with local over-expansions and over-contractions of the jet.}. We found that the data are described well by a semi-parabolic shape (with a power-law index of $0.58\pm0.05$) and a conical/hyperbolic shape (with a power-law index of $1.16\pm0.01$) inside and outside the transition distance at $z_b = (1.1\pm0.2)\times10^5\ R_g$, respectively. Therefore, we confirm that the "jet collimation break" exists in NGC 315 (see Section~\ref{dis:collimation} for more discussion).

\subsection{Limb-brightening}
\label{sec:limb_brightening}
\begin{figure*}[t!]
\centering
\includegraphics[trim=0mm -5mm 0mm 0mm, width = 0.6\textwidth]{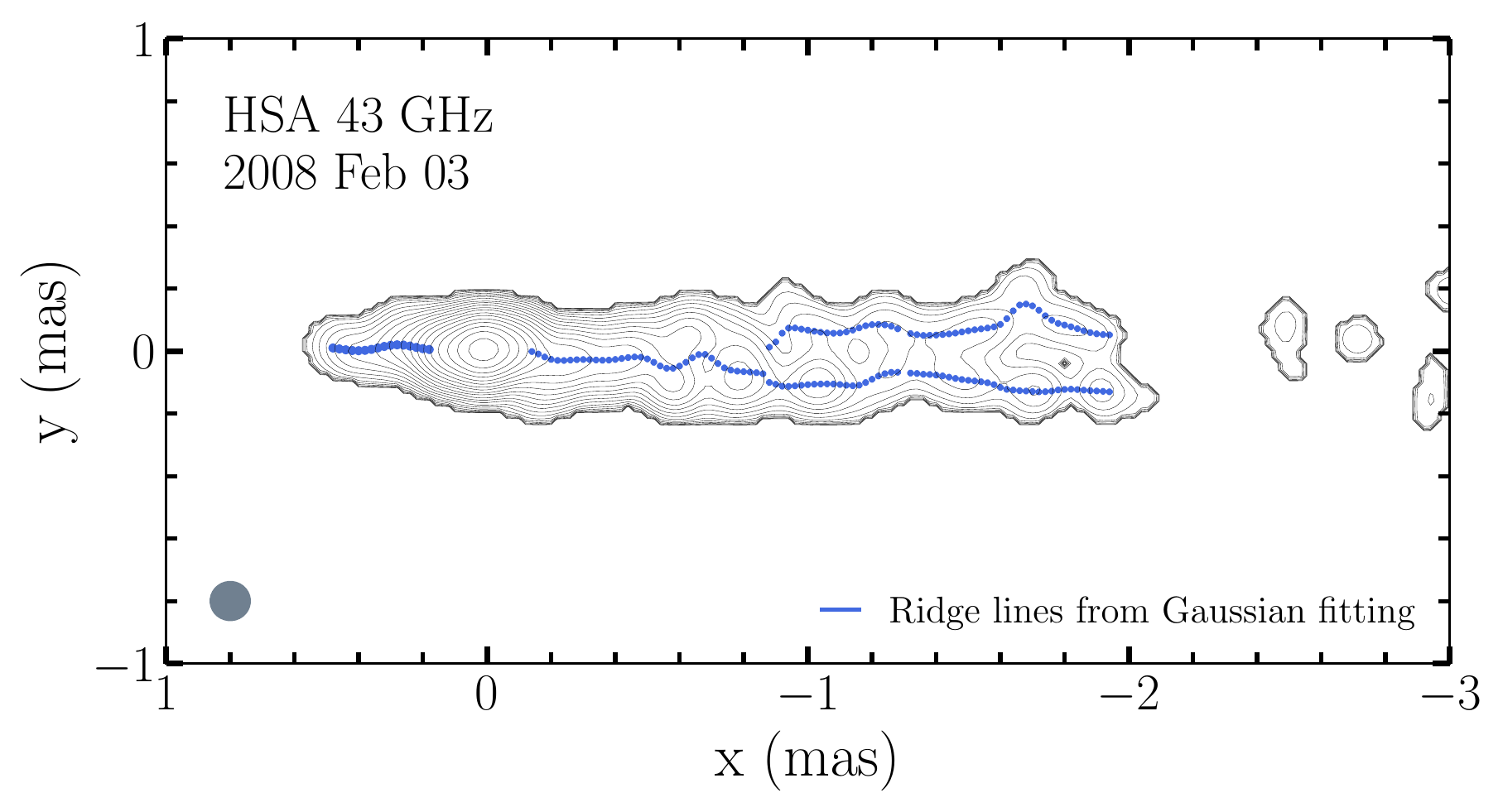}
\includegraphics[width = 0.39\textwidth]{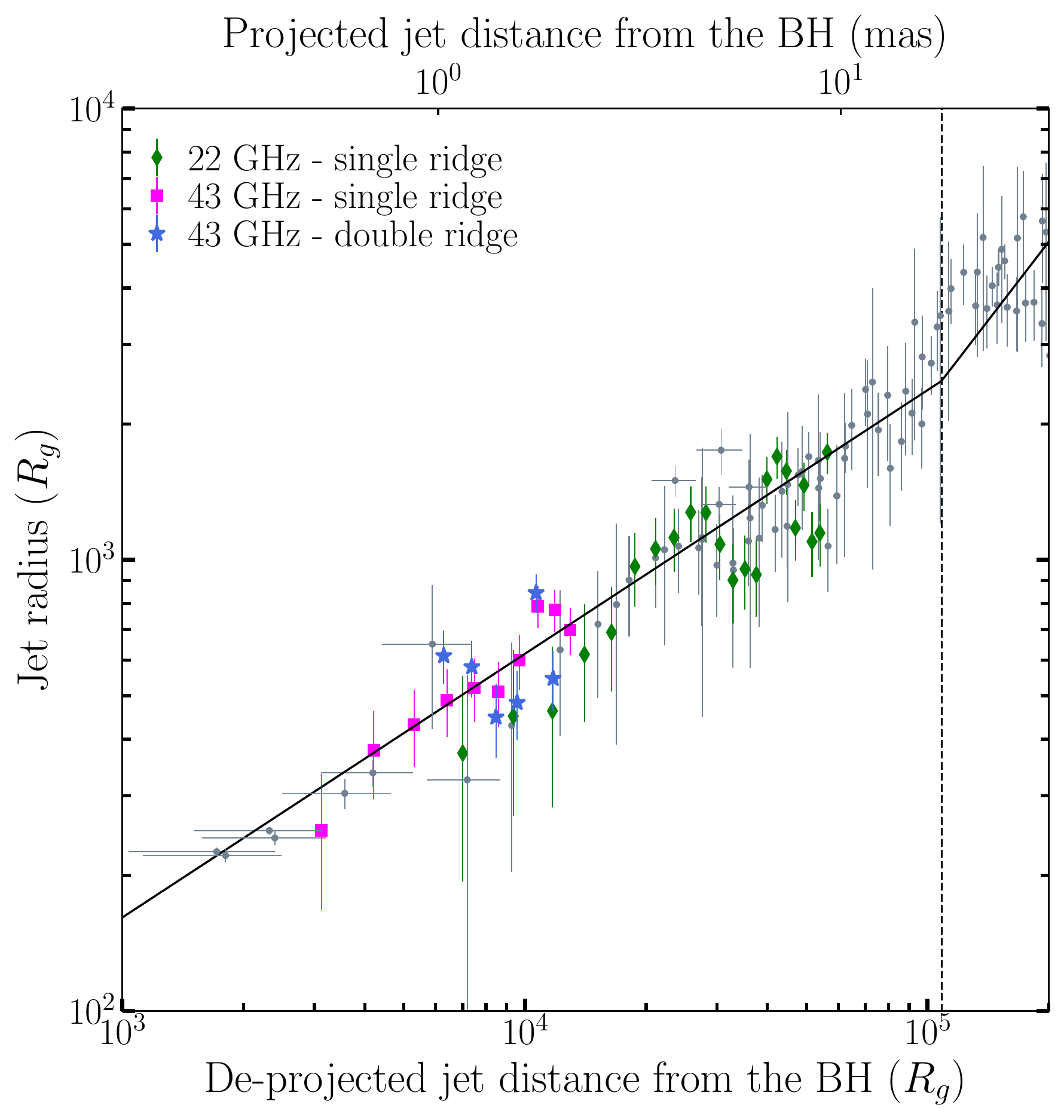}
\caption{Left: contour of total intensity observed with the HSA at 43 GHz, convolved with a circular beam with the size of the minor axis of the synthesized beam shown on the bottom left. The image is rotated clockwise by $40^\circ$. The blue lines show the positions of the peaks of the Gaussian functions fitted to the transverse intensity profiles. At distances $\gtrsim0.8$ mas, double Gaussian functions are needed to fit the data and there are two ridge lines in that region. Right: comparison of jet radii obtained from single Gaussian fitting (magenta squares) and double Gaussian fitting (blue stars) for the transverse intensity profile analysis of the HSA 43 GHz data. The jet radii obtained from single Gaussian fitting at 22 GHz are shown with the green diamonds for a reference. For single (double) Gaussian fitting, the map convolved with the major (minor) axis circular beam is used. For the double Gaussian fitting, the edge-to-edge widths divided by two are measured (see text). Grey circles show the other frequencies/methods data points presented in Figure~\ref{fig:jet_radius}. \label{fig:doubleridge}}
\end{figure*}

The jet structure appears to consist of a single ridge on pc-scales (Figure~\ref{fig:images}) and we found that a single Gaussian function can describe the transverse intensity profiles well in most cases. However, we found an indication of the jet transverse structures resolved at 43 GHz. This is because the 43 GHz data includes the Effelsberg 100m telescope, which provides very long baselines in the direction nearly perpendicular to the jet axis (the minor axis beam size is 0.17 mas). In the left panel of Figure~\ref{fig:doubleridge}, we show the 43 GHz image convolved with a circular beam with a size of the minor axis of the synthesized beam. The jet transverse intensity profiles could not be described well by a single Gaussian function at distances larger than $\approx0.8$ mas from the apparent core. We fitted double Gaussian functions for this distance range and obtained the edge-to-edge jet widths (the distances between the outer edges of the FWHMs of the two ridges), and compared them with the jet widths derived from the single Gaussian fits to the 43 GHz image convolved with a circular beam with the size of the major axis of the synthesized beam (as done for the VLBA images at $\lesssim22$ GHz) in the right panel of Figure~\ref{fig:doubleridge}. The jet radii from the two estimates are in good agreement with each other and with the radii at 22 GHz, which suggests that the observed limb-brightening in this region may be robust.

We show half of the beam FWHM sizes along the transverse jet direction for each VLBA/HSA data on the left y-axis in Figure~\ref{fig:jet_radius}. We find that the maximum jet radii are comparable to the corresponding angular resolutions at all frequencies but at 43 GHz. Therefore, there is a possibility that the jet in NGC 315 is intrinsically limb-brightened on pc-scales, similar to M87, but this feature could not be resolved in the previous and our VLBA observations. We will test this possibility with future observations with the HSA at multiple frequencies.

\subsection{Jet velocity field}
\label{sec:velocity}

As both the jet and counterjet were detected at all VLBA/HSA frequencies, we could infer the jet velocity field on pc-scales by assuming that they are intrinsically the same but the jet is brighter than the counterjet due to relativistic aberration. This approach was adopted to derive the jet velocity field, viewing angle, and other quantities of NGC 315 on kpc-scales \citep{Canvin2005, LB2014}. They could disentangle the degeneracy between the velocity and viewing angle by using both the total intensity and linear polarization data. The jet velocity field was derived also on pc-scales by combining the jet-to-counterjet brightness ratios and the jet kinematic results from multi-epoch monitoring observations \citep{Cotton1999}. 

The jet-to-counterjet intensity ratio is related to $\beta$ and the viewing angle ($\theta$) via
\begin{equation}
\label{eq:ratio}
    R\equiv\frac{I_{\rm jet}}{I_{\rm cjet}} = \left(\frac{1+\beta\cos\theta}{1-\beta\cos\theta}\right)^{2-\alpha},
\end{equation}
\noindent where $I_{\rm jet}$ and $I_{\rm cjet}$ are the intensities of the jet and counterjet at the same jet distance, respectively, and $\alpha$ is the spectral index defined as $I_\nu \propto \nu^\alpha$. We define the origin of the jets as $\approx0.11$ mas upstream of the 22 GHz physical core position from our core-shift result (Section~\ref{sec:coreshift}), from which we calculate the distance for the jet and counterjet. We obtained $R$ for the distances separated from the apparent cores by more than the major axes of the synthesized beams to avoid possible contamination from the convolution of the bright core emission.

For each $R$ derived at each frequency, we need a corresponding spectral index $\alpha$ to derive $\beta$. We obtained $\alpha$ by using an adjacent (higher) frequency map after the image registration. For example, for each $R$ measured at 1.5 GHz, we derived a corresponding $\alpha$ from the 1.5 and 2.3 GHz maps. This approach would minimize the distortions in $\alpha$ from different synthesized beams at different frequencies. 

Two approaches are available to disentangle $\beta$ and $\theta$ (Equation~\ref{eq:ratio}). One is to obtain the apparent jet speed from VLBI monitoring observations, which is also a function of $\beta$ and $\theta$. Combining the apparent speed with the measured $R$ and $\alpha$ at a similar distance, we can solve for $\beta$ and $\theta$. The other approach is to use the jet viewing angle constrained from modelling of the kpc-scale jets \citep{LB2014}, assuming that the viewing angles on pc and kpc scales are the same.

It is known that jet kinematic analysis for radio galaxies is generally much more difficult than for blazars. The main cause is presumably that the jets of radio galaxies have smooth brightness distributions over distances and it is difficult to identify the same jet regions (or "brightness patterns") in different epochs. This issue was addressed well in our previous study of the jet kinematics of M87 \citep{Park2019a}. We argued that a reliable jet kinematic analysis is possible only when high-resolution and high-cadence monitoring data having similar uv-coverages in the sampled epochs\footnote{Significantly different uv-coverages in different epochs can result in artificial jet motions. See also \cite{Walker2018} for a related discussion.} are available. The M87 jet also has re-brightened regions\footnote{These are the regions of local brightness enhancement. The jets of nearby radio galaxies usually show gradually decreasing intensity with increasing distance except in the re-brightened regions.} at several locations, which makes the jet apparently look stationary when observed with a low-resolution and at a low-cadence.

A similar issue could exist for jet kinematics of NGC 315. \cite{Lister2019} found much slower apparent speeds of $\beta_{\rm app}\lesssim0.05c$, using the VLBA monitoring data over nearly 20 years at 15 GHz with an average interval of more than one year, than \cite{Cotton1999} at a similar distance range. The significant difference in the jet kinematic results of different studies could originate from the low cadence ($\gtrsim1$ year) in the previous observations and the re-brightened regions possibly existing in the jet of NGC 315. Motivated by these controversial results, we have performed dense monitoring observations with the KVN and VERA array (KaVA, \citealt{Niinuma2014, Oh2015, Wajima2016, Asada2017, Cho2017, Hada2017, An2018, Lee2019, Zhao2019}). We present the details of our observations, data reduction, kinematic analysis and results in Appendix~\ref{appendix}. 

In summary, we found nearly stationary motions with the observed speeds being consistent with zero within 1--2$\sigma$ at distances less than $\approx5$ mas from the core, while a fast outward motion of $\beta_{\rm app} = 1.85\pm0.44c$ is observed at a distance of $\approx8$ mas. The inferred jet viewing angle by combining the observed apparent speed with the jet-to-counterjet brightness ratio at a similar distance is $\theta = 52.8\pm8.0^\circ$, which is in good agreement with the kpc-scale viewing angle constraint \citep{LB2014}. We found that the stationary motions at $\lesssim5$ mas may be associated with the re-brightened regions, which demonstrates the difficulty of obtaining a robust velocity field from a jet kinematic analysis for our source, similar to M87.

Therefore, we conclude that obtaining the jet velocity field from the jet-to-counterjet intensity ratio, by using the kpc-scale jet viewing angle of $\theta=49.8^\circ$ \citep{LB2014}, is more robust. Assuming the same viewing angle on pc and kpc scales is reasonable because the jet morphology appears to be very straight until the jet reaches a few tens to hundreds of kpc \citep[][see also Figure~\ref{fig:images}]{Laing2006} and the pc-scale viewing angle constrained from our jet kinematic analysis is indeed consistent with the kpc-scale one, although it has a substantially larger uncertainty.

A careful analysis of uncertainties in $R$ and $\alpha$ is necessary for a robust estimation of $\beta$. We found that the biggest uncertainties in $R$ and $\alpha$ originate from the uncertainties in the core-shift. We adopted a Monte-Carlo approach to take these uncertainties into account. We draw 1,000 random core-shifts along the jet direction for each frequency from normal distributions with means and standard deviations determined from the best-fit apparent core positions and their $1\sigma$ uncertainties (Table~\ref{tab:fitted_coreshift} and Table~\ref{tab:app_coreshift}). We obtained corresponding 1,000 realizations for $R$ and $\alpha$ as a function of projected jet distance at each frequency. The intensities $I_{\rm jet}$ and $I_{\rm cjet}$, which comprise $R$, are obtained by fitting a single Gaussian function to the transverse jet intensity profile for each distance (Section~\ref{sec:collimation}). We obtained a representative $\alpha$ for each jet distance from the intensity-weighted average of the spectral indices along the transverse jet direction. We binned $R$ and $\alpha$ from the 1,000 realizations in distance with a bin size of half of the major axis of the synthesized beam and obtained a representative value and $1\sigma$ uncertainty for each bin from the median and standard deviation of the data points within the bin.

\begin{figure}[t!]
\centering
\includegraphics[width = 0.49\textwidth]{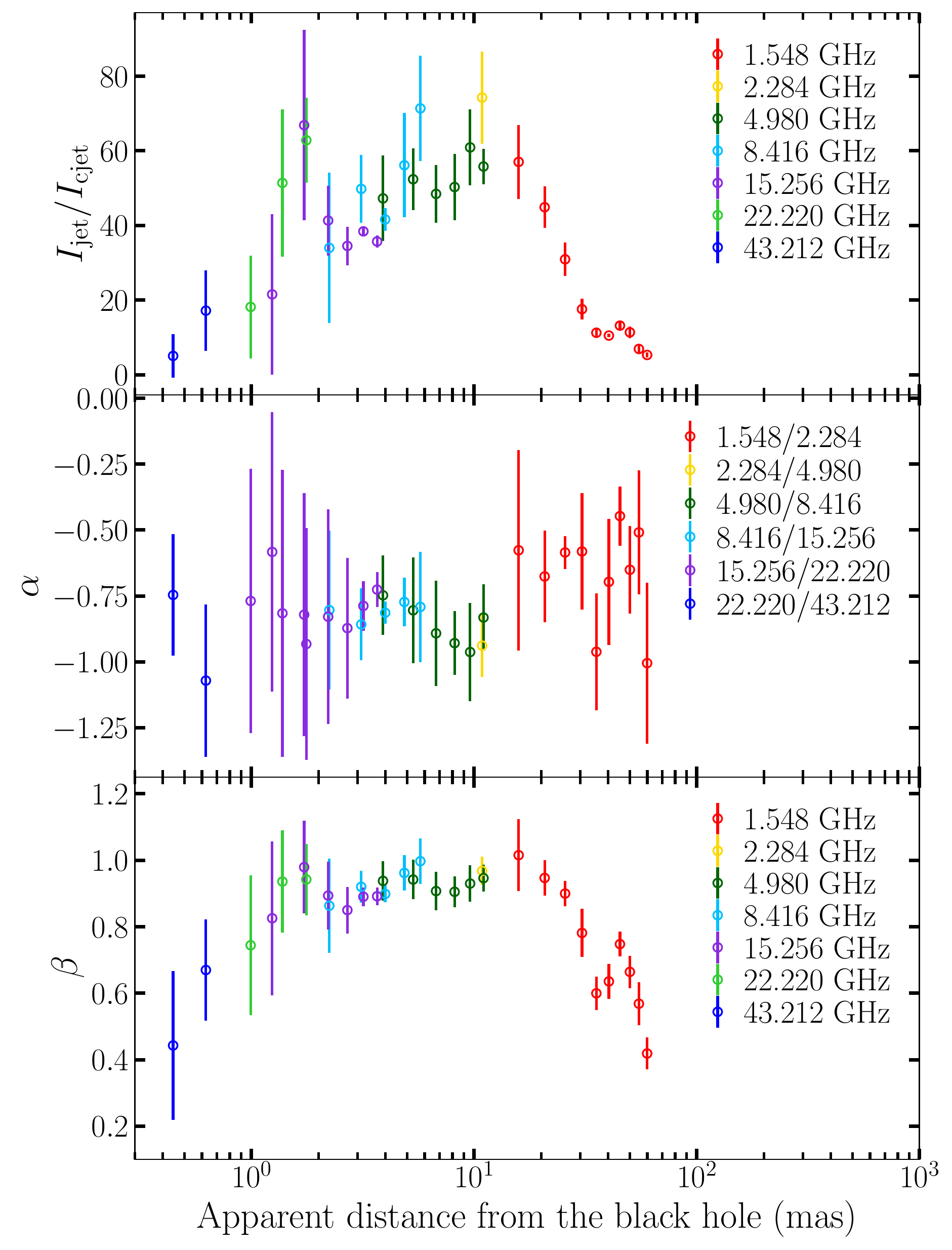}
\caption{Jet-to-counterjet intensity ratio (top), spectral index (middle), and jet speed in units of the speed of light (bottom) as functions of apparent (projected) distance from the black hole. \label{fig:properties_binned}}
\end{figure}

We present $R$, $\alpha$, and $\beta$ as a function of apparent projected distance from the black hole in Figure~\ref{fig:properties_binned}. The brightness ratios gradually increase at distances from $\approx0.4$ to $\approx10$ mas and then decrease at larger distances. The spectral indices are nearly constant over distance. Thus, the derived jet speeds follow the same trend as the jet brightness ratios: showing gradual acceleration and deceleration in the inner and outer regions of the distance of $\approx10$ mas. Our conclusion of jet acceleration and deceleration rely on $\beta$ at the shortest and longest distance bins, where the data from only a single frequency are available. We note that the 1.5 GHz data at the longest distance bin has a SNR of $\approx 7$ for the counterjet intensity, while the other data points have SNRs larger than 10, indicating that these measurements are robust. The 43 GHz data also seem to have large enough SNRs ($\gtrsim10$) but the large difference in observing epochs between 43 GHz and other frequencies could cause additional uncertainty. We address this issue using the VLBA 43 GHz data, which was observed nearly simultaneously to the VLBA data at other frequencies but not used for our main analysis due to the limited data quality (Section~\ref{sec:data}), in Appendix~\ref{appendix:beta}. We conclude that $\beta$ at 43 GHz in Figure~\ref{fig:properties_binned} is not underestimated, implying that the NGC 315 jets do accelerate on this scale.

We note that we could not find a significant difference in the spectral indices between the jet and counterjet at the same jet distance in the regions where $R$ and $\alpha$ were measured, i.e., separated from the apparent core by more than one beam size. Therefore, we assume that the effects of possible free-free absorption by ionized material near the central engine on the velocity field derivation is insignificant and that the observed core-shifts are dominated by synchrotron self-absorption. A strong indication of significant free-free absorption is the presence of the "emission gap" between the jet and counterjet at low frequencies (e.g., \citealt{Walker2000, Kameno2001, Baczko2019}) and of the cores of both the jet and counterjet shifting towards each other with increasing frequency (e.g., \citealt{Haga2015}). We could find neither of these indications for NGC 315 (Figure~\ref{fig:contours_shift}).

\section{Discussion}
\label{sec:discussion}

\subsection{Jet Collimation}
\label{dis:collimation}
\begin{figure*}[t!]
\centering
\includegraphics[width = 0.75\textwidth]{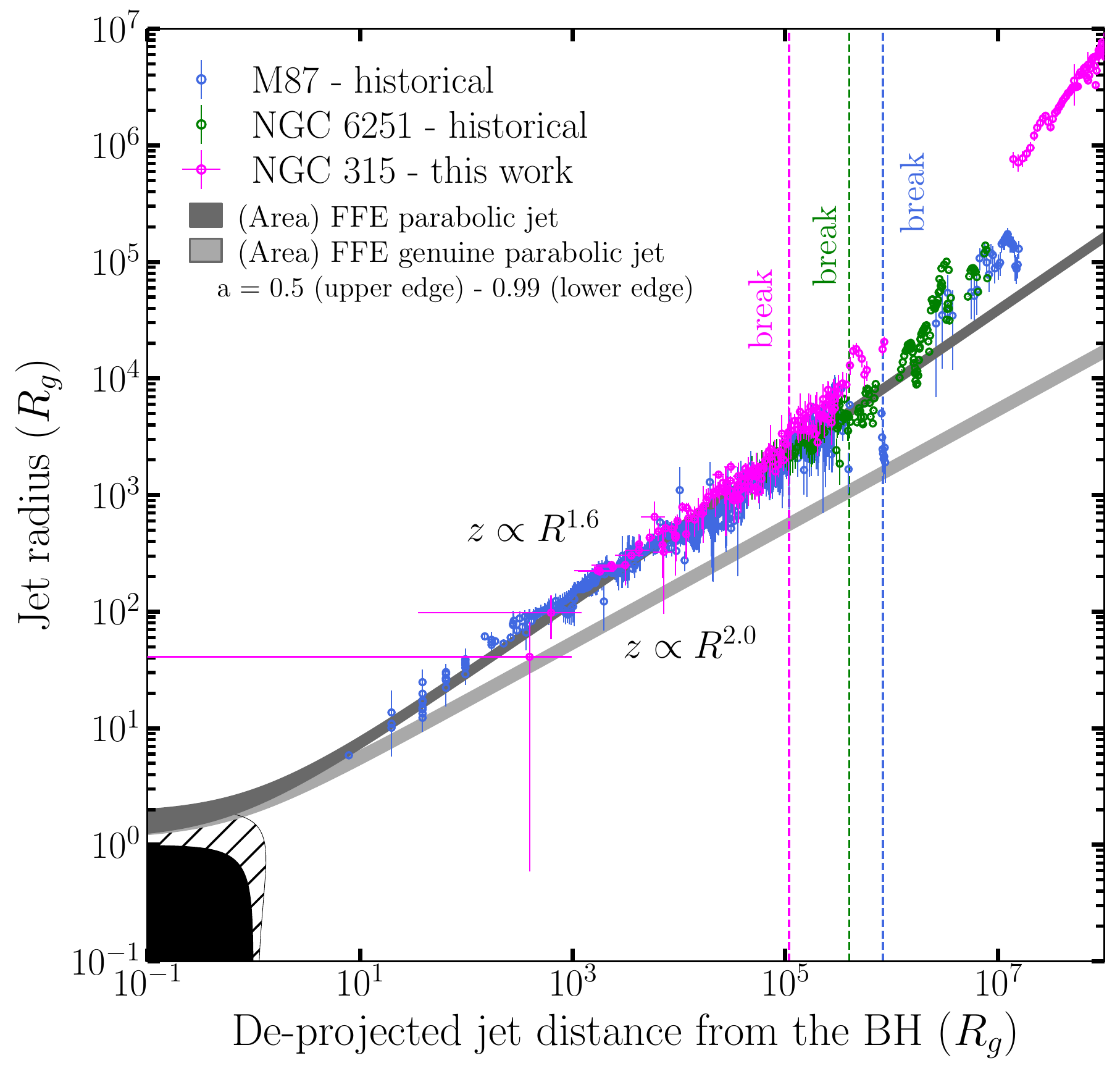}
\caption{Jet radius as a function of de-projected jet distance from the black hole in units of $R_g$ for M87 \citep[blue, ][]{AN2012, Doeleman2012, Hada2013, NA2013, Akiyama2015, Nakamura2018}, NGC 6251 \citep[green, ][]{Tseng2016, Nakamura2018}, and NGC 315 (magenta). The dotted vertical lines show the locations of the jet collimation breaks. The dark (light) grey area shows the outermost poloidal field line anchored to the equator of the event horizon obtained from the FFE solution for $\kappa=0.75$ ($\kappa=1$) for the black hole spin range of 0.5--0.99, which corresponds to $z\propto R^{1.6}$ ($z\propto R^2$) asymptotically. The filled black region on the bottom left denotes the event horizon, and the shaded region shows the ergosphere for the spin parameter a = 0.99 (see \citealt{Nakamura2018} for more details). \label{fig:jet_width_comparison}}
\end{figure*}

In Figure~\ref{fig:jet_width_comparison}, we compare the jet radius of NGC 315 as a function of distance with the nearby FR I radio galaxies M87 \citep{AN2012, Doeleman2012, NA2013, Hada2013, Hada2016, Akiyama2015, Nakamura2018} and NGC 6251 \citep{Tseng2016}. We also show the steady axisymmetric force-free electrodynamic (FFE) solution \citep{Narayan2007, Tchekhovskoy2008} for the outermost poloidal field line anchored to the black hole event horizon on the equatorial plane with $\kappa=0.75$\footnote{$\kappa$ is related to the radial power-law index in the poloidal flux function of the approximate FFE solution, which describes the asymptotic shape of the field line (see \citealt{Narayan2007, Tchekhovskoy2008, Nakamura2018} for more details).}, which describes the observed parabolic jet collimation profiles of M87 and NGC 6251 well \citep{Nakamura2018}. The jet collimation profiles of NGC 315, NGC 6251, and M87 in the parabolic regions (before the jet collimation breaks) are remarkably similar, although the locations of the breaks are different.

\cite{Nakamura2018} showed that AGN jets may be collimated by the pressure of winds, which are non-relativistic and moderately magnetized gas outflows launched from the accretion flows \citep{Sadowski2013, Yuan2015}, on scales of $\lesssim100\ R_g$ from GRMHD simulations. The boundary shape between the jet and wind is consistent with the observed jet collimation profile for M87 \citep{AN2012, NA2013, Hada2013} and with the FFE solution for the outermost poloidal field line anchored to the event horizon on the equatorial plane. This result was confirmed on scales down to $\approx10^5 \ R_g$ by recent GRMHD simulations \citep{Chatterjee2019}. Also, \cite{Park2019b} showed that the inferred pressure profile of an external confining medium is flat enough to collimate the jet \citep{Komissarov2009} from the Faraday rotation observations of the jet collimation region in M87. Given that the parabolic jet shape of NGC 315 is consistent with the M87 jet (Figure~\ref{fig:jet_width_comparison}), a similar scenario can be applied to the NGC 315 jet.

However, there are two major differences in the jet radius profiles between NGC 315 and M87. One is the location of the jet collimation break and the other is the existence of a recollimation feature. The jet collimation break in M87 was suggested to occur near the Bondi radius \citep[\rb,][]{AN2012, NA2013, Nakamura2018}, within which the dynamics of materials is thought to be governed by the black hole gravity. We infer the Bondi radius from the temperature of $\approx0.44$ keV of the X-ray emitting gas in the core of NGC 315 (within 1 arcsec, \citealt{Worrall2007}), which is $R_B \approx 1.5\times10^6\ R_g$\footnote{1 arcsec corresponds to the physical scale of $\approx4.6\times10^6\ R_g$ and the estimate of \rb is based on the assumption that the temperature is the same between \rb and 1 arcsec. A flat temperature profile is generally expected in the cool cores of elliptical galaxies \citep[e.g.,][]{Hudson2010, Werner2019} including NGC 315 \citep{Sun2009}, but the temperature may not be exactly the same in the considered distance range \citep[e.g.,][]{Gaspari2013}, which is a source of uncertainty in the \rb estimate. See \cite{Tseng2016} for a related discussion.}. We can also infer the sphere of influence of the black hole gravity via $R_{\rm SOI} \equiv GM/\sigma_v^2$ \citep{Peebles1972}, where $\sigma_v$ is the stellar velocity dispersion of the host bulge measured near the black hole. \cite{Ene2020} obtained $\sigma_v\approx350\ {\rm km\ s^{-1}}$ at a radius of $\approx0.1$ arcsec in NGC 315, which gives $R_{\rm SOI} \approx 7.3\times10^5\ R_g$. These estimates suggest that the location of jet collimation break in NGC 315, $z_b \approx (1.1\pm0.2)\times10^5\ R_g$, is an order of magnitude smaller than \rb and \rsoi, unlike M87\footnote{We note that \rb and \rsoi scale with the black hole mass, while $z_b$ does not. This indicates that our conclusion that $z_b$ is much smaller than \rb and \rsoi may depend on how accurate the black hole mass we used, which is based on the $M_{\rm BH}$-$\sigma_v$ relation, is. It is believed that the black hole mass estimates based on the $M_{\rm BH}$-$\sigma_v$ relation can be uncertain typically by a factor of two \citep[e.g.,][]{KH2013}. Therefore, it is unlikely that the black hole mass is overestimated by an order of magnitude and our conclusion may be robust against the mass uncertainty.}.

Also, the M87 jet has a dip in the jet width near the jet geometry transition region, at the location of a jet feature known as HST-1 \citep{AN2012, NA2013, Nakamura2018}. This feature is associated with the strong multiwavelength flare in 2005 including X-rays and even TeV $\gamma$-rays (e.g., \citealt{Aharonian2006, Cheung2007, Harris2009}), shows both superluminal and quasi-stationary knots \citep[e.g.,][]{Biretta1999, Cheung2007, Giroletti2012, Nakamura2010, NM2014}, and shows enhanced linealy polarized emission and Faraday rotation measure as compared with the neighboring inner and outer jets \citep{Chen2011, Park2019b}. HST-1 was suggested as a recollimation shock \citep{Stawarz2006, BL2009, LG2017} and the inferred jet pressure is orders of magnitude higher than the surrounding medium \citep[e.g.,][]{AN2012}. Thus, \cite{AN2012} suggested that the M87 jet can conically expand into the interstellar medium having a flat pressure profile with distance \citep[e.g.,][]{Russell2015, Russell2018} because of the high jet internal pressure caused by the recollimation shock. However, there is no indication of recollimation nor linear polarization detected at the jet collimation break point in NGC 315.

Therefore, different mechanisms are needed to explain the geometrical transition of NGC 315. If the jet is collimated by the winds from hot accretion flows as suggested above, the collimation break location being an order of magnitude smaller than \rb and \rsoi may indicate that the winds may not reach down to the Bondi radius. It is yet unclear how far the winds can reach away from the black hole \citep[e.g.,][]{Yuan2015}; it may depend on the global geometry of the hot accretion flows \citep{Chatterjee2019} and the effects of the gravitational potential of the nuclear star clusters \citep{Bu2016a, Bu2016b}. There is a growing evidence that AGN jet collimation breaks do not necessarily occur at the Bondi radii. The break locations appear to be smaller than the Bondi radii (or \rsoi) in NGC 6251, NGC 4261, NGC 1052 \citep{Tseng2016, Nakahara2018, Nakahara2020}, while it is possibly much larger in Cygnus A\footnote{No transition from a parabolic to conical geometry was observed at distances $\lesssim10^9\ R_g$.} \citep{Nakahara2019}.

\begin{figure}[t!]
\centering
\includegraphics[width = 0.47\textwidth]{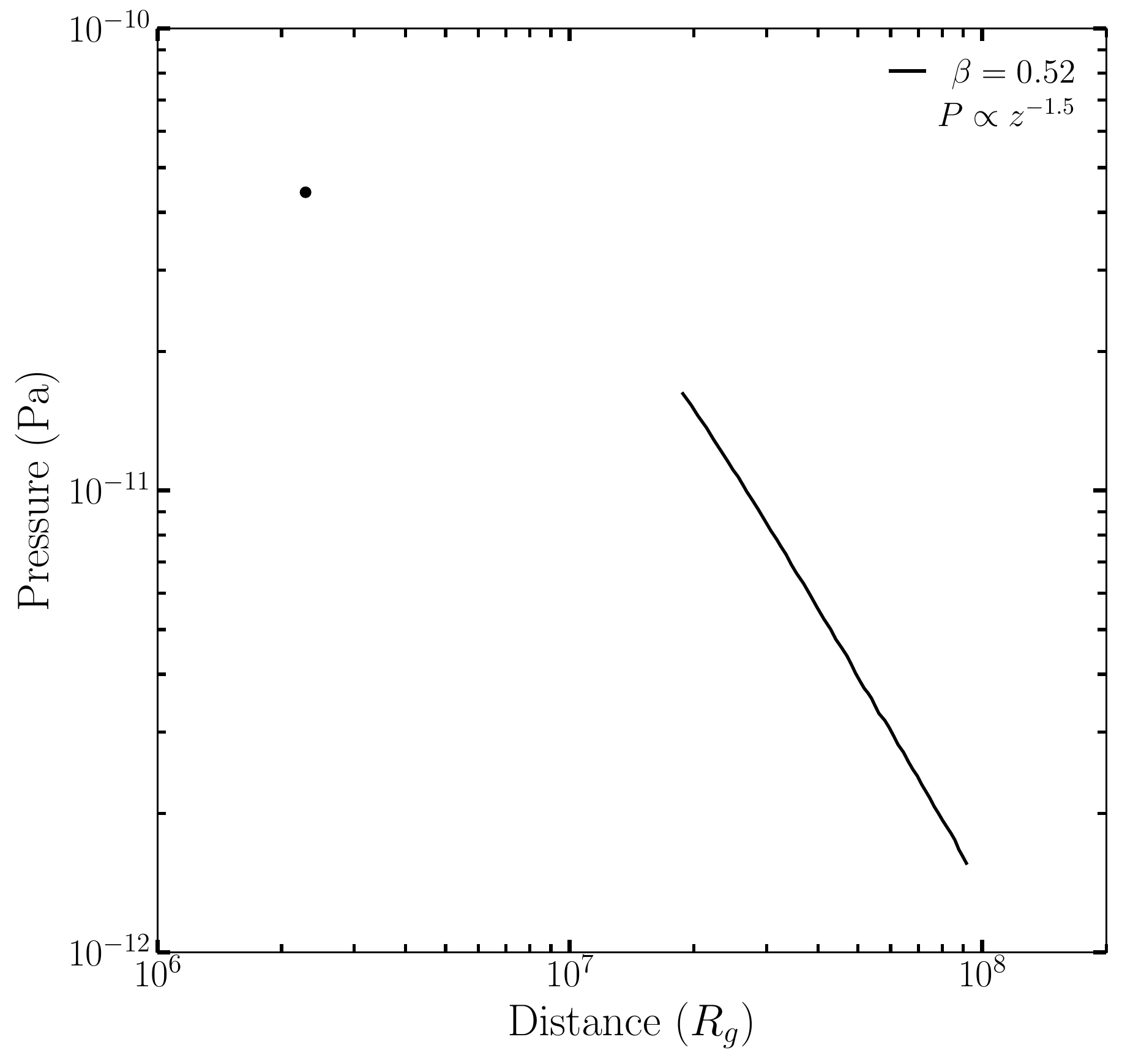}
\caption{Pressure of the X-ray emitting gas in NGC 315 as a function of distance in units of $R_g$, read off from Figure 13 of \cite{Worrall2007}. The black dot is the pressure inferred from the thermal component in the X-ray spectral fit to the core, while the black solid curve is obtained from the best-fit $\beta$ model, which describes the observed X-ray surface brightness as being proportional to $[1 + (\theta/\theta_{\rm CX}^2)]^{0.5-3\beta}$, where $\theta$ is the angular radius and $\theta_{\rm CX}$ is the angular core radius \citep[e.g.,][]{CF1978, BW1993}. This model allows to infer the radial pressure profile of the X-ray emitting gas \citep{BW1993}, which is $P \propto z^{-1.5}$ for NGC 315 \citep{Worrall2007}. \label{fig:worrall}}
\end{figure}

How the jet in NGC 315 expands in a conical/hyperbolic shape after the collimation break also seems puzzling. Two scenarios have been broadly considered to explain the observed conical/hyperbolic expansions. One is that the pressure of an external confining medium decreases with distance and the asymptotic jet shape becomes conical \citep[e.g.,][]{BL1994, Zakamska2008, Komissarov2009, Lyubarsky2009, Vlahakis2015}. \cite{LB2014} indeed showed that the jets of many nearby radio galaxies, including NGC 315, expand rapidly in conical/hyperbolic shapes on kpc-scales. These expansions occur in the regions of steeply falling external pressure gradients. However, the jet geometry transition of NGC 315 occurs at a shorter distance where the pressure gradient is expected to be much flatter. We present the radial pressure profile of the X-ray emitting hot gas in NGC 315 estimated from Chandra observations \citep{Worrall2007} in Figure~\ref{fig:worrall}. The pressure decreases with distance as $P_{\rm ext}\propto z^{-1.5}$ in the regions where the jet expands conically at $z\gtrsim10^7\ R_g$ (Figure~\ref{fig:jet_radius}). However, the pressure profile becomes flatter in the inner region at $z \lesssim 10^7\ R_g$, similar to M87 \citep[e.g.,][]{Russell2015, Russell2018} and NGC 6251 \citep{Evans2005}. This indicates that the NGC 315 jet can maintain its conical/hyperbolic shape over a large range of distance (z $\approx10^5$--$10^8\ R_g$) regardless of the change in the external medium's pressure profile in the same region.

Another explanation is that the jet is overpressured due to a recollimation shock occuring near the jet geometry transition point and can freely expand into the interstellar medium having a nearly flat pressure profile. This was applied to the M87 jet \citep{AN2012}, but the absence of a recollimation feature nor significantly enhanced linear polarization emission at the jet collimation break site in NGC 315 makes it difficult to apply this scenario. A similar case was seen for NGC 6251 \citep{Tseng2016} and the authors suggested that an in-situ energy dissipation by converting the jet bulk kinetic energy into the jet internal energy may take place. The increased internal energy may be responsible for the free jet expansion into an external medium having a flat pressure profile. We found that the bulk jet speeds systematically decrease with distance right after the jet collimation break (Section~\ref{dis:acceleration}), which makes this explanation plausible.

\subsection{Jet Acceleration}
\label{dis:acceleration}

\begin{figure*}[t!]
\centering
\includegraphics[width = 0.75\textwidth]{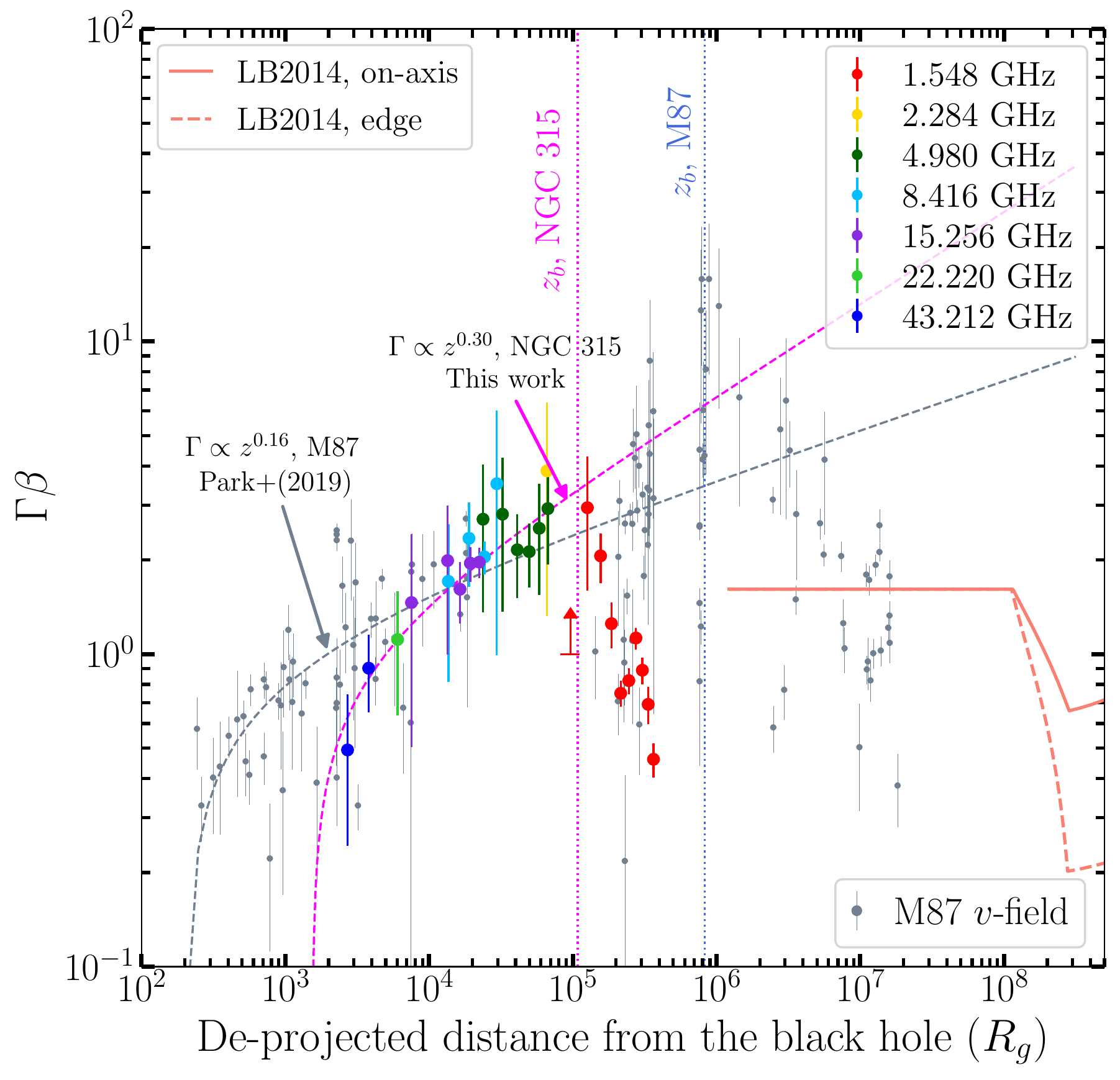}
\caption{$\Gamma\beta$ as a function of de-projected jet distance from the black hole in units of $R_g$, constrained by the jet-to-counterjet brightness ratio analysis (Section~\ref{sec:velocity}). The lower limit of the data point having $\beta\ge1$ is indicated. The grey small circles are the results for M87 obtained in previous studies \citep{Biretta1995, Biretta1999, Cheung2007, Ly2007, Giroletti2012, Meyer2013, Asada2014, Hada2015, Hada2016, Hada2017, Mertens2016, Kim2018, Walker2018, Park2019a}, which were compiled in \cite{Park2019a}. The dashed vertical lines show the locations of the jet collimation breaks. The magenta dashed line is the best-fit power-law function to the NGC 315 data points before the jet collimation break, which is $\Gamma\propto z^{0.30\pm0.04}$. The best-fit jet acceleration profile obtained for M87 by \cite{Park2019a} is presented with the grey dashed line. The jet velocity field of NGC 315 constrained on kpc-scales by \cite{LB2014} for the jet on-axis and edge are shown by the salmon solid and dashed lines, respectively. \label{fig:fourvel}}
\end{figure*}

In Figure~\ref{fig:fourvel}, we present $\Gamma\beta$, where $\Gamma \equiv 1 / \sqrt{1 - \beta^2}$ is the bulk Lorentz factor, derived from the jet-to-counterjet brightness ratio analysis (Section~\ref{sec:velocity}), as a function of de-projected distance from the black hole. We obtained $\beta>1$ at the innermost distance bin at 1.5 GHz (Figure~\ref{fig:properties_binned}), which is due to the uncertainty, and we cannot convert this data point into $\Gamma$. Thus, we instead plot a lower limit by taking a $\beta-3\sigma$ value for this data point, where $\sigma$ is the uncertainty in $\beta$. We also did not include the data points having uncertainties larger than 90\% of their data values. We fitted a simple power-law for $\Gamma$ to the data points at distances smaller than the jet collimation break position ($z_b$), and obtained $\Gamma\propto z^{0.30\pm0.04}$. We overplot $\Gamma\beta$ for M87 observed in previous studies which were compiled in \cite{Park2019a} and their best-fit power-law function of $\Gamma\propto z^{0.16\pm0.01}$ in the jet acceleration zone.

The MHD jet acceleration model predicts that jets can be efficiently accelerated to relativistic speeds when the jets are collimated. More specifically, an efficient jet acceleration occurs when the inner poloidal magnetic field lines close to the jet axis are collimated more than the outer field lines, which is called the "differential bunching/collimation" of the poloidal field lines \citep[Also known as the "magnetic nozzle effect", e.g.,][]{Li1992, BL1994, Vlahakis2004, Vlahakis2015, Komissarov2009, Tchekhovskoy2009}. This model is characterized by the acceleration zone spanning a large distance range, which is thought to be coincident with the jet collimation zone \citep[e.g.,][]{VK2004, Lyubarsky2009}. It is remarkable that the gradual jet acceleration of NGC 315 continues exactly until the jet geometry maintains a parabolic shape. The jet speeds start to decrease right after the jet collimation break. The same trend is observed in M87. These findings indicate that a gradual jet acceleration through the Poynting flux conversion takes place in these sources.

The observed jet acceleration profile of $\Gamma \propto z^{0.30}$ is much flatter than the efficient "linear acceleration" of $\Gamma \propto R \propto z^{0.58}$, expected for the initial acceleration region of parabolic outflows in the models of highly magnetized jets \citep{Tchekhovskoy2008, Tchekhovskoy2009, Komissarov2009, Lyubarsky2009}. The "slow acceleration" was also observed in M87 in previous studies \citep{Mertens2016, Park2019a}. This result indicates that jet acceleration is not simply determined by the jet collimation profile but may be determined by the interplay between (i) the degree of jet magnetization near the jet base, (ii) the differential collimation of poloidal magnetic field lines, which can be different in different sources even if they show the same jet geometries\footnote{The observed jet collimation profile might reflect only parts of the field lines (also called streamlines), while the jet acceleration efficiency is associated with the behaviors of multiple streamlines.}, and (iii) the interaction between the jet and the ambient medium. In fact, the maximum Lorentz factor achieved for NGC 315 is at most $\Gamma\approx3$. If we assume that the jet reaches equipartition between Poynting and matter energy flux at the end of the jet acceleration zone ($\sigma_m\approx1$, where $\sigma_m$ is the Poynting flux per unit matter energy flux), then the total energy flux per unit rest-mass energy flux is $\mu = \Gamma(1+\sigma_m) \approx 6$ \citep[e.g.,][]{TT2013}. If this is the case, the NGC 315 jet may not be highly magnetized at its base. Also, the observed rapid deceleration right after the collimation break suggests that there could be an active interaction of the jets with the surrounding medium on pc-scales. The interaction can result in gas entrainment from surrounding material and substantial deceleration of the jet, as suggested by observations of the kpc-scale jets of many radio galaxies \citep{LB2014}. Also, substantial jet deceleration due to entrainment of surrounding winds in the jet acceleration zone was shown in recent GRMHD simulations \citep{Chatterjee2019}.

In order for the differential collimation of poloidal field lines to occur, and thus for an efficient jet acceleration to occur, the field lines must be able to communicate to other regions of the jet. In other words, they must be causally connected  with regions near the jet axis \citep[e.g.,][]{Tchekhovskoy2009, Komissarov2009, Clausen-Brown2013}. This condition is satisfied if the jet half-opening angle $\theta_j$ is smaller than the Mach cone half-opening angle $\theta_M$, which can be translated into the condition $\Gamma\theta_j\lesssim \sqrt{\sigma_m}$ \citep{Komissarov2009}. We found that $\theta_j$ gradually decreases with distance in the parabolic jet region down to $\approx1^\circ$ and the maximum Lorentz factor is $\Gamma\approx3$, which indicates that the condition is satisfied when assuming $\sigma_m\gtrsim1$ in the jet acceleration zone.

We note that the lowest jet speed after the jet deceleration is $\beta\approx0.4c$. However, the inferred jet speed from the kpc-scale observations at distances $\lesssim10^8\ R_g$ is $\beta\approx0.85c$ \citep{LB2014}, suggesting that there must be an additional jet acceleration zone in the conically expanding jet region. We note that \cite{LB2014} used a model which assumes a constant speed at short jet distances, which corresponds to the distances of $10^6$--$10^8\ R_g$ for NGC 315. A more accurate jet velocity field in this region, which we plan to investigate with future observations, would help to identify where the additional jet acceleration zone is located and how it is related to the jet geometry.

\subsection{Jet Velocity Stratification}
\label{dis:stratification}

It is commonly accepted that AGN jets are stratified in velocity. A faster-spine and slower-sheath jet structure, i.e., the inner layers closer to the jet axis (spine) being faster than the outer layers (sheath), has been considered in many previous studies. For example, \cite{Clausen-Brown2013} showed that the model with velocity shear can explain the observed apparent sizes of many blazars better than the model without shear (i.e., constant speed across jet layers). \cite{LB2014} showed that the jets of ten nearby radio galaxies indeed have faster inner layers and slower outer layers on kpc-scales. This velocity structure was also used for modelling the high-energy emission up to TeV energies observed in blazars and radio galaxies \citep[e.g.,][]{Ghisellini2005, TG2008, TG2014, Marscher2010, MacDonald2015, Park2019c}. Also, \cite{Nakahara2018} explained the observed jet radii for NGC 4261 \citep{Nakahara2018} and Cygnus A \citep{Boccardi2016b, Nakahara2019} being systematically larger than those for M87 and NGC 6251 (see Figure 8 in \citealt{Nakahara2018}) with the spine-sheath scenario.

On the other hand, a relativistic jet launched by a rotating black hole and accelerated by the Poynting flux conversion is expected to have a slower-spine and faster-sheath structure \citep[e.g.,][]{Komissarov2007, Komissarov2009, Tchekhovskoy2008, Tchekhovskoy2009, Penna2013, Nakamura2018, PT2020}. If the jet is viewed at a small angle, then the brightness of the outer boundary layers can be much more enhanced than the inner layers due to the relativistic Doppler boosting effect. \cite{Nakamura2018} suggested that this velocity structure can naturally produce the observed limb-brightening of the M87 jet in the collimation zone. 

\begin{figure}[t!]
\centering
\includegraphics[width = 0.47\textwidth]{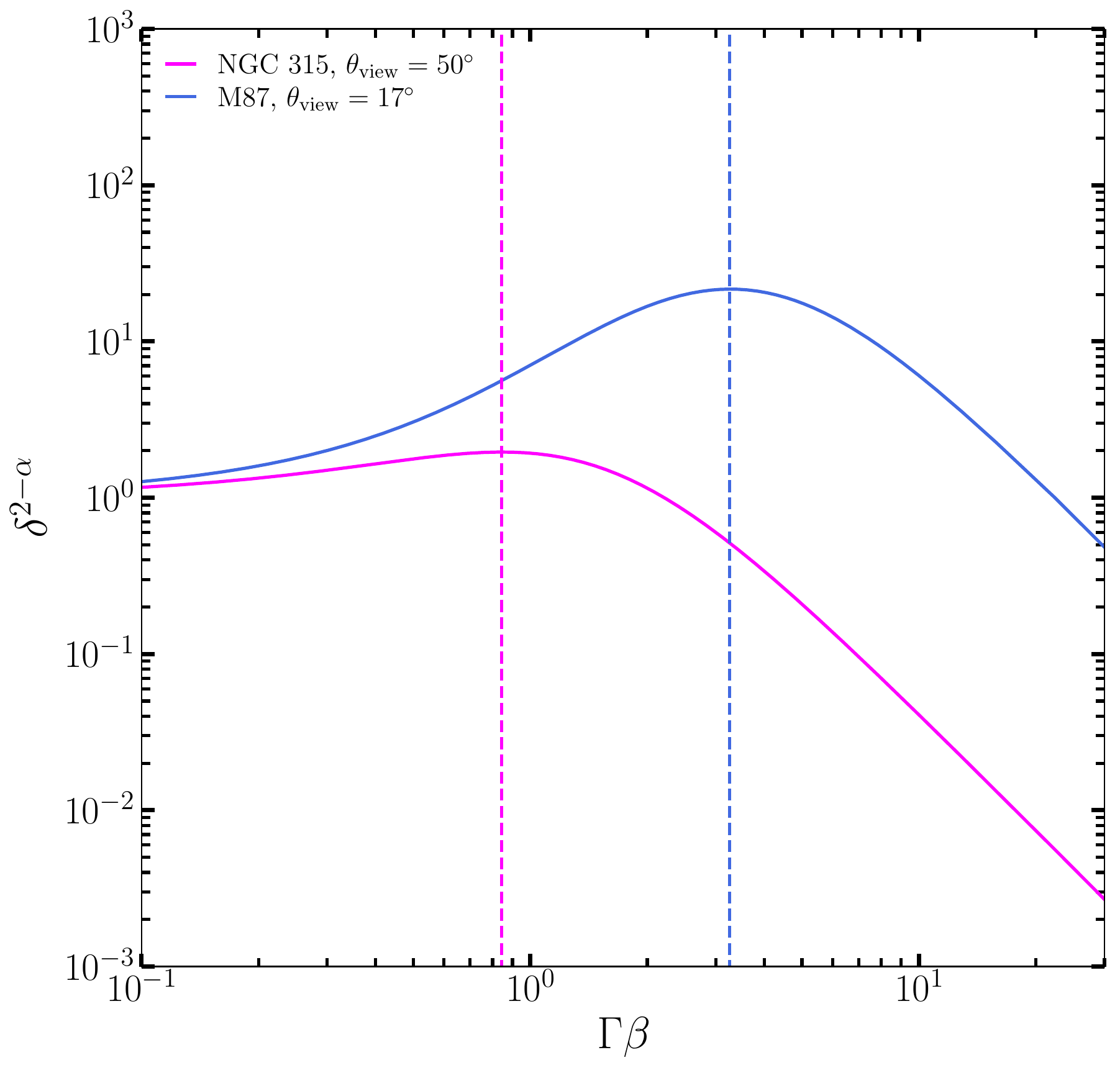}
\caption{Flux enhancement factor by the Doppler boosting effect, $\delta^{2-\alpha}$, as a function of $\Gamma\beta$ for NGC 315 (magenta) and M87 (blue). $\alpha = -0.5$ is assumed for both sources. The vertical dashed lines show the positions of the maximum enhancement factors. \label{fig:doppler}}
\end{figure}

Interestingly, we found an indication of limb-brightening in our HSA image of NGC 315 at 43 GHz (Figure~\ref{fig:doubleridge}). However, the jet viewing angle of NGC 315 is quite large and one cannot expect much flux enhancement from the Doppler boosting effect for this source. Figure~\ref{fig:doppler} shows the expected flux enhancement factor of $\delta^{2-\alpha}$, where $\delta\equiv1/\Gamma(1-\beta\cos\theta)$ is the Doppler factor, for M87 \citep[$\theta=17^\circ$,][]{Mertens2016, Walker2018} and NGC 315 \citep[$\theta=50^\circ$,][See also Appendix~\ref{appendix}]{LB2014}. We assumed $\alpha = -0.5$, which is a good approximation for those sources \citep[See Figure~\ref{fig:properties_binned} and][]{Hada2016}.

The expected maximum flux enhancement is at most less than by a factor of two for NGC 315. For the fast speeds of $\Gamma \beta \gtrsim 1$, the flux can even be suppressed because the jet beaming cone moves away from our line of sight. Therefore, the observed limb-brightening of the jet in NGC 315 cannot be attributed to the Doppler-boosted emission of the fast moving jet sheath. We consider two possible scenarios. One is that the jet spine is much faster than the jet sheath on pc-scales, similar to the jet lateral velocity structure on kpc-scales \citep{LB2014}, resulting in the spine emission significantly De-boosted. A limb-brightening has also been observed in the nearby radio galaxy Cygnus A on pc-scales, known to have a large viewing angle of $\theta \approx 75^\circ$, and the faster-spine and slower-sheath structure was employed to explain the observed jet kinematics in this source \citep{Boccardi2016a, Boccardi2016b}. However, it is questionable how the jet spine attains very fast speeds at such a short distance, which appears difficult to achieve in the MHD jet acceleration models. We compare the radii of the jet and counterjet, which can provide some hints for jet velocity stratification but does not seem to be conclusive with our data, in Appendix~\ref{appendix:radii}. 

The other scenario is that the jet sheath is intrinsically brighter (has a higher emissivity) than the spine. The consistency between the observed jet collimation profile and the FFE solution for the outermost poloidal field line anchored to the equator of the event horizon (Figure~\ref{fig:jet_width_comparison}) implies that the jet limbs may follow the boundary against the ambient medium, which is presumably non-relativistic winds \citep[e.g.,][]{Sadowski2013}. The velocity shear near the boundary layers can result in efficient particle acceleration \citep[e.g.,][]{Ostrowski1998, SO2002, Kataoka2006}. Also, recent GRMHD simulations showed that pinch instabilities can be developed near the jet-wind boundary, which can produce radiating superluminal knots \citep{Nakamura2018} and efficient particle acceleration through magnetic reconnection \citep{Chatterjee2019}. We note that confirming the existence of limb-brightening at other distance ranges with future observations at high resolutions and with high sensitivity, combining with the jet velocity field, can be critical to distinguish these scenarios. Probing the innermost jet region with millimeter VLBI arrays, where the jet speeds are presumably small and less Doppler effect is expected, can be especially critical.

\section{Conclusions}
\label{sec:summary}

We have studied the collimation and acceleration in the jets of the nearby FR I radio galaxy NGC 315 with our multifrequency VLBA observations and archival HSA and VLA data. We also have performed complementary monitoring observations with KaVA to study the jet kinematics. Our work leads us to the following principal conclusions:

\begin{enumerate}
    \item We measured the frequency dependent position of the core from the 2D cross-correlation analysis, which follows a power-law relation of $r_{\rm core}(\nu) \propto \nu^{-1.39 \pm 0.20}$. The core-shift measurements allow us to infer the origin of the jets by extrapolating the power-law relation to $\nu \xrightarrow{} \infty$, which is $\approx0.11$ mas upstream of the 22 GHz physical core. This is crucial to accurately measure the "jet distance" for both the jet and counterjet and thus to derive accurate jet collimation and acceleration profiles.
    \item We found that the jet geometry transitions from a semi-parabolic shape ($R\propto z^{0.58\pm0.05}$) into a conical/hyperbolical shape ($R \propto z^{1.16\pm0.01}$) at a distance of $z_b=(1.1\pm0.2)\times10^5\ R_g$. The jet collimation profile in the parabolic region is consistent with the profiles of the FR I radio galaxies M87 and NGC 6251 and with the FFE solution for the outermost poloidal magnetic field line anchored to the event horizon on the equatorial plane. We conclude that the jet may be collimated by the pressure of winds, non-relativistic gas outflows launched from hot accretion flows. The jet collimation break occurs at a distance an order of magnitude smaller than the Bondi radius and the black hole sphere of influence radius, which indicates that the winds may not reach down to the Bondi radius. Also, neither a recollimation feature nor significant linear polarization was detected at the jet geometry transition point. This implies that other mechanisms to increase the jet internal pressure than a recollimation shock are needed so that the jet expands conically through the surrounding hot gas having a flat pressure profile.
    \item We derived the jet velocity field at distances of $\approx3,000$--$300,000\ R_g$ based on the assumption that the observed asymmetry in brightness between the jet and counterjet is due to relativistic aberration. We found that the jet gradually accelerates up to the bulk Lorentz factor of $\Gamma \sim 3$ with an acceleration profile of $\Gamma \propto z^{0.30\pm0.04}$ in the same region as the jet collimation zone. The jet decelerates right after the jet collimation break, similar to M87. We conclude that the jet is accelerated to relativistic speeds by converting the electromagnetic energy of the flow to its kinetic energy through the magnetic nozzle effect.
    \item We found an indication of limb-brightening in the jet only in the HSA 43 GHz image. The angular resolution of our VLBI data is significantly smaller than the maximum observed jet radii only at this frequency, indicating that there is a possibility that the jet is intrinsically limb-brightened at other distances as well on pc-scales. As the NGC 315 jets have a relatively large viewing angle of $\approx50^\circ$, the flux enhancement expected from the Doppler boosting effect is at most by a factor of about two. This implies that either (i) the jet spine is much faster than the jet sheath on pc-scales, even though this velocity structure is difficult to reproduce with the MHD jet acceleration model, or (ii) the jet sheath has much higher emissivity than the spine due to the interaction with the surrounding medium.
    \item Our monitogring observations with KaVA have shown that the jet structure at distances $\lesssim5$ mas from the apparent core appear very stationary over eight months. We argue that this does not mean that the jet is actually stationary but is because of the re-brightened regions in the jet, similar to M87, which can make the jet appear stationary. We could detect an outward motion with an apparent jet speed of $\beta_{\rm app} = 1.85\pm0.44c$ at a distance of $\approx8$ mas. Combining this speed with the observed jet-to-counterjet brightness ratio at a similar distance, we derive a jet viewing angle of $\theta\approx52.8\pm8.0^\circ$, which is in good agreement with the viewing angle constrained on kpc-scales.
\end{enumerate}

We finally remark that NGC 315 is the third radio-loud AGN for which a firm evidence for jet acceleration and collimation occurring simultaneously over a large range of jet distance (after M87 and 1H 0323+342), as the MHD jet acceleration model has predicted. The indication of limb-brightening in the jet also provides hints for velocity stratification and non-thermal particle acceleration mechanisms in AGN jets. We plan to search other AGNs for jet acceleration, collimation, and limb-brightening to have a more complete view in the near future.

\vspace{-0.1cm}
\acknowledgments

We thank the anonymous ApJ referee for detailed comments that improved the manuscript. J.P. thanks Hung-Yi Pu for useful discussions. J.P. acknowledges financial support from the Korean National Research Foundation (NRF) via Global PhD Fellowship Grant 2014H1A2A1018695 and support through the EACOA Fellowship awarded by the East Asia Core Observatories Association, which consists of the Academia Sinica Institute of Astronomy and Astrophysics, the National Astronomical Observatory of Japan, Center for Astronomical Mega-Science, Chinese Academy of Sciences, and the Korea Astronomy and Space Science Institute. This work is supported by the Ministry of Science and Technology of Taiwan grant MOST 109-2112-M-001-025 (K.A). The VLBA is an instrument of the National Radio Astronomy Observatory. The National Radio Astronomy Observatory is a facility of the National Science Foundation operated by Associated Universities, Inc. This work is based in part on observations made with the KaVA, which is operated by the the Korea Astronomy and Space Science Institute and the National Astronomical Observatory of Japan.

%\vspace{5mm}
\facilities{VLBA (NRAO), HSA (NRAO), VLA (NRAO), KaVA (KASI/NAOJ)}

\software{AIPS \citep{Greisen2003}, Difmap \citep{Shepherd1997}, GPCAL \citep{Park2020}, Numpy \citep{Numpy2011}, Scipy \citep{Scipy2020}, Pandas \citep{Pandas2010}, Astropy \citep{Astropy2013, Astropy2018}, Matplotlib \citep{Matplotlib2007}}

% $\quad$
\appendix

\section{Identification of Apparent Core Position}
\label{appendix:core_identification}

In this appendix, we derive the distance between the physical core, i.e., the $\tau=1$ surface, and the apparent core, which is defined as the brightest pixel in the core region in the image (Section~\ref{sec:coreshift}). The apparent core can be separated from the physical core to the extended jet side due to the finite beam size and the core-jet blending effect. We extend the approach used by \cite{Hada2014}. For each frequency, we increase the convolving beam size in steps of $10\%$ of the original beam size up to 2.5 times the beam size and found that the brightest pixel position progressively moves to the extended jet side.

\begin{figure}[t!]
\centering
\includegraphics[width = 0.49\textwidth]{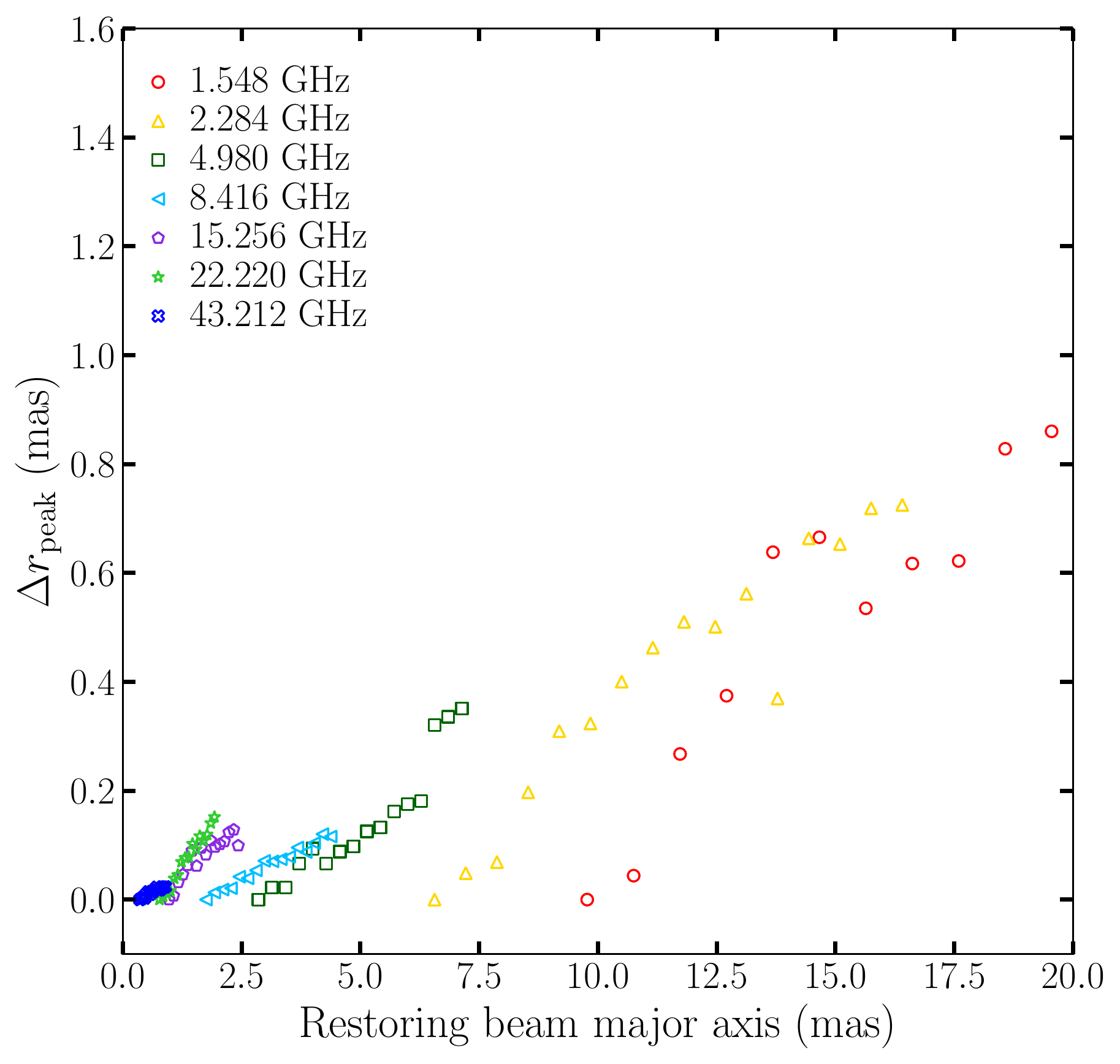}
\includegraphics[width = 0.49\textwidth]{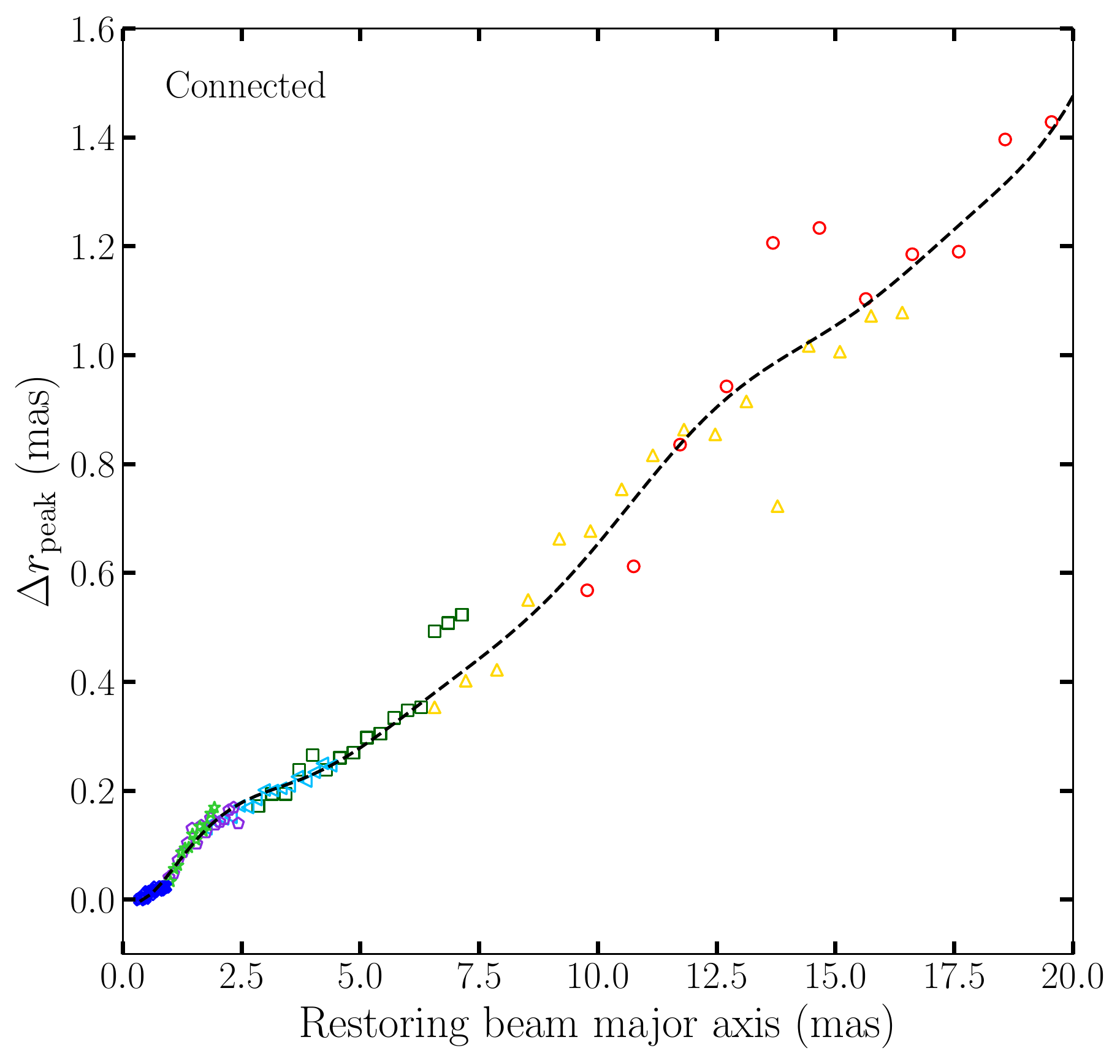}
\caption{The amount of shift of the apparent core as a function of convolving beam major axis. The beam size spans $1-2.5$ times the nominal beam size and the leftmost data point at each frequency has a zero shift (left). The data points at lower frequencies are moved upwards by certain amounts to match the data points at higher frequencies, connecting the data points at different frequencies (right). The black dashed line is a high order polynomial function fitted to the data points. \label{fig:core_identification}}
\end{figure}

The left panel of Figure~\ref{fig:core_identification} shows the amount of shift of the apparent core ($\Delta r_{\rm peak}$) as a function of major axis of the convolving beam for different frequencies. The leftmost data point for each frequency is for the nominal beam and has a zero shift by construction. $\Delta r_{\rm peak}$ nearly linearly increases with convolving beam size for each frequency. We assume that $\Delta r_{\rm peak}$ can solely be determined by convolving beam size, regardless of observing frequency. If this assumption holds, then we can infer $\Delta r_{\rm peak}$ at beam sizes smaller than the nominal beam size at a certain frequency from $\Delta r_{\rm peak}$ obtained at beam sizes larger than the nominal beam size at higher frequencies. In other words, we can connect $\Delta r_{\rm peak}$ at different frequencies by moving the data points at lower frequencies upwards by certain amounts to match the data points at higher frequencies. The right panel is after connecting the data points at different frequencies. We moved the data points upwards to minimize the scatters of the overlapping data points at adjacent frequencies. We found an abrupt increase in $\Delta r_{\rm peak}$ at 5 GHz for large restoring beam sizes and did not consider the outermost three data points robust.

We fitted a high order polynomial function and obtained a smooth curve which describes the data well (the black dashed line). We obtain the representative distance between the apparent core and the physical core ($r_{\rm peak}-r_{\rm core}$) from the value on the function at the nominal beam size for each frequency. We could not obtain the distance between the cores at 43 GHz because there is no higher frequency data having smaller beam sizes. We infer the distance by extrapolating a linear function fitted to the 43 GHz data down to a zero beam size, which is $\approx0.018$ mas.

We note that this approach is based on the assumption that the separation of the apparent core from the physical core depends solely on convolving beam size. This assumption holds only when the source has a flat radial spectral index profile. In this case, the intensities scale with frequency by the same factor at all jet distances and $r_{\rm peak}-r_{\rm core}$ does not depend on frequency. However, in reality, the physical core is optically thick and the extended jet is optically thin. Thus, the contribution of the extended jet emission to the shift of the apparent core can be larger at lower frequencies and one cannot simply connect $\Delta r_{\rm peak}$ at different frequencies. Nevertheless, we found that the trends of $\Delta r_{\rm peak}$ at different frequencies are quite similar in the overlapping regions (the right panel of Figure~\ref{fig:core_identification}), which indicates that our assumption is reasonably good.

\begin{figure}[t!]
\centering
\includegraphics[width = 0.49\textwidth]{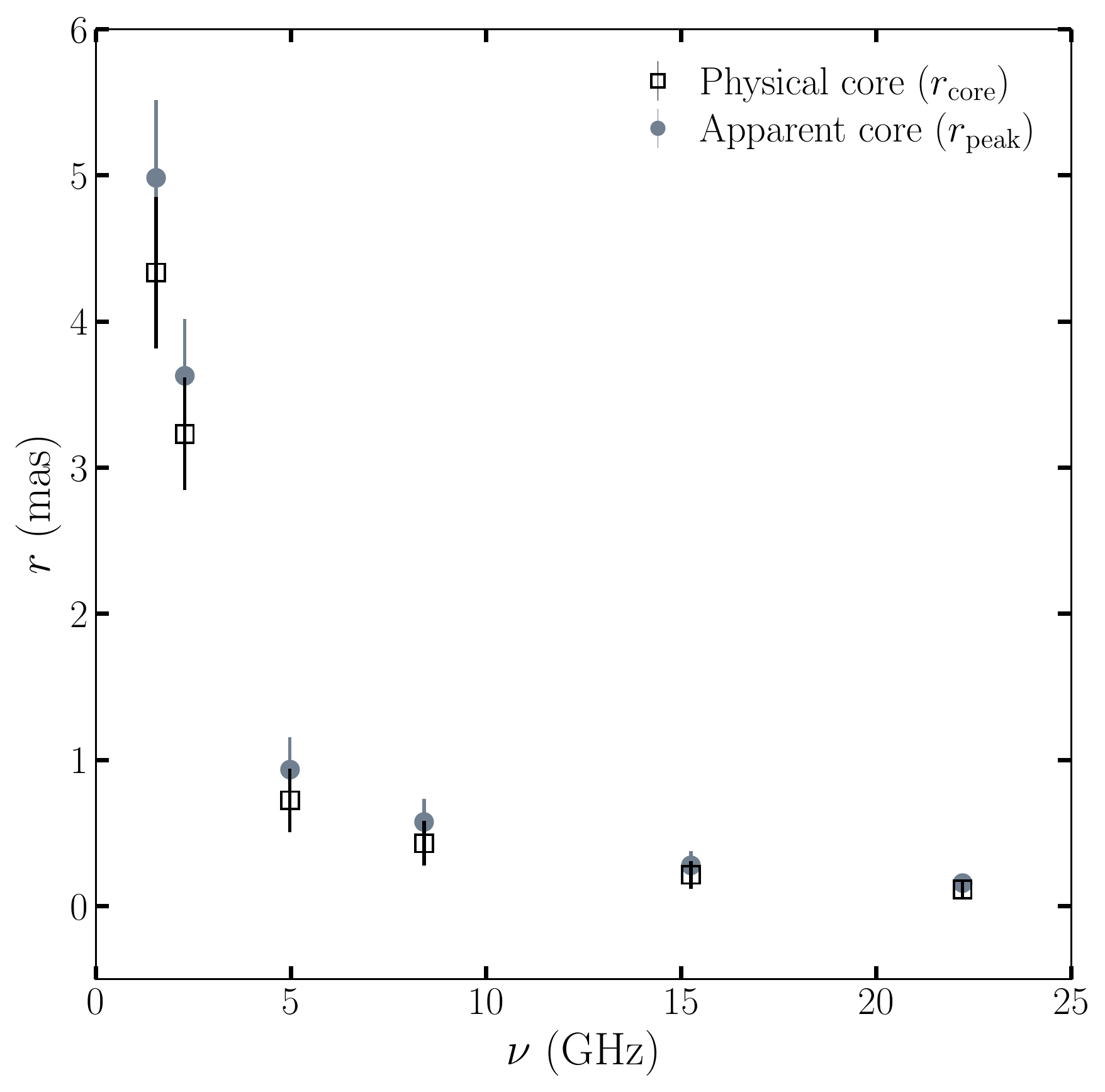}
\caption{Core position with respect to the inferred jet base as a function of frequency. The black open squares denote the positions of the physical cores ($r_{\rm core}$), presented in Figure~\ref{fig:coreshift}, but are shifted upwards by $\approx0.11$ mas, which is the distance between the 22 GHz physical core and the inferred jet base. The grey filled circles are the positions of the apparent cores ($r_{\rm peak}$). \label{fig:app_coreshift}}
\end{figure}

We estimate the uncertainties in $r_{\rm peak}-r_{\rm core}$ by comparing them with the amount of shift expected when extrapolating the linear trend of $\Delta r_{\rm peak}$ down to a zero beam size at each frequency (before connecting $\Delta r_{\rm peak}$). The shift derived in this way may be at the extreme end of the allowed range of $r_{\rm peak}-r_{\rm core}$ because Figure~\ref{fig:core_identification} already shows that a single linear function cannot describe the data in a wide range of convolving beam size well. Nevertheless, those estimates may serve as a good tool to infer the uncertainties. We present $r_{\rm peak}-r_{\rm core}$ and their uncertainties in Table~\ref{tab:app_coreshift} and the positions of apparent cores with respect to the inferred jet base in Figure~\ref{fig:app_coreshift}. The derived $r_{\rm peak}-r_{\rm core}$ values are around 1/10 of the minor axis of the synthesized beams and are comparable to the uncertainties in the physical core positions (Table~\ref{tab:coreshift}). We tested that our results and main conclusions are not changed even when we assume zero $r_{\rm peak}-r_{\rm core}$ for all frequencies.

\begin{deluxetable}{ccccccc}
\tablecaption{Apparent core positions \label{tab:app_coreshift}}
\tablewidth{0pt}
\tablehead{
 & \colhead{1.548 GHz} & \colhead{2.284 GHz} & \colhead{4.980 GHz} & \colhead{8.416 GHz} & \colhead{15.256 GHz} & \colhead{22.220 GHz}
}
\startdata
$r_{\rm peak}-r_{\rm core}$ & $0.648\pm0.120$ & $0.398\pm0.063$ & $0.210\pm0.027$ & $0.147\pm0.032$ & $0.065\pm0.015$ & $0.043\pm0.028$
\enddata
\tablecomments{Separation between the apparent core ($r_{\rm peak}$) and the physical core ($r_{\rm core}$).}
\end{deluxetable}

\section{Error Analysis for Core-shift}
\label{appendix:2dcc_error}

We obtained the core-shifts between frequencies by performing two-dimensional cross-correlation of the optically thin jet emission in the VLBA images (Section~\ref{sec:coreshift}). We derive the uncertainties of the core-shifts in the following way. We found that the spectral index maps after the image registration show flat radial profiles on top of which small and gradual changes exist in the optically thin jet regions. If the image registration is made using wrong core-shift values, then we expect that the resulting radial $\alpha$ profiles will show notable deviations from the flat and smooth profiles that are obtained with the best core-shift estimates.

\begin{figure}[t!]
\centering
\includegraphics[width = 0.49\textwidth]{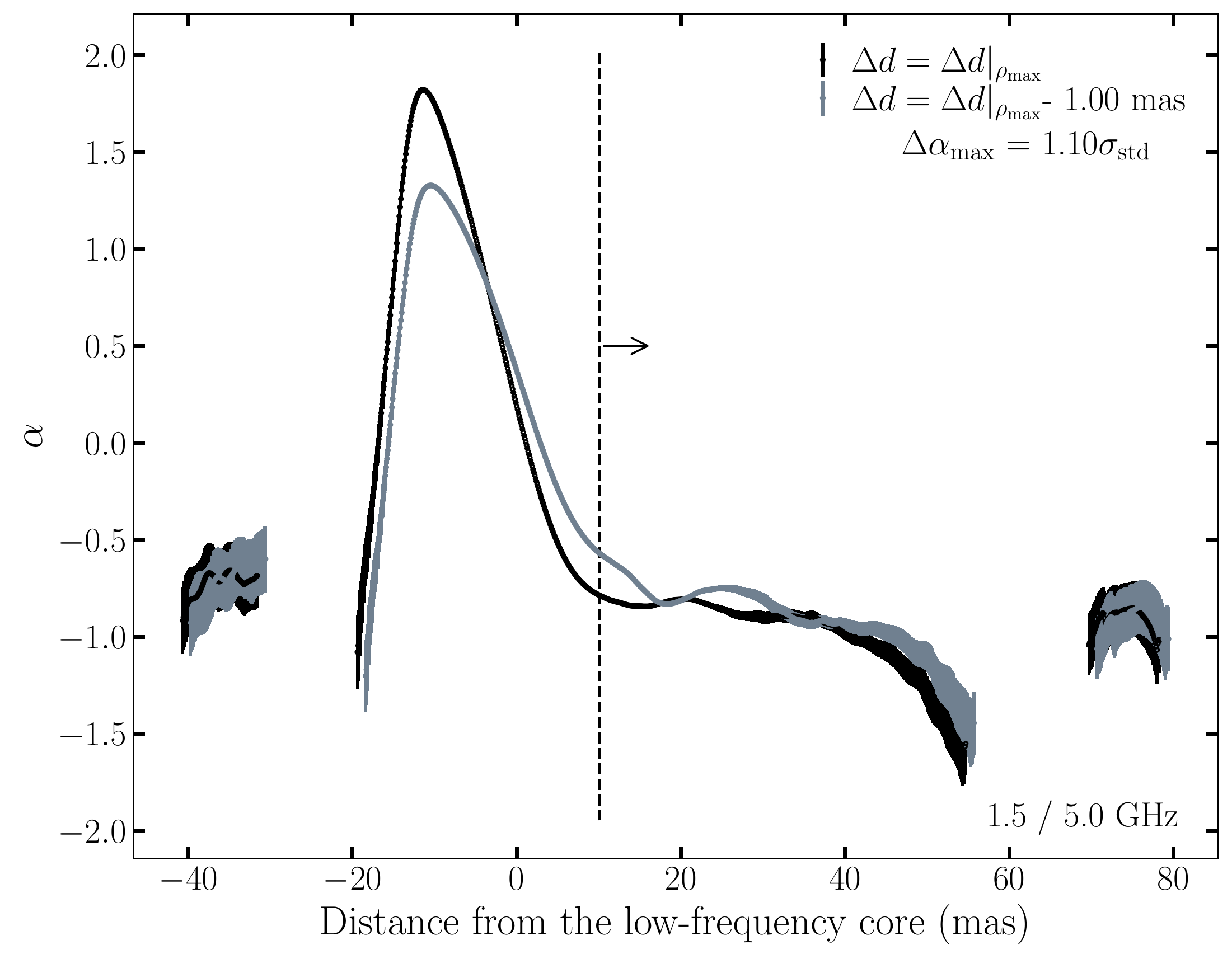}
\includegraphics[width = 0.49\textwidth]{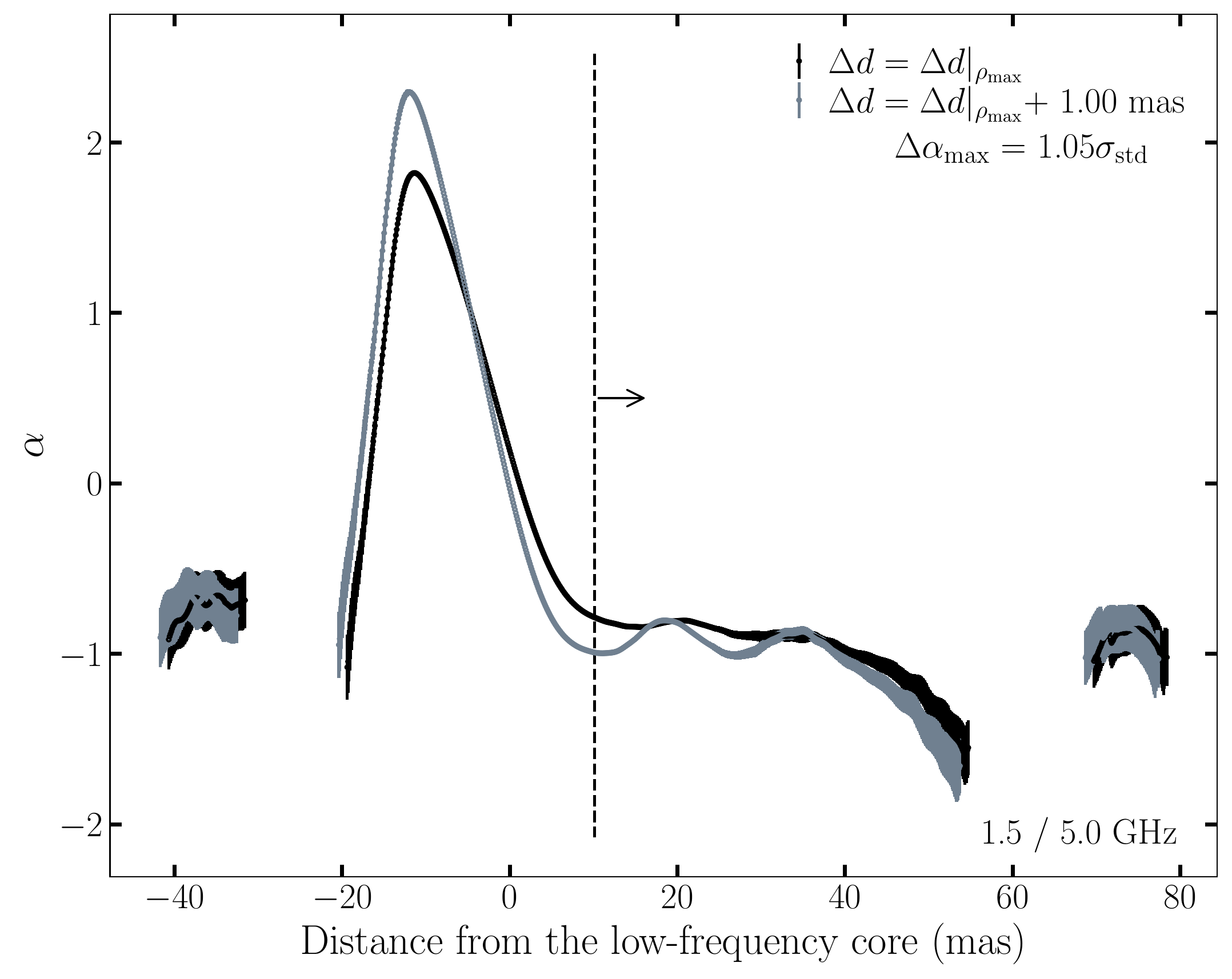}
\caption{Radial spectral index profiles for an example frequency pair of 1.5 and 5.0 GHz. The black curves are obtained with a shift of the high-frequency map using the best core-shift estimate (with the shift of $\Delta d = \Delta d|_{\rho_{\rm max}}$, where $\Delta d|_{\rho_{\rm max}}$ is the amount of the shift that provides the highest cross-correlation coefficient) and the grey curves are obtained with additional shifts along the jet direction by $-1$ (left) and $+1$ mas (right). $\sigma_{\rm std}$ is the standard deviation of $\alpha$ in the black curve in the optically thin regions at distances larger than that marked by the vertical black dashed line. $\Delta\alpha_{\rm max}$ is the maximum absolute deviation between the two profiles in the optically thin regions. $\Delta\alpha_{\rm max}$ is derived using the pixels for which the deviations between the two curves are larger than their $1\sigma$ uncertainties. \label{fig:2dcc_error}}
\end{figure}

Based on this argument, we first obtain a radial $\alpha$ profile for each frequency pair using our best core-shift estimate (the black curves in Figure~\ref{fig:2dcc_error}). We estimate the standard deviation of $\alpha$ ($\sigma_{\rm std}$) in the optically thin region of the radial profile (marked as the vertical dashed lines and the rightward arrows), at distances separated from the core by more than $\approx1.5-2$ synthesized beam size, and regard $\sigma_{\rm std}$ as the natural fluctuation in $\alpha$ expected in the optically thin region. We introduce additional shifts along the jet direction to the higher frequency map and obtain the corresponding radial $\alpha$ profile for each additional shift (the grey curves). The artificial profiles with the additional shifts indeed show larger oscillations in $\alpha$ than those in the original profile. We compare the profiles for many additional shifts with the amounts of the shifts being integers of the image pixel size. If the maximum absolute deviation between the profiles ($\Delta\alpha_{\rm max}$) at any distance in the optically thin region exceeds\footnote{We consider the deviation significant when it is larger than the $1\sigma$ errors of the two profiles.} $\sigma_{\rm std}$, we regard the amount of the additional shift as a $1\sigma$ uncertainty in the core-shift, which introduces a notable artificial fluctuation in the radial $\alpha$ profile. The additional shifts were introduced in both directions along the jet axis. We take an average of the $1\sigma$ uncertainties obtained from the shifts along the two directions, which are nearly identical to each other for most frequency pairs. We add half of the pixel sizes to the uncertainties in quadrature to take into account additional (small) errors caused by the finite pixel sizes. We present the uncertainties in the core-shifts in Table~\ref{tab:coreshift}.

\section{Jet Kinematics on Scales of $\lesssim10$ mas Based on KaVA Observations}
\label{appendix}

In this Appendix, we present the results of our monitoring observations of NGC 315 with KaVA. The observations were performed at 22 GHz at a recording rate of 1 Gbps in eight epochs from late 2019 to mid-2020 with a monitoring interval of about a month. The data were correlated by the Daejeon correlator at the Korea–Japan Correlation Center \citep{Lee2014, Lee2015}. We performed a standard data reduction by following our previous study of the M87 jet with KaVA observations \citep{Park2019a}. We summarize the basic information of our observations in Table~\ref{tab:kava_data}.

\begin{deluxetable}{cccccc}
\tablecaption{KaVA Monitoring Observations of NGC 315\label{tab:kava_data}}
\tablewidth{0pt}
\tablehead{
Proj. Code & Obs. Date & \colhead{Freq.} & \colhead{Beam Size} & \colhead{$I_{\rm p}$} & \colhead{$I_{\rm rms}$} \\
&& \colhead{(GHZ)} & \colhead{(mas $\times$ mas, degree)} & \colhead{(Jy/B)} & \colhead{(mJy/B)}
}
\startdata
\hline
\multicolumn{6}{c}{KaVA}\\
\hline
k19jp01c & 2019 Nov 18 & \multirow{ 8}{*}{22.227} & $1.468\times1.348$, -76.544 & 0.473 & 0.123 \\
k19jp01d & 2019 Dec 16 & & $1.543\times1.177$, -45.036 & 0.451 & 0.197 \\
k19jp01e & 2020 Jan 11 & & $1.551\times1.342$, 34.424 & 0.435 & 0.179 \\
a2006a & 2020 Jan 31 & & $1.430\times1.262$, -59.899 & 0.465 & 0.119 \\
a2006b & 2020 Feb 23 & & $1.345\times1.195$, -49.210 & 0.457 & 0.176 \\
a2006c & 2020 Mar 23 & & $1.411\times1.285$, -43.754 & 0.453 & 0.125 \\
a2006d$^1$ & 2020 Apr 20 & & $1.436\times1.297$, -33.684 & 0.461 & 0.141 \\
a2006e & 2020 May 13 & & $1.264\times1.182$, -38.913 & 0.435 & 0.192 \\
\enddata
\tablecomments{Same as Table~\ref{tab:data} but for monitoring observations with KaVA. $^1$KUS (KVN Ulsan station) missing.}
\end{deluxetable}

\begin{figure}[t!]
\centering
\includegraphics[width = 0.49\textwidth]{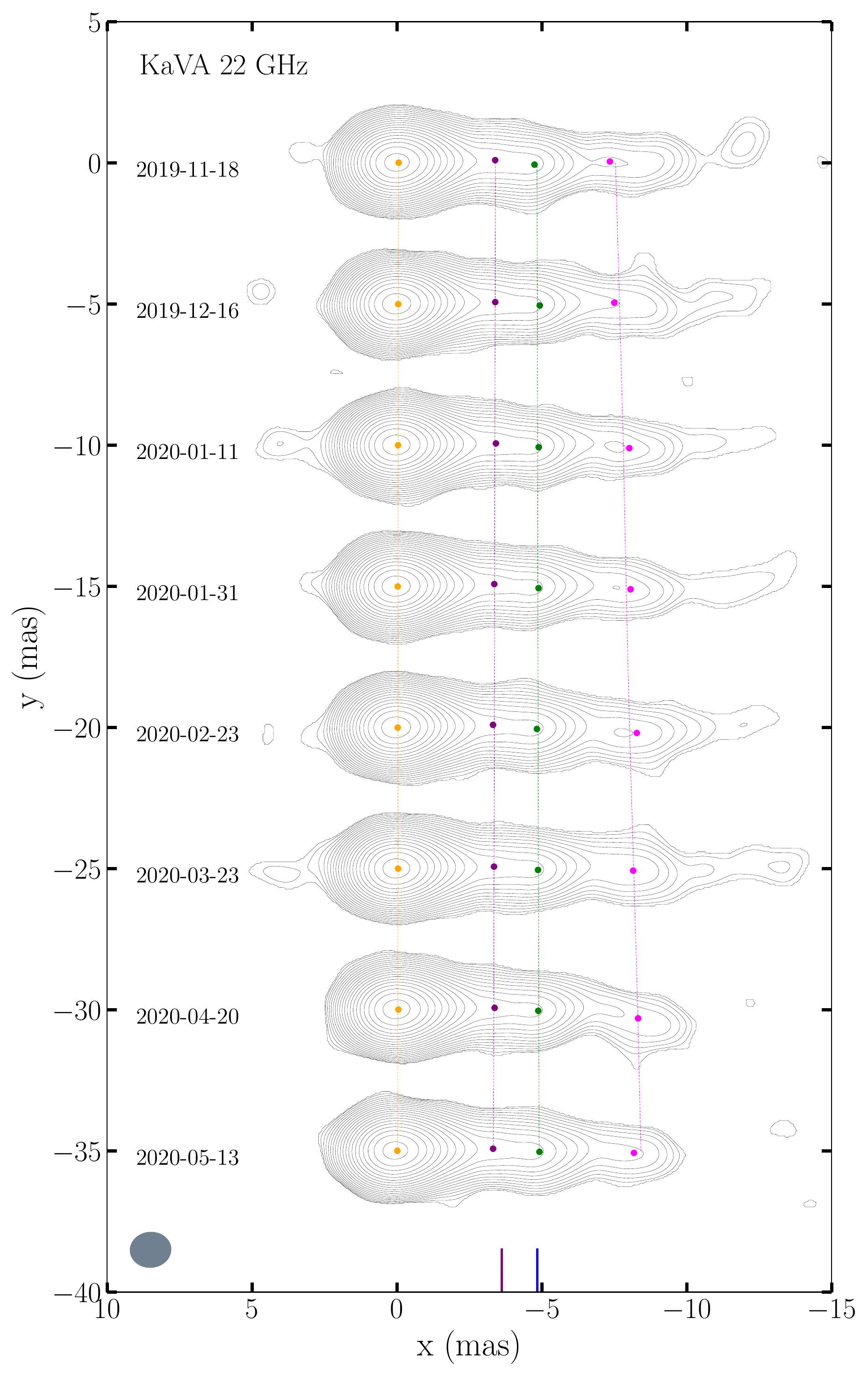}
\caption{Contours show total intensity of the NGC 315 jet observed with KaVA at 22 GHz in eight epochs from late 2019 to mid-2020. The maps are rotated clockwise by $40^\circ$. The observation date is indicated in the left of each map. The small colored circles show the positions of the SSPs identified in more than four successive epochs from the WISE analysis for the SWD scale of 0.48. The dotted lines are the best-fit linear functions to the identified SSP positions over time. The purple and blue solid vertical lines on the bottom x-axis indicate the average positions of the \texttt{modelfit} components in a similar distance range obtained in the MOJAVE program over $\approx18$ years \citep{Lister2019}. \label{fig:kava_contours}}
\end{figure}

The observed jet structures in different epochs are shown in Figure~\ref{fig:kava_contours}. The structures appear to be nearly identical in different epochs at distances less than $\approx5$ mas, while significant changes in the structures could be seen in the outer regions. We first attempted to derive the apparent jet speeds using the standard \texttt{modelfit} analysis in Difmap by fitting multiple circular Gaussian components to the visibility data. However, we found that the numbers of fitted components in different epochs are not the same, which prevents a reliable kinematic analysis using this approach \citep[See][for a related discussion]{Park2019a}.

We instead used the Wavelet-based Image Segmentation and Evaluation (WISE) analysis technique \citep{ML2015, Mertens2016}. This technique allows us to identify local jet brightness distributions with arbitrary patterns (not predefined templates such as 2D Gaussian distributions) in different epochs, which is suitable for jet kinematic studies of nearby radio galaxies \citep{Mertens2016, Boccardi2019, Park2019a}. We applied WISE to our KaVA images in a similar manner to our previous jet kinematic study of M87 \citep{Park2019a}. We detected significant structural patterns (SSPs) with a $3\sigma$ detection threshold on scales of 0.24, 0.36, 0.48, 0.72, 0.96, and 1.44 mas determined by the segmented wavelet decomposition (SWD) and intermediate wavelet decomposition (IWD) methods. We identify SSPs in adjacent epochs by using the multiscale cross-correlation (MCC) method with a tolerance factor of 1.5 and a correlation threshold of 0.7. We search for velocity vectors with symmetric search windows of [$-20$, $+20$] and [$-5$, $+5$] $\rm mas\ yr^{-1}$ in a longitudinal and transverse direction relative to the jet axis, respectively. We obtained the apparent speeds from a line fitting to the separation from the core with time for each identified SSP.

The obtained speeds at distances $\lesssim5$ mas from the core are all consistent with zero within 1--2$\sigma$ on all SWD/IWD scales, as expected from the very similar distributions of jet brightness in different epochs (Figure~\ref{fig:kava_contours}). Outward motions are detected at $\approx$8 mas, but the results are not consistent between different SWD/IWD scales. On the smallest scale of 0.24 mas, we found that the positions of the SSPs at $\approx$7--8 mas abruptly shift inward between the fifth and sixth epochs by more than 1 mas; this is likely because WISE tries to split the region into two separate SSPs in the sixth and seventh epochs, which could result in the inward shift due to possible mis-identification of SSPs. On the three largest scales of 0.72, 0.96, and 1.44 mas, WISE detects a single SSP at $\approx8$ mas in the first two epochs, while significant extended jet emission at distances down to $\approx10$ mas is detected in those epochs. This results in nearly stationary apparent motions detected on those SWD/IWD scales, while it is evident from the maps that the positions of a locally bright region at $\approx$8 mas gradually moves outward over time. This apparent motion was captured by WISE on the intermediate scales of 0.36 and 0.48 mas and the observed apparent speeds are $1.19\pm0.71$ and $1.81\pm0.47\rm\ mas\ yr^{-1}$, which corresponds to $1.36\pm0.81$ and $2.06\pm0.53c$, respectively. We present the positions of the SSPs identified over more than four successive epochs on the SWD scale of 0.48 mas in Figure~\ref{fig:kava_contours}. The weighted average of the two apparent speeds is $1.85\pm0.44c$ at the mean distance of $\approx7.95$ mas from the core.

The results of jet kinematic analysis from our dense monitoring observations with KaVA are quite different from previous monitoring observations. \cite{Cotton1999} found many fast apparent motions with the apparent speeds of 0.81--$1.79c$ at distances of $\approx$3--12 mas from the core from observations in three epochs with an interval of about one year. We note that different sensitivities in the different epochs and the large time separation between epochs can easily lead to artificial motions of the jets. Their observed apparent motions may reflect a long-term evolution of the jet brightness distribution, as seen between our HSA 43 GHz image and the VLBA images at $\lesssim22$ GHz (Figure~\ref{fig:contours_shift}). This evolution, if it exists, occurs on time scales of years, as we could not see an evidence for such evolution in our KaVA images spanning about eight months in time. \cite{Lister2018} found sub-luminal motions at distances $\lesssim7$ mas from the core from the Monitoring of Jets in AGNs with VLBA Experiments (MOJAVE) program, which is not consistent with the fast motion detected in our observations at $\approx7$--8 mas. This is also possibly due to the large monitoring interval and reminiscent of the sub-luminal motions detected by the MOJAVE observations of the M87 jet \citep{Kovalev2007}, while much faster motions have been observed from higher cadence observations at similar jet distances (e.g., \citealt{Mertens2016, Hada2016, Hada2017, Walker2018, Park2019a}).

The positions where the stationary motions are detected, at $\approx3.5$ and $5$ mas from the core, correspond to the re-brightened regions seen in our higher-resolution VLBA maps (see the bottom right panel of Figure~\ref{fig:contours_shift}). These re-brightened regions appear to be stationary when observed with a low-resolution, even if there are fast jet motions moving into and out from the regions, as demonstrated in our previous study of M87 \citep{Park2019a}. NGC 315 is more distant than M87 by a factor of four and has a less massive black hole by a factor of three, which makes it challenging to detect fast motions near the re-brightened regions even if they exist. We note that the locations of the re-brightened regions appear to be similar to the positions of the \texttt{modelfit} components in a similar distance range obtained in the MOJAVE program \citep{Lister2019}. These positions have been changed by less than $\approx0.6$ mas over $\approx18$ years, which indicates that the NGC 315 jet shows a long-term evolution of the brightness distribution.

The fast apparent motion of $\beta_{\rm app} = 1.85\pm0.44c$ allows us to constrain the jet viewing angle through the relation of $\beta_{\rm app} = \beta\sin\theta / (1-\beta\cos\theta)$ and by combining with Equation~\ref{eq:ratio}. We take $R$ and $\alpha$ detected at the nearest distance to the observed motion (Figure~\ref{fig:properties_binned}, see also Section~\ref{sec:velocity}) and derived the jet viewing angle of $\theta = 52.8\pm8.0^\circ$. This value is in good agreement with the viewing angle constrained by modelling the jet and counterjet on kpc-scales, which is $\theta=49.8^{\circ+0.5}_{-0.2}$ \citep{LB2014}.

\section{Verification of jet speeds at short distance}
\label{appendix:beta}

\begin{figure}[t!]
\centering
\includegraphics[width = 0.49\textwidth]{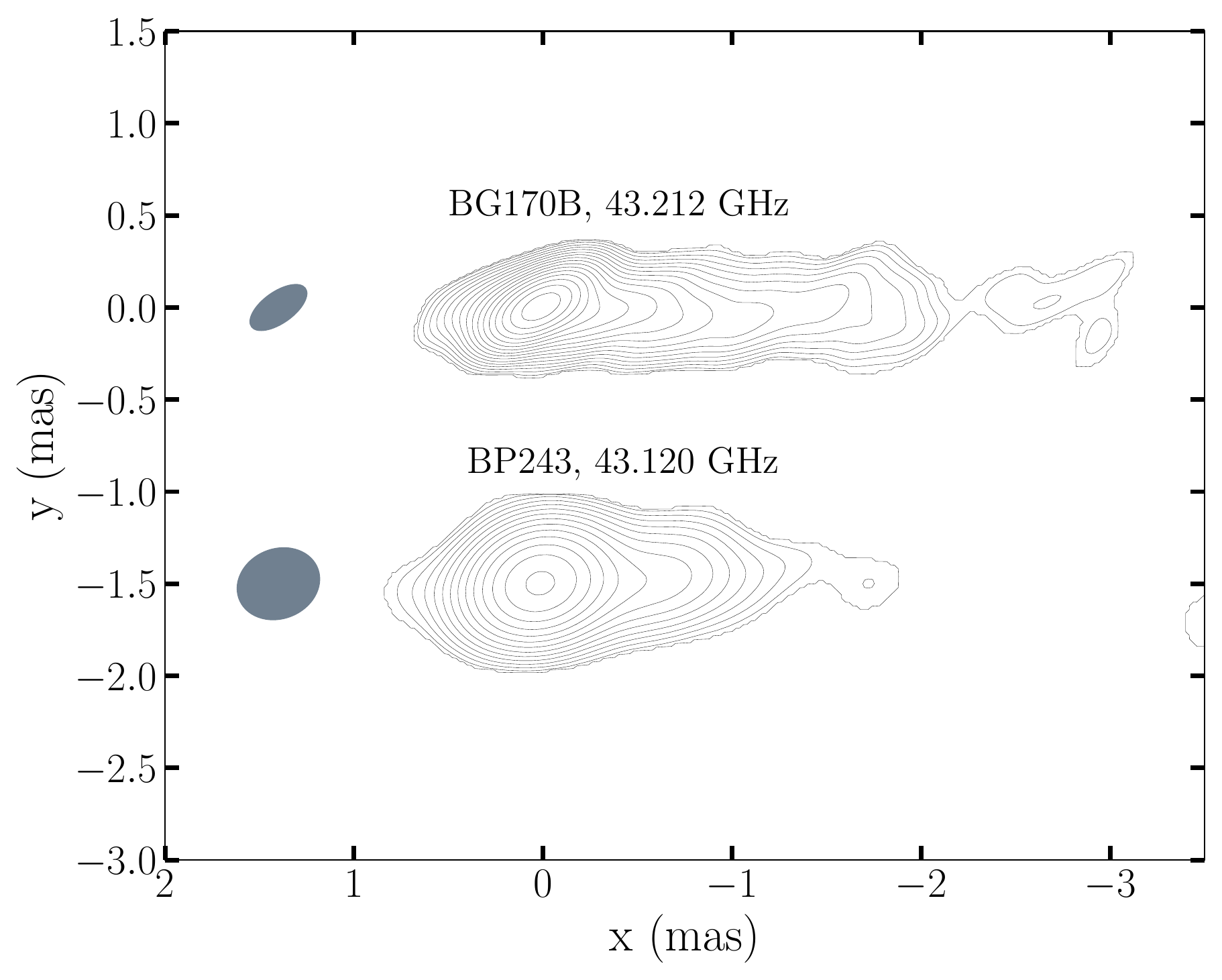}
\caption{CLEAN maps of NGC 315 from the archival HSA data at 43 GHz (upper, used in our main analysis) and from the VLBA observation at 43 GHz (lower). The maps are rotated clockwise by $40^\circ$. Contours start at 0.6 and 2.4 mJy/beam for the upper and lower maps, respectively, and increase by factors of the square root of two. The grey shaded ellipses show the synthesized beams under the natural weighting of the data. \label{fig:map_q2}}
\end{figure}

\begin{figure}[t!]
\centering
\includegraphics[width = 0.49\textwidth]{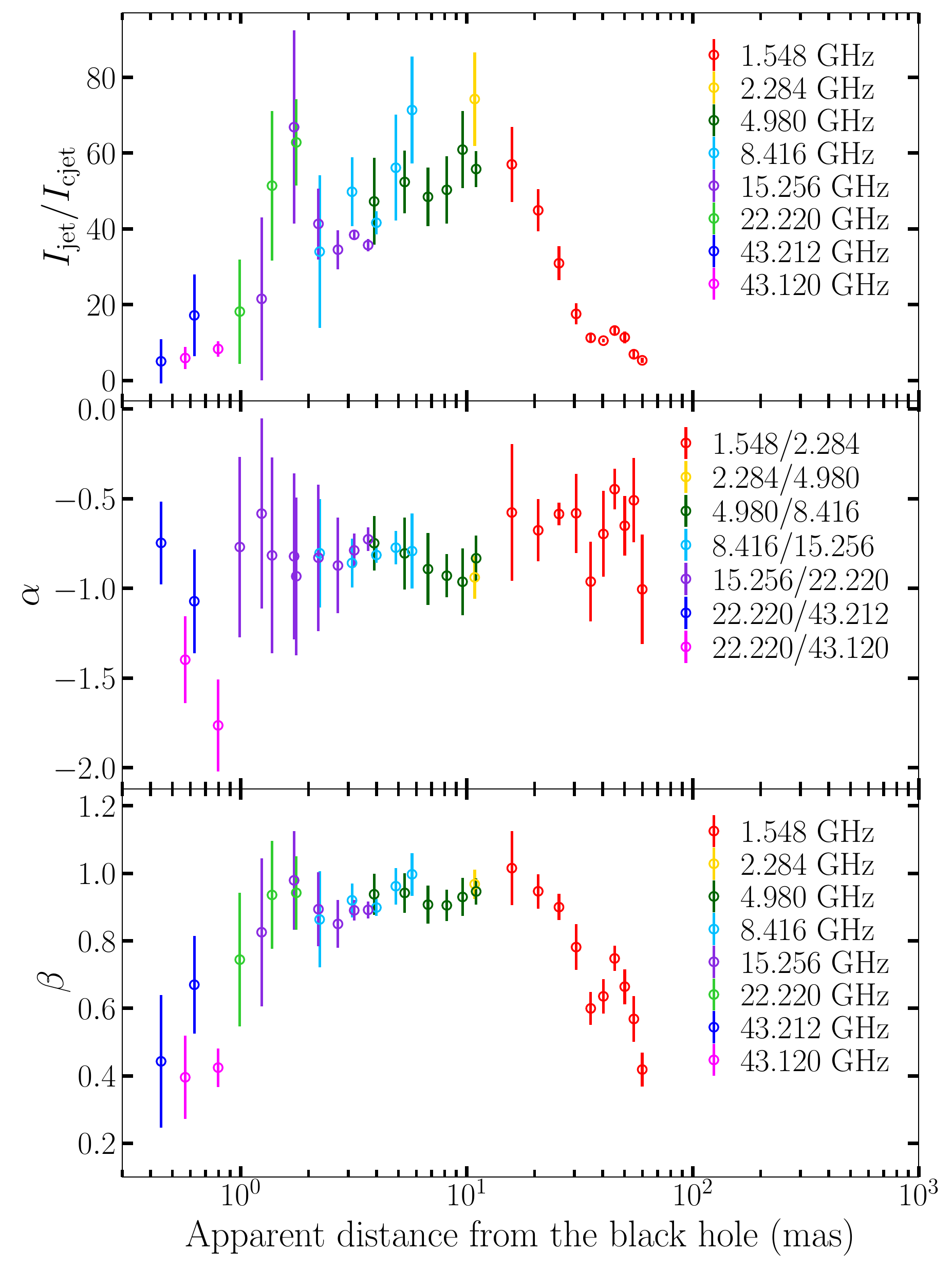}
\caption{Same as Figure~\ref{fig:properties_binned} but including the results from the VLBA data at 43.120 GHz (magenta data points). \label{fig:properties_binned_q2}}
\end{figure}

In this appendix, we present the jet speeds measured from the VLBA 43 GHz data, which was observed nearly simultaneously to the VLBA data at other frequencies but not used for our main analysis due to the limited data quality (Section~\ref{sec:data}). Although the data quality is not very good, this is the only data with which we could test whether our conclusion of significant jet acceleration on pc-scales (Section~\ref{sec:velocity}), which relies on the HSA 43 GHz data, is robust against the large time difference between 43 GHz and other VLBA frequencies. We present the CLEAN map obtained from the VLBA 43 GHz data and compare it with the HSA 43 GHz map in Figure~\ref{fig:map_q2}. We detected the jet and counterjet using this data extending down to $\approx1.4$ and 0.7 mas from the core, respectively. We obtained $R$, $\alpha$, and $\beta$ in the same way as in Section~\ref{sec:velocity}, and present the results in Figure~\ref{fig:properties_binned_q2}.

We found that the jet intensity ratios are consistent with the HSA 43 GHz data. However, the spectral indices appear to slightly deviate from the other data points, which show a nearly constant $\alpha\approx-0.8$. This results in lower $\beta$ than the HSA 43 GHz data. We speculate that the flux in the VLBA 43 GHz data may be underestimated due to a relatively large antenna pointing error, which can easily lead to underestimation of $\alpha$ (and thus $\beta$ as well). The total CLEAN fluxes obtained with the VLBA data at 8.4, 15.3, 22.2, and 43.1 GHz are 0.63, 0.56, 0.52, and 0.31 Jy, respectively, showing an abrupt drop between 22 and 43 GHz as compared to the trend at lower frequencies. Furthermore, the gain correction factors obtained during self-calibration at $\lesssim22$ GHz are within $\approx20\%$ from unity for most scans, while the VLBA 43 GHz data shows correction factors larger than $\approx30$--40\% for many scans, suggesting that there could be severe pointing errors, which always leads to underestimation (not overestimation) of visibility amplitudes. Nevertheless, this result implies that $\beta$ at the shortest distance bins from the HSA 43 GHz data may not be underestimated and we conclude that the NGC 315 jets do accelerate from non-relativistic to relativistic speeds on pc-scales.

\section{Comparison between Jet Radius and Counterjet Radius}
\label{appendix:radii}

\begin{figure}[t!]
\centering
\includegraphics[width = 0.49\textwidth]{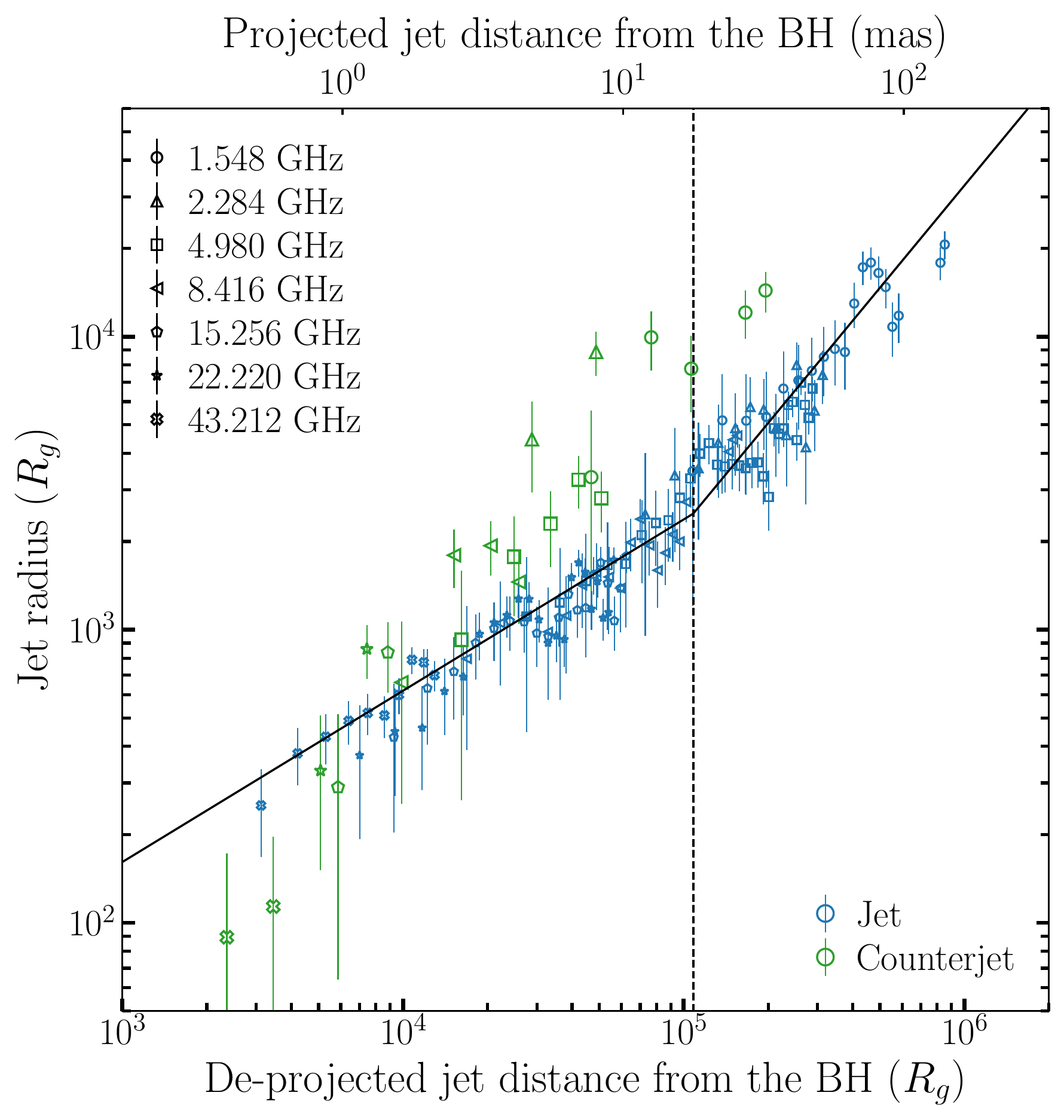}
\caption{Same as Figure~\ref{fig:jet_radius} but showing the radii of the jet (blue) and counterjet (green) obtained by the VLBA/HSA data. The data at different frequencies are shown in different symbols. The counterjet radius is generally larger than the jet radius at the same distance, which may imply that the jet velocity field is stratified. \label{fig:cjet_width}}
\end{figure}

In this appendix, we compare the radii of the jet and counterjet. We obtained the radius of the counterjet as a function of distance in the same way as for the jet (Section~\ref{sec:collimation}) and present them in Figure~\ref{fig:cjet_width}. We found that the radius of the counterjet is generally larger than that of the jet at the same distance. This can be explained if the jet spine is much faster than the jet sheath. In this case, the brightness ratio between the jet and counterjet for the spine emission becomes larger than that for the sheath (Equation~\ref{eq:ratio}). We measure the jet radii from the FWHMs of the transverse intensity profiles. Therefore, our results may support the fast-spine and slower-sheath scenario for the NGC 315 jets on pc-scales, even though how this is physically possible is still questionable (Section~\ref{dis:acceleration}). We note, however, that jet radii measurements can be subject to large uncertainties because the FWHMs (not only intensity peaks) have to be constrained accurately. One can assure that the results are robust if we find similar counterjet radii measured at the same distance but at different frequencies, as we did for the jet. Unfortunately, this was not possible with our data and requires future observations with high sensitivity for further investigation.

\bibliography{AAS27832R1.bib}{}

\begin{thebibliography}{}
\expandafter\ifx\csname natexlab\endcsname\relax\def\natexlab#1{#1}\fi
\providecommand{\url}[1]{\href{#1}{#1}}
\providecommand{\dodoi}[1]{doi:~\href{http://doi.org/#1}{\nolinkurl{#1}}}
\providecommand{\doeprint}[1]{\href{http://ascl.net/#1}{\nolinkurl{http://ascl.net/#1}}}
\providecommand{\doarXiv}[1]{\href{https://arxiv.org/abs/#1}{\nolinkurl{https://arxiv.org/abs/#1}}}

\bibitem[{{Aharonian} {et~al.}(2006){Aharonian}, {Akhperjanian}, {Bazer-Bachi},
  {Beilicke}, {Benbow}, {Berge}, {Bernl{\"o}hr}, {Boisson}, {Bolz}, {Borrel},
  {Braun}, {Brown}, {B{\"u}hler}, {B{\"u}sching}, {Carrigan}, {Chadwick},
  {Chounet}, {Coignet}, {Cornils}, {Costamante}, {Degrange}, {Dickinson},
  {Djannati-Ata{\"\i}}, {Drury}, {Dubus}, {Egberts}, {Emmanoulopoulos},
  {Espigat}, {Feinstein}, {Ferrero}, {Fiasson}, {Fontaine}, {Funk}, {Funk},
  {F{\"u}{\ss}ling}, {Gallant}, {Giebels}, {Glicenstein}, {Goret},
  {Hadjichristidis}, {Hauser}, {Hauser}, {Heinzelmann}, {Henri}, {Hermann},
  {Hinton}, {Hoffmann}, {Hofmann}, {Holleran}, {Hoppe}, {Horns},
  {Jacholkowska}, {de Jager}, {Kendziorra}, {Kerschhaggl}, {Kh{\'e}lifi},
  {Komin}, {Konopelko}, {Kosack}, {Lamanna}, {Latham}, {Le Gallou},
  {Lemi{\`e}re}, {Lemoine-Goumard}, {Lenain}, {Lohse}, {Martin},
  {Martineau-Huynh}, {Marcowith}, {Masterson}, {Maurin}, {McComb}, {Moulin},
  {de Naurois}, {Nedbal}, {Nolan}, {Noutsos}, {Orford}, {Osborne}, {Ouchrif},
  {Panter}, {Pelletier}, {Pita}, {P{\"u}hlhofer}, {Punch}, {Ranchon},
  {Raubenheimer}, {Raue}, {Rayner}, {Reimer}, {Ripken}, {Rob}, {Rolland},
  {Rosier-Lees}, {Rowell}, {Sahakian}, {Santangelo}, {Saug{\'e}}, {Schlenker},
  {Schlickeiser}, {Schr{\"o}der}, {Schwanke}, {Schwarzburg}, {Schwemmer},
  {Shalchi}, {Sol}, {Spangler}, {Spanier}, {Steenkamp}, {Stegmann}, {Superina},
  {Tam}, {Tavernet}, {Terrier}, {Tluczykont}, {van Eldik}, {Vasileiadis},
  {Venter}, {Vialle}, {Vincent}, {V{\"o}lk}, {Wagner}, \&
  {Ward}}]{Aharonian2006}
{Aharonian}, F., {Akhperjanian}, A.~G., {Bazer-Bachi}, A.~R., {et~al.} 2006,
  Science, 314, 1424, \dodoi{10.1126/science.1134408}

\bibitem[{{Akiyama} {et~al.}(2018){Akiyama}, {Asada}, {Fish}, {Nakamura},
  {Hada}, {Nagai}, \& {Lonsdale}}]{Akiyama2018}
{Akiyama}, K., {Asada}, K., {Fish}, V., {et~al.} 2018, Galaxies, 6, 15,
  \dodoi{10.3390/galaxies6010015}

\bibitem[{{Akiyama} {et~al.}(2015){Akiyama}, {Lu}, {Fish}, {Doeleman},
  {Broderick}, {Dexter}, {Hada}, {Kino}, {Nagai}, {Honma}, {Johnson}, {Algaba},
  {Asada}, {Brinkerink}, {Blundell}, {Bower}, {Cappallo}, {Crew}, {Dexter},
  {Dzib}, {Freund}, {Friberg}, {Gurwell}, {Ho}, {Inoue}, {Krichbaum},
  {Loinard}, {MacMahon}, {Marrone}, {Moran}, {Nakamura}, {Nagar}, {Ortiz-Leon},
  {Plambeck}, {Pradel}, {Primiani}, {Rogers}, {Roy}, {SooHoo}, {Tavares},
  {Tilanus}, {Titus}, {Wagner}, {Weintroub}, {Yamaguchi}, {Young}, {Zensus}, \&
  {Ziurys}}]{Akiyama2015}
{Akiyama}, K., {Lu}, R.-S., {Fish}, V.~L., {et~al.} 2015, \apj, 807, 150,
  \dodoi{10.1088/0004-637X/807/2/150}

\bibitem[{{An} {et~al.}(2018){An}, {Sohn}, \& {Imai}}]{An2018}
{An}, T., {Sohn}, B.~W., \& {Imai}, H. 2018, Nature Astronomy, 2, 118,
  \dodoi{10.1038/s41550-017-0277-z}

\bibitem[{{Asada} \& {Nakamura}(2012)}]{AN2012}
{Asada}, K., \& {Nakamura}, M. 2012, \apjl, 745, L28,
  \dodoi{10.1088/2041-8205/745/2/L28}

\bibitem[{{Asada} {et~al.}(2014){Asada}, {Nakamura}, {Doi}, {Nagai}, \&
  {Inoue}}]{Asada2014}
{Asada}, K., {Nakamura}, M., {Doi}, A., {Nagai}, H., \& {Inoue}, M. 2014,
  \apjl, 781, L2, \dodoi{10.1088/2041-8205/781/1/L2}

\bibitem[{{Asada} {et~al.}(2017){Asada}, {Kino}, {Honma}, {Hirota}, {Lu},
  {Inoue}, {Sohn}, {Shen}, {Ho}, {Akiyama}, {Algaba}, {An}, {Bower}, {Byun},
  {Dodson}, {Doi}, {Edwards}, {Fujisawa}, {Gu}, {Hada}, {Hagiwara},
  {Jaroenjittichai}, {Jung}, {Kawashima}, {Koyama}, {Lee}, {Matsushita},
  {Nagai}, {Nakamura}, {Niinuma}, {Phillips}, {Park}, {Pu}, {Ro}, {Stevens},
  {Trippe}, {Wajima}, \& {Zhao}}]{Asada2017}
{Asada}, K., {Kino}, M., {Honma}, M., {et~al.} 2017, arXiv e-prints,
  arXiv:1705.04776.
\newblock \doarXiv{1705.04776}

\bibitem[{{Astropy Collaboration} {et~al.}(2013){Astropy Collaboration},
  {Robitaille}, {Tollerud}, {Greenfield}, {Droettboom}, {Bray}, {Aldcroft},
  {Davis}, {Ginsburg}, {Price-Whelan}, {Kerzendorf}, {Conley}, {Crighton},
  {Barbary}, {Muna}, {Ferguson}, {Grollier}, {Parikh}, {Nair}, {Unther},
  {Deil}, {Woillez}, {Conseil}, {Kramer}, {Turner}, {Singer}, {Fox}, {Weaver},
  {Zabalza}, {Edwards}, {Azalee Bostroem}, {Burke}, {Casey}, {Crawford},
  {Dencheva}, {Ely}, {Jenness}, {Labrie}, {Lim}, {Pierfederici}, {Pontzen},
  {Ptak}, {Refsdal}, {Servillat}, \& {Streicher}}]{Astropy2013}
{Astropy Collaboration}, {Robitaille}, T.~P., {Tollerud}, E.~J., {et~al.} 2013,
  \aap, 558, A33, \dodoi{10.1051/0004-6361/201322068}

\bibitem[{{Astropy Collaboration} {et~al.}(2018){Astropy Collaboration},
  {Price-Whelan}, {Sip{\H{o}}cz}, {G{\"u}nther}, {Lim}, {Crawford}, {Conseil},
  {Shupe}, {Craig}, {Dencheva}, {Ginsburg}, {Vand erPlas}, {Bradley},
  {P{\'e}rez-Su{\'a}rez}, {de Val-Borro}, {Aldcroft}, {Cruz}, {Robitaille},
  {Tollerud}, {Ardelean}, {Babej}, {Bach}, {Bachetti}, {Bakanov}, {Bamford},
  {Barentsen}, {Barmby}, {Baumbach}, {Berry}, {Biscani}, {Boquien}, {Bostroem},
  {Bouma}, {Brammer}, {Bray}, {Breytenbach}, {Buddelmeijer}, {Burke},
  {Calderone}, {Cano Rodr{\'\i}guez}, {Cara}, {Cardoso}, {Cheedella}, {Copin},
  {Corrales}, {Crichton}, {D'Avella}, {Deil}, {Depagne}, {Dietrich}, {Donath},
  {Droettboom}, {Earl}, {Erben}, {Fabbro}, {Ferreira}, {Finethy}, {Fox},
  {Garrison}, {Gibbons}, {Goldstein}, {Gommers}, {Greco}, {Greenfield},
  {Groener}, {Grollier}, {Hagen}, {Hirst}, {Homeier}, {Horton}, {Hosseinzadeh},
  {Hu}, {Hunkeler}, {Ivezi{\'c}}, {Jain}, {Jenness}, {Kanarek}, {Kendrew},
  {Kern}, {Kerzendorf}, {Khvalko}, {King}, {Kirkby}, {Kulkarni}, {Kumar},
  {Lee}, {Lenz}, {Littlefair}, {Ma}, {Macleod}, {Mastropietro}, {McCully},
  {Montagnac}, {Morris}, {Mueller}, {Mumford}, {Muna}, {Murphy}, {Nelson},
  {Nguyen}, {Ninan}, {N{\"o}the}, {Ogaz}, {Oh}, {Parejko}, {Parley}, {Pascual},
  {Patil}, {Patil}, {Plunkett}, {Prochaska}, {Rastogi}, {Reddy Janga},
  {Sabater}, {Sakurikar}, {Seifert}, {Sherbert}, {Sherwood-Taylor}, {Shih},
  {Sick}, {Silbiger}, {Singanamalla}, {Singer}, {Sladen}, {Sooley},
  {Sornarajah}, {Streicher}, {Teuben}, {Thomas}, {Tremblay}, {Turner},
  {Terr{\'o}n}, {van Kerkwijk}, {de la Vega}, {Watkins}, {Weaver}, {Whitmore},
  {Woillez}, {Zabalza}, \& {Astropy Contributors}}]{Astropy2018}
{Astropy Collaboration}, {Price-Whelan}, A.~M., {Sip{\H{o}}cz}, B.~M., {et~al.}
  2018, \aj, 156, 123, \dodoi{10.3847/1538-3881/aabc4f}

\bibitem[{{Baczko} {et~al.}(2019){Baczko}, {Schulz}, {Kadler}, {Ros},
  {Perucho}, {Fromm}, \& {Wilms}}]{Baczko2019}
{Baczko}, A.~K., {Schulz}, R., {Kadler}, M., {et~al.} 2019, \aap, 623, A27,
  \dodoi{10.1051/0004-6361/201833828}

\bibitem[{{Begelman} \& {Li}(1994)}]{BL1994}
{Begelman}, M.~C., \& {Li}, Z.-Y. 1994, \apj, 426, 269, \dodoi{10.1086/174061}

\bibitem[{{Beskin} \& {Nokhrina}(2006)}]{BN2006}
{Beskin}, V.~S., \& {Nokhrina}, E.~E. 2006, \mnras, 367, 375,
  \dodoi{10.1111/j.1365-2966.2006.09957.x}

\bibitem[{{Biretta} {et~al.}(1999){Biretta}, {Sparks}, \&
  {Macchetto}}]{Biretta1999}
{Biretta}, J.~A., {Sparks}, W.~B., \& {Macchetto}, F. 1999, \apj, 520, 621,
  \dodoi{10.1086/307499}

\bibitem[{{Biretta} {et~al.}(1995){Biretta}, {Zhou}, \& {Owen}}]{Biretta1995}
{Biretta}, J.~A., {Zhou}, F., \& {Owen}, F.~N. 1995, \apj, 447, 582,
  \dodoi{10.1086/175901}

\bibitem[{{Birkinshaw} \& {Worrall}(1993)}]{BW1993}
{Birkinshaw}, M., \& {Worrall}, D.~M. 1993, \apj, 412, 568,
  \dodoi{10.1086/172944}

\bibitem[{{Blandford} {et~al.}(2019){Blandford}, {Meier}, \&
  {Readhead}}]{Blandford2019}
{Blandford}, R., {Meier}, D., \& {Readhead}, A. 2019, \araa, 57, 467,
  \dodoi{10.1146/annurev-astro-081817-051948}

\bibitem[{{Blandford} \& {Payne}(1982)}]{BP1982}
{Blandford}, R.~D., \& {Payne}, D.~G. 1982, \mnras, 199, 883,
  \dodoi{10.1093/mnras/199.4.883}

\bibitem[{{Blandford} \& {Znajek}(1977)}]{BZ1977}
{Blandford}, R.~D., \& {Znajek}, R.~L. 1977, \mnras, 179, 433,
  \dodoi{10.1093/mnras/179.3.433}

\bibitem[{{Boccardi} {et~al.}(2016{\natexlab{a}}){Boccardi}, {Krichbaum},
  {Bach}, {Bremer}, \& {Zensus}}]{Boccardi2016a}
{Boccardi}, B., {Krichbaum}, T.~P., {Bach}, U., {Bremer}, M., \& {Zensus},
  J.~A. 2016{\natexlab{a}}, \aap, 588, L9, \dodoi{10.1051/0004-6361/201628412}

\bibitem[{{Boccardi} {et~al.}(2016{\natexlab{b}}){Boccardi}, {Krichbaum},
  {Bach}, {Mertens}, {Ros}, {Alef}, \& {Zensus}}]{Boccardi2016b}
{Boccardi}, B., {Krichbaum}, T.~P., {Bach}, U., {et~al.} 2016{\natexlab{b}},
  \aap, 585, A33, \dodoi{10.1051/0004-6361/201526985}

\bibitem[{{Boccardi} {et~al.}(2019){Boccardi}, {Migliori}, {Grandi}, {Torresi},
  {Mertens}, {Karamanavis}, {Angioni}, \& {Vignali}}]{Boccardi2019}
{Boccardi}, B., {Migliori}, G., {Grandi}, P., {et~al.} 2019, \aap, 627, A89,
  \dodoi{10.1051/0004-6361/201935183}

\bibitem[{{Bower} {et~al.}(2017){Bower}, {Dexter}, {Markoff}, {Rao}, \&
  {Plambeck}}]{Bower2017}
{Bower}, G.~C., {Dexter}, J., {Markoff}, S., {Rao}, R., \& {Plambeck}, R.~L.
  2017, \apjl, 843, L31, \dodoi{10.3847/2041-8213/aa7b2e}

\bibitem[{{Bridle} {et~al.}(1976){Bridle}, {Davis}, {Meloy}, {Fomalont},
  {Strom}, \& {Willis}}]{Bridle1976}
{Bridle}, A.~H., {Davis}, M.~M., {Meloy}, D.~A., {et~al.} 1976, \nat, 262, 179,
  \dodoi{10.1038/262179a0}

\bibitem[{{Bromberg} \& {Levinson}(2009)}]{BL2009}
{Bromberg}, O., \& {Levinson}, A. 2009, \apj, 699, 1274,
  \dodoi{10.1088/0004-637X/699/2/1274}

\bibitem[{{Bu} {et~al.}(2016{\natexlab{a}}){Bu}, {Yuan}, {Gan}, \&
  {Yang}}]{Bu2016a}
{Bu}, D.-F., {Yuan}, F., {Gan}, Z.-M., \& {Yang}, X.-H. 2016{\natexlab{a}},
  \apj, 818, 83, \dodoi{10.3847/0004-637X/818/1/83}

\bibitem[{{Bu} {et~al.}(2016{\natexlab{b}}){Bu}, {Yuan}, {Gan}, \&
  {Yang}}]{Bu2016b}
---. 2016{\natexlab{b}}, \apj, 823, 90, \dodoi{10.3847/0004-637X/823/2/90}

\bibitem[{{Camenzind}(1987)}]{Camenzind1987}
{Camenzind}, M. 1987, \aap, 184, 341

\bibitem[{{Canvin} {et~al.}(2005){Canvin}, {Laing}, {Bridle}, \&
  {Cotton}}]{Canvin2005}
{Canvin}, J.~R., {Laing}, R.~A., {Bridle}, A.~H., \& {Cotton}, W.~D. 2005,
  \mnras, 363, 1223, \dodoi{10.1111/j.1365-2966.2005.09537.x}

\bibitem[{{Cavaliere} \& {Fusco-Femiano}(1978)}]{CF1978}
{Cavaliere}, A., \& {Fusco-Femiano}, R. 1978, \aap, 70, 677

\bibitem[{{Chatterjee} {et~al.}(2019){Chatterjee}, {Liska}, {Tchekhovskoy}, \&
  {Markoff}}]{Chatterjee2019}
{Chatterjee}, K., {Liska}, M., {Tchekhovskoy}, A., \& {Markoff}, S.~B. 2019,
  \mnras, 490, 2200, \dodoi{10.1093/mnras/stz2626}

\bibitem[{{Chen} {et~al.}(2011){Chen}, {Zhao}, \& {Shen}}]{Chen2011}
{Chen}, Y.~J., {Zhao}, G.~Y., \& {Shen}, Z.~Q. 2011, \mnras, 416, L109,
  \dodoi{10.1111/j.1745-3933.2011.01110.x}

\bibitem[{{Cheung} {et~al.}(2007){Cheung}, {Harris}, \& {Stawarz}}]{Cheung2007}
{Cheung}, C.~C., {Harris}, D.~E., \& {Stawarz}, {\L}. 2007, \apjl, 663, L65,
  \dodoi{10.1086/520510}

\bibitem[{{Cho} {et~al.}(2017){Cho}, {Jung}, {Zhao}, {Akiyama}, {Sawada-Satoh},
  {Kino}, {Byun}, {Sohn}, {Shibata}, {Hirota}, {Niinuma}, {Yonekura},
  {Fujisawa}, \& {Oyama}}]{Cho2017}
{Cho}, I., {Jung}, T., {Zhao}, G.-Y., {et~al.} 2017, \pasj, 69, 87,
  \dodoi{10.1093/pasj/psx090}

\bibitem[{{Clausen-Brown} {et~al.}(2013){Clausen-Brown}, {Savolainen},
  {Pushkarev}, {Kovalev}, \& {Zensus}}]{Clausen-Brown2013}
{Clausen-Brown}, E., {Savolainen}, T., {Pushkarev}, A.~B., {Kovalev}, Y.~Y., \&
  {Zensus}, J.~A. 2013, \aap, 558, A144, \dodoi{10.1051/0004-6361/201322203}

\bibitem[{{Contopoulos}(1995)}]{Contopoulos1995}
{Contopoulos}, J. 1995, \apj, 450, 616, \dodoi{10.1086/176170}

\bibitem[{{Cotton} {et~al.}(1999){Cotton}, {Feretti}, {Giovannini}, {Lara}, \&
  {Venturi}}]{Cotton1999}
{Cotton}, W.~D., {Feretti}, L., {Giovannini}, G., {Lara}, L., \& {Venturi}, T.
  1999, \apj, 519, 108, \dodoi{10.1086/307358}

\bibitem[{{Croke} \& {Gabuzda}(2008)}]{CG2008}
{Croke}, S.~M., \& {Gabuzda}, D.~C. 2008, \mnras, 386, 619,
  \dodoi{10.1111/j.1365-2966.2008.13087.x}

\bibitem[{{Doeleman} {et~al.}(2012){Doeleman}, {Fish}, {Schenck}, {Beaudoin},
  {Blundell}, {Bower}, {Broderick}, {Chamberlin}, {Freund}, {Friberg},
  {Gurwell}, {Ho}, {Honma}, {Inoue}, {Krichbaum}, {Lamb}, {Loeb}, {Lonsdale},
  {Marrone}, {Moran}, {Oyama}, {Plambeck}, {Primiani}, {Rogers}, {Smythe},
  {SooHoo}, {Strittmatter}, {Tilanus}, {Titus}, {Weintroub}, {Wright}, {Young},
  \& {Ziurys}}]{Doeleman2012}
{Doeleman}, S.~S., {Fish}, V.~L., {Schenck}, D.~E., {et~al.} 2012, Science,
  338, 355, \dodoi{10.1126/science.1224768}

\bibitem[{{Eichler}(1993)}]{Eichler1993}
{Eichler}, D. 1993, \apj, 419, 111, \dodoi{10.1086/173464}

\bibitem[{{Ene} {et~al.}(2020){Ene}, {Ma}, {Walsh}, {Greene}, {Thomas}, \&
  {Blakeslee}}]{Ene2020}
{Ene}, I., {Ma}, C.-P., {Walsh}, J.~L., {et~al.} 2020, \apj, 891, 65,
  \dodoi{10.3847/1538-4357/ab7016}

\bibitem[{{Evans} {et~al.}(2005){Evans}, {Hardcastle}, {Croston}, {Worrall}, \&
  {Birkinshaw}}]{Evans2005}
{Evans}, D.~A., {Hardcastle}, M.~J., {Croston}, J.~H., {Worrall}, D.~M., \&
  {Birkinshaw}, M. 2005, \mnras, 359, 363,
  \dodoi{10.1111/j.1365-2966.2005.08900.x}

\bibitem[{{Faber} {et~al.}(1989){Faber}, {Wegner}, {Burstein}, {Davies},
  {Dressler}, {Lynden-Bell}, \& {Terlevich}}]{Faber1989}
{Faber}, S.~M., {Wegner}, G., {Burstein}, D., {et~al.} 1989, \apjs, 69, 763,
  \dodoi{10.1086/191327}

\bibitem[{{Fabian}(2012)}]{Fabian2012}
{Fabian}, A.~C. 2012, \araa, 50, 455,
  \dodoi{10.1146/annurev-astro-081811-125521}

\bibitem[{{Fanaroff} \& {Riley}(1974)}]{FR1974}
{Fanaroff}, B.~L., \& {Riley}, J.~M. 1974, \mnras, 167, 31P,
  \dodoi{10.1093/mnras/167.1.31P}

\bibitem[{{Fomalont}(1999)}]{Fomalont1999}
{Fomalont}, E.~B. 1999, in Astronomical Society of the Pacific Conference
  Series, Vol. 180, Synthesis Imaging in Radio Astronomy II, ed. G.~B.
  {Taylor}, C.~L. {Carilli}, \& R.~A. {Perley}, 301

\bibitem[{{Fromm} {et~al.}(2013){Fromm}, {Ros}, {Perucho}, {Savolainen},
  {Mimica}, {Kadler}, {Lobanov}, \& {Zensus}}]{Fromm2013}
{Fromm}, C.~M., {Ros}, E., {Perucho}, M., {et~al.} 2013, \aap, 557, A105,
  \dodoi{10.1051/0004-6361/201321784}

\bibitem[{{Gaspari} {et~al.}(2013){Gaspari}, {Ruszkowski}, \&
  {Oh}}]{Gaspari2013}
{Gaspari}, M., {Ruszkowski}, M., \& {Oh}, S.~P. 2013, \mnras, 432, 3401,
  \dodoi{10.1093/mnras/stt692}

\bibitem[{{Ghisellini} {et~al.}(2005){Ghisellini}, {Tavecchio}, \&
  {Chiaberge}}]{Ghisellini2005}
{Ghisellini}, G., {Tavecchio}, F., \& {Chiaberge}, M. 2005, \aap, 432, 401,
  \dodoi{10.1051/0004-6361:20041404}

\bibitem[{{Giovannini} {et~al.}(1994){Giovannini}, {Feretti}, {Venturi},
  {Lara}, {Marcaide}, {Rioja}, {Spangler}, \& {Wehrle}}]{Giovannini1994}
{Giovannini}, G., {Feretti}, L., {Venturi}, T., {et~al.} 1994, \apj, 435, 116,
  \dodoi{10.1086/174799}

\bibitem[{{Giovannini} {et~al.}(2018){Giovannini}, {Savolainen}, {Orienti},
  {Nakamura}, {Nagai}, {Kino}, {Giroletti}, {Hada}, {Bruni}, {Kovalev},
  {Anderson}, {D'Ammando}, {Hodgson}, {Honma}, {Krichbaum}, {Lee}, {Lico},
  {Lisakov}, {Lobanov}, {Petrov}, {Sohn}, {Sokolovsky}, {Voitsik}, {Zensus}, \&
  {Tingay}}]{Giovannini2018}
{Giovannini}, G., {Savolainen}, T., {Orienti}, M., {et~al.} 2018, Nature
  Astronomy, 2, 472, \dodoi{10.1038/s41550-018-0431-2}

\bibitem[{{Giroletti} {et~al.}(2012){Giroletti}, {Hada}, {Giovannini},
  {Casadio}, {Beilicke}, {Cesarini}, {Cheung}, {Doi}, {Krawczynski}, {Kino},
  {Lee}, \& {Nagai}}]{Giroletti2012}
{Giroletti}, M., {Hada}, K., {Giovannini}, G., {et~al.} 2012, \aap, 538, L10,
  \dodoi{10.1051/0004-6361/201218794}

\bibitem[{{Greisen}(2003)}]{Greisen2003}
{Greisen}, E.~W. 2003, Astrophysics and Space Science Library, Vol. 285, {AIPS,
  the VLA, and the VLBA}, ed. A.~{Heck}, 109, \dodoi{10.1007/0-306-48080-8_7}

\bibitem[{{Hada}(2019)}]{Hada2019}
{Hada}, K. 2019, Galaxies, 8, 1, \dodoi{10.3390/galaxies8010001}

\bibitem[{{Hada} {et~al.}(2011){Hada}, {Doi}, {Kino}, {Nagai}, {Hagiwara}, \&
  {Kawaguchi}}]{Hada2011}
{Hada}, K., {Doi}, A., {Kino}, M., {et~al.} 2011, \nat, 477, 185,
  \dodoi{10.1038/nature10387}

\bibitem[{{Hada} {et~al.}(2013){Hada}, {Doi}, {Nagai}, {Inoue}, {Honma},
  {Giroletti}, \& {Giovannini}}]{Hada2013}
{Hada}, K., {Doi}, A., {Nagai}, H., {et~al.} 2013, \apj, 779, 6,
  \dodoi{10.1088/0004-637X/779/1/6}

\bibitem[{{Hada} {et~al.}(2014){Hada}, {Giroletti}, {Kino}, {Giovannini},
  {D'Ammando}, {Cheung}, {Beilicke}, {Nagai}, {Doi}, {Akiyama}, {Honma},
  {Niinuma}, {Casadio}, {Orienti}, {Krawczynski}, {G{\'o}mez}, {Sawada-Satoh},
  {Koyama}, {Cesarini}, {Nakahara}, \& {Gurwell}}]{Hada2014}
{Hada}, K., {Giroletti}, M., {Kino}, M., {et~al.} 2014, \apj, 788, 165,
  \dodoi{10.1088/0004-637X/788/2/165}

\bibitem[{{Hada} {et~al.}(2015){Hada}, {Giroletti}, {Giovannini}, {Casadio},
  {Beilicke}, {Cesarini}, {Cheung}, {Doi}, {G{\'o}mez}, {Krawczynski}, {Kino},
  \& {Nagai}}]{Hada2015}
{Hada}, K., {Giroletti}, M., {Giovannini}, G., {et~al.} 2015, arXiv e-prints,
  arXiv:1504.01808.
\newblock \doarXiv{1504.01808}

\bibitem[{{Hada} {et~al.}(2016){Hada}, {Kino}, {Doi}, {Nagai}, {Honma},
  {Akiyama}, {Tazaki}, {Lico}, {Giroletti}, {Giovannini}, {Orienti}, \&
  {Hagiwara}}]{Hada2016}
{Hada}, K., {Kino}, M., {Doi}, A., {et~al.} 2016, \apj, 817, 131,
  \dodoi{10.3847/0004-637X/817/2/131}

\bibitem[{{Hada} {et~al.}(2017){Hada}, {Park}, {Kino}, {Niinuma}, {Sohn}, {Ro},
  {Jung}, {Algaba}, {Zhao}, {Lee}, {Akiyama}, {Trippe}, {Wajima},
  {Sawada-Satoh}, {Tazaki}, {Cho}, {Hodgson}, {Lee}, {Hagiwara}, {Honma},
  {Koyama}, {Oh}, {Lee}, {Yoo}, {Kawaguchi}, {Roh}, {Oh}, {Yeom}, {Jung}, {Oh},
  {Kim}, {Hwang}, {Byun}, {Cho}, {Kim}, {Kobayashi}, \& {Shibata}}]{Hada2017}
{Hada}, K., {Park}, J.~H., {Kino}, M., {et~al.} 2017, \pasj, 69, 71,
  \dodoi{10.1093/pasj/psx054}

\bibitem[{{Hada} {et~al.}(2018){Hada}, {Doi}, {Wajima}, {D'Ammand o},
  {Orienti}, {Giroletti}, {Giovannini}, {Nakamura}, \& {Asada}}]{Hada2018}
{Hada}, K., {Doi}, A., {Wajima}, K., {et~al.} 2018, \apj, 860, 141,
  \dodoi{10.3847/1538-4357/aac49f}

\bibitem[{{Haga} {et~al.}(2015){Haga}, {Doi}, {Murata}, {Sudou}, {Kameno}, \&
  {Hada}}]{Haga2015}
{Haga}, T., {Doi}, A., {Murata}, Y., {et~al.} 2015, \apj, 807, 15,
  \dodoi{10.1088/0004-637X/807/1/15}

\bibitem[{{Hardcastle} \& {Croston}(2020)}]{HC2020}
{Hardcastle}, M.~J., \& {Croston}, J.~H. 2020, \nar, 88, 101539,
  \dodoi{10.1016/j.newar.2020.101539}

\bibitem[{{Harris} {et~al.}(2009){Harris}, {Cheung}, {Stawarz}, {Biretta}, \&
  {Perlman}}]{Harris2009}
{Harris}, D.~E., {Cheung}, C.~C., {Stawarz}, {\L}., {Biretta}, J.~A., \&
  {Perlman}, E.~S. 2009, \apj, 699, 305, \dodoi{10.1088/0004-637X/699/1/305}

\bibitem[{{Hirotani}(2005)}]{Hirotani2005}
{Hirotani}, K. 2005, \apj, 619, 73, \dodoi{10.1086/426497}

\bibitem[{{Hudson} {et~al.}(2010){Hudson}, {Mittal}, {Reiprich}, {Nulsen},
  {Andernach}, \& {Sarazin}}]{Hudson2010}
{Hudson}, D.~S., {Mittal}, R., {Reiprich}, T.~H., {et~al.} 2010, \aap, 513,
  A37, \dodoi{10.1051/0004-6361/200912377}

\bibitem[{{Hunter}(2007)}]{Matplotlib2007}
{Hunter}, J.~D. 2007, Computing in Science Engineering, 9, 90

\bibitem[{{Jorstad} {et~al.}(2017){Jorstad}, {Marscher}, {Morozova},
  {Troitsky}, {Agudo}, {Casadio}, {Foord}, {G{\'o}mez}, {MacDonald}, {Molina},
  {L{\"a}hteenm{\"a}ki}, {Tammi}, \& {Tornikoski}}]{Jorstad2017}
{Jorstad}, S.~G., {Marscher}, A.~P., {Morozova}, D.~A., {et~al.} 2017, \apj,
  846, 98, \dodoi{10.3847/1538-4357/aa8407}

\bibitem[{{Junor} {et~al.}(1999){Junor}, {Biretta}, \& {Livio}}]{Junor1999}
{Junor}, W., {Biretta}, J.~A., \& {Livio}, M. 1999, \nat, 401, 891,
  \dodoi{10.1038/44780}

\bibitem[{{Kameno} {et~al.}(2001){Kameno}, {Sawada-Satoh}, {Inoue}, {Shen}, \&
  {Wajima}}]{Kameno2001}
{Kameno}, S., {Sawada-Satoh}, S., {Inoue}, M., {Shen}, Z.-Q., \& {Wajima}, K.
  2001, \pasj, 53, 169, \dodoi{10.1093/pasj/53.2.169}

\bibitem[{{Kataoka} {et~al.}(2006){Kataoka}, {Stawarz}, {Aharonian},
  {Takahara}, {Ostrowski}, \& {Edwards}}]{Kataoka2006}
{Kataoka}, J., {Stawarz}, {\L}., {Aharonian}, F., {et~al.} 2006, \apj, 641,
  158, \dodoi{10.1086/500407}

\bibitem[{{Kim} {et~al.}(2018){Kim}, {Krichbaum}, {Lu}, {Ros}, {Bach},
  {Bremer}, {de Vicente}, {Lindqvist}, \& {Zensus}}]{Kim2018}
{Kim}, J.~Y., {Krichbaum}, T.~P., {Lu}, R.~S., {et~al.} 2018, \aap, 616, A188,
  \dodoi{10.1051/0004-6361/201832921}

\bibitem[{{Komissarov} {et~al.}(2007){Komissarov}, {Barkov}, {Vlahakis}, \&
  {K{\"o}nigl}}]{Komissarov2007}
{Komissarov}, S.~S., {Barkov}, M.~V., {Vlahakis}, N., \& {K{\"o}nigl}, A. 2007,
  \mnras, 380, 51, \dodoi{10.1111/j.1365-2966.2007.12050.x}

\bibitem[{{Komissarov} {et~al.}(2009){Komissarov}, {Vlahakis}, {K{\"o}nigl}, \&
  {Barkov}}]{Komissarov2009}
{Komissarov}, S.~S., {Vlahakis}, N., {K{\"o}nigl}, A., \& {Barkov}, M.~V. 2009,
  \mnras, 394, 1182, \dodoi{10.1111/j.1365-2966.2009.14410.x}

\bibitem[{{Konigl}(1981)}]{Konigl1981}
{Konigl}, A. 1981, \apj, 243, 700, \dodoi{10.1086/158638}

\bibitem[{{Kormendy} \& {Ho}(2013)}]{KH2013}
{Kormendy}, J., \& {Ho}, L.~C. 2013, \araa, 51, 511,
  \dodoi{10.1146/annurev-astro-082708-101811}

\bibitem[{{Kovalev} {et~al.}(2007){Kovalev}, {Lister}, {Homan}, \&
  {Kellermann}}]{Kovalev2007}
{Kovalev}, Y.~Y., {Lister}, M.~L., {Homan}, D.~C., \& {Kellermann}, K.~I. 2007,
  \apjl, 668, L27, \dodoi{10.1086/522603}

\bibitem[{{Kovalev} {et~al.}(2020){Kovalev}, {Pushkarev}, {Nokhrina}, {Plavin},
  {Beskin}, {Chernoglazov}, {Lister}, \& {Savolainen}}]{Kovalev2020}
{Kovalev}, Y.~Y., {Pushkarev}, A.~B., {Nokhrina}, E.~E., {et~al.} 2020, \mnras,
  495, 3576, \dodoi{10.1093/mnras/staa1121}

\bibitem[{{Laing} \& {Bridle}(2002)}]{LB2002}
{Laing}, R.~A., \& {Bridle}, A.~H. 2002, \mnras, 336, 328,
  \dodoi{10.1046/j.1365-8711.2002.05756.x}

\bibitem[{{Laing} \& {Bridle}(2014)}]{LB2014}
---. 2014, \mnras, 437, 3405, \dodoi{10.1093/mnras/stt2138}

\bibitem[{{Laing} {et~al.}(2006){Laing}, {Canvin}, {Cotton}, \&
  {Bridle}}]{Laing2006}
{Laing}, R.~A., {Canvin}, J.~R., {Cotton}, W.~D., \& {Bridle}, A.~H. 2006,
  \mnras, 368, 48, \dodoi{10.1111/j.1365-2966.2006.10099.x}

\bibitem[{{Lee} {et~al.}(2008){Lee}, {Lobanov}, {Krichbaum}, {Witzel},
  {Zensus}, {Bremer}, {Greve}, \& {Grewing}}]{Lee2008}
{Lee}, S.-S., {Lobanov}, A.~P., {Krichbaum}, T.~P., {et~al.} 2008, \aj, 136,
  159, \dodoi{10.1088/0004-6256/136/1/159}

\bibitem[{{Lee} {et~al.}(2014){Lee}, {Petrov}, {Byun}, {Kim}, {Jung}, {Song},
  {Oh}, {Roh}, {Je}, {Wi}, {Sohn}, {Oh}, {Kim}, {Yeom}, {Chung}, {Kang}, {Han},
  {Lee}, {Kim}, {Chung}, {Kim}, {Ryoung Kim}, {Kang}, \& {Cho}}]{Lee2014}
{Lee}, S.-S., {Petrov}, L., {Byun}, D.-Y., {et~al.} 2014, \aj, 147, 77,
  \dodoi{10.1088/0004-6256/147/4/77}

\bibitem[{{Lee} {et~al.}(2015){Lee}, {Oh}, {Roh}, {Oh}, {Kim}, {Yeom}, {Kim},
  {Jung}, {Byun}, {Jung}, {Kawaguchi}, {Shibata}, \& {Wajima}}]{Lee2015}
{Lee}, S.-S., {Oh}, C.~S., {Roh}, D.-G., {et~al.} 2015, Journal of Korean
  Astronomical Society, 48, 125, \dodoi{10.5303/JKAS.2015.48.2.125}

\bibitem[{{Lee} {et~al.}(2019){Lee}, {Trippe}, {Kino}, {Sohn}, {Park}, {Oh},
  {Hada}, {Niinuma}, {Ro}, {Jung}, {Zhao}, {Lee}, {Algaba}, {Akiyama},
  {Wajima}, {Sawada-Satoh}, {Tazaki}, {Cho}, {Hodgson}, {Lee}, {Hagiwara},
  {Honma}, {Koyama}, {An}, {Cui}, {Yoo}, {Kawaguchi}, {Roh}, {Oh}, {Yeom},
  {Jung}, {Oh}, {Kim}, {Hwang}, {Byun}, {Cho}, {Kim}, {Kobayashi}, {Shibata},
  {Shen}, {Jiang}, \& {Lee}}]{Lee2019}
{Lee}, T., {Trippe}, S., {Kino}, M., {et~al.} 2019, \mnras, 486, 2412,
  \dodoi{10.1093/mnras/stz970}

\bibitem[{{Levinson} \& {Globus}(2017)}]{LG2017}
{Levinson}, A., \& {Globus}, N. 2017, \mnras, 465, 1608,
  \dodoi{10.1093/mnras/stw2902}

\bibitem[{{Li} {et~al.}(1992){Li}, {Chiueh}, \& {Begelman}}]{Li1992}
{Li}, Z.-Y., {Chiueh}, T., \& {Begelman}, M.~C. 1992, \apj, 394, 459,
  \dodoi{10.1086/171597}

\bibitem[{{Lister} {et~al.}(2018){Lister}, {Aller}, {Aller}, {Hodge}, {Homan},
  {Kovalev}, {Pushkarev}, \& {Savolainen}}]{Lister2018}
{Lister}, M.~L., {Aller}, M.~F., {Aller}, H.~D., {et~al.} 2018, \apjs, 234, 12,
  \dodoi{10.3847/1538-4365/aa9c44}

\bibitem[{{Lister} {et~al.}(2019){Lister}, {Homan}, {Hovatta}, {Kellermann},
  {Kiehlmann}, {Kovalev}, {Max-Moerbeck}, {Pushkarev}, {Readhead}, {Ros}, \&
  {Savolainen}}]{Lister2019}
{Lister}, M.~L., {Homan}, D.~C., {Hovatta}, T., {et~al.} 2019, \apj, 874, 43,
  \dodoi{10.3847/1538-4357/ab08ee}

\bibitem[{{Lobanov}(1998)}]{Lobanov1998}
{Lobanov}, A.~P. 1998, \aap, 330, 79.
\newblock \doarXiv{astro-ph/9712132}

\bibitem[{{Ly} {et~al.}(2007){Ly}, {Walker}, \& {Junor}}]{Ly2007}
{Ly}, C., {Walker}, R.~C., \& {Junor}, W. 2007, \apj, 660, 200,
  \dodoi{10.1086/512846}

\bibitem[{{Lyubarsky}(2009)}]{Lyubarsky2009}
{Lyubarsky}, Y. 2009, \apj, 698, 1570, \dodoi{10.1088/0004-637X/698/2/1570}

\bibitem[{{MacDonald} {et~al.}(2015){MacDonald}, {Marscher}, {Jorstad}, \&
  {Joshi}}]{MacDonald2015}
{MacDonald}, N.~R., {Marscher}, A.~P., {Jorstad}, S.~G., \& {Joshi}, M. 2015,
  \apj, 804, 111, \dodoi{10.1088/0004-637X/804/2/111}

\bibitem[{{Marscher} {et~al.}(2008){Marscher}, {Jorstad}, {D'Arcangelo},
  {Smith}, {Williams}, {Larionov}, {Oh}, {Olmstead}, {Aller}, {Aller},
  {McHardy}, {L{\"a}hteenm{\"a}ki}, {Tornikoski}, {Valtaoja}, {Hagen-Thorn},
  {Kopatskaya}, {Gear}, {Tosti}, {Kurtanidze}, {Nikolashvili}, {Sigua},
  {Miller}, \& {Ryle}}]{Marscher2008}
{Marscher}, A.~P., {Jorstad}, S.~G., {D'Arcangelo}, F.~D., {et~al.} 2008, \nat,
  452, 966, \dodoi{10.1038/nature06895}

\bibitem[{{Marscher} {et~al.}(2010){Marscher}, {Jorstad}, {Larionov}, {Aller},
  {Aller}, {L{\"a}hteenm{\"a}ki}, {Agudo}, {Smith}, {Gurwell}, {Hagen-Thorn},
  {Konstantinova}, {Larionova}, {Larionova}, {Melnichuk}, {Blinov},
  {Kopatskaya}, {Troitsky}, {Tornikoski}, {Hovatta}, {Schmidt}, {D'Arcangelo},
  {Bhattarai}, {Taylor}, {Olmstead}, {Manne-Nicholas}, {Roca-Sogorb},
  {G{\'o}mez}, {McHardy}, {Kurtanidze}, {Nikolashvili}, {Kimeridze}, \&
  {Sigua}}]{Marscher2010}
{Marscher}, A.~P., {Jorstad}, S.~G., {Larionov}, V.~M., {et~al.} 2010, \apjl,
  710, L126, \dodoi{10.1088/2041-8205/710/2/L126}

\bibitem[{{McKinney}(2006)}]{McKinney2006}
{McKinney}, J.~C. 2006, \mnras, 368, 1561,
  \dodoi{10.1111/j.1365-2966.2006.10256.x}

\bibitem[{{Meier}(2012)}]{Meier2012}
{Meier}, D.~L. 2012, {Black Hole Astrophysics: The Engine Paradigm}

\bibitem[{{Mertens} \& {Lobanov}(2015)}]{ML2015}
{Mertens}, F., \& {Lobanov}, A. 2015, \aap, 574, A67,
  \dodoi{10.1051/0004-6361/201424566}

\bibitem[{{Mertens} {et~al.}(2016){Mertens}, {Lobanov}, {Walker}, \&
  {Hardee}}]{Mertens2016}
{Mertens}, F., {Lobanov}, A.~P., {Walker}, R.~C., \& {Hardee}, P.~E. 2016,
  \aap, 595, A54, \dodoi{10.1051/0004-6361/201628829}

\bibitem[{{Meyer} {et~al.}(2013){Meyer}, {Sparks}, {Biretta}, {Anderson},
  {Sohn}, {van der Marel}, {Norman}, \& {Nakamura}}]{Meyer2013}
{Meyer}, E.~T., {Sparks}, W.~B., {Biretta}, J.~A., {et~al.} 2013, \apjl, 774,
  L21, \dodoi{10.1088/2041-8205/774/2/L21}

\bibitem[{{Nakahara} {et~al.}(2018){Nakahara}, {Doi}, {Murata}, {Hada},
  {Nakamura}, \& {Asada}}]{Nakahara2018}
{Nakahara}, S., {Doi}, A., {Murata}, Y., {et~al.} 2018, \apj, 854, 148,
  \dodoi{10.3847/1538-4357/aaa45e}

\bibitem[{{Nakahara} {et~al.}(2019){Nakahara}, {Doi}, {Murata}, {Nakamura},
  {Hada}, \& {Asada}}]{Nakahara2019}
---. 2019, \apj, 878, 61, \dodoi{10.3847/1538-4357/ab1b0e}

\bibitem[{{Nakahara} {et~al.}(2020){Nakahara}, {Doi}, {Murata}, {Nakamura},
  {Hada}, {Asada}, {Sawada-Satoh}, \& {Kameno}}]{Nakahara2020}
---. 2020, \aj, 159, 14, \dodoi{10.3847/1538-3881/ab465b}

\bibitem[{{Nakamura} \& {Asada}(2013)}]{NA2013}
{Nakamura}, M., \& {Asada}, K. 2013, \apj, 775, 118,
  \dodoi{10.1088/0004-637X/775/2/118}

\bibitem[{{Nakamura} {et~al.}(2010){Nakamura}, {Garofalo}, \&
  {Meier}}]{Nakamura2010}
{Nakamura}, M., {Garofalo}, D., \& {Meier}, D.~L. 2010, \apj, 721, 1783,
  \dodoi{10.1088/0004-637X/721/2/1783}

\bibitem[{{Nakamura} \& {Meier}(2014)}]{NM2014}
{Nakamura}, M., \& {Meier}, D.~L. 2014, \apj, 785, 152,
  \dodoi{10.1088/0004-637X/785/2/152}

\bibitem[{{Nakamura} {et~al.}(2018){Nakamura}, {Asada}, {Hada}, {Pu}, {Noble},
  {Tseng}, {Toma}, {Kino}, {Nagai}, {Takahashi}, {Algaba}, {Orienti},
  {Akiyama}, {Doi}, {Giovannini}, {Giroletti}, {Honma}, {Koyama}, {Lico},
  {Niinuma}, \& {Tazaki}}]{Nakamura2018}
{Nakamura}, M., {Asada}, K., {Hada}, K., {et~al.} 2018, \apj, 868, 146,
  \dodoi{10.3847/1538-4357/aaeb2d}

\bibitem[{{Narayan} {et~al.}(2007){Narayan}, {McKinney}, \&
  {Farmer}}]{Narayan2007}
{Narayan}, R., {McKinney}, J.~C., \& {Farmer}, A.~J. 2007, \mnras, 375, 548,
  \dodoi{10.1111/j.1365-2966.2006.11272.x}

\bibitem[{{Narayan} \& {Quataert}(2005)}]{NQ2005}
{Narayan}, R., \& {Quataert}, E. 2005, Science, 307, 77,
  \dodoi{10.1126/science.1105746}

\bibitem[{{Netzer}(2015)}]{Netzer2015}
{Netzer}, H. 2015, \araa, 53, 365, \dodoi{10.1146/annurev-astro-082214-122302}

\bibitem[{{Niinuma} {et~al.}(2014){Niinuma}, {Lee}, {Kino}, {Sohn}, {Akiyama},
  {Zhao}, {Sawada-Satoh}, {Trippe}, {Hada}, {Jung}, {Hagiwara}, {Dodson},
  {Koyama}, {Honma}, {Nagai}, {Chung}, {Doi}, {Fujisawa}, {Han}, {Kim}, {Lee},
  {Lee}, {Miyazaki}, {Oyama}, {Sorai}, {Wajima}, {Bae}, {Byun}, {Cho}, {Choi},
  {Chung}, {Chung}, {Han}, {Hirota}, {Hwang}, {Je}, {Jike}, {Jung}, {Jung},
  {Kang}, {Kang}, {Kang}, {Kan-ya}, {Kanaguchi}, {Kawaguchi}, {Kim}, {Kim},
  {Kim}, {Kim}, {Kim}, {Kim}, {Kim}, {Kobayashi}, {Kono}, {Kurayama}, {Lee},
  {Lee}, {Lee}, {Minh}, {Matsumoto}, {Nakagawa}, {Oh}, {Oh}, {Park}, {Roh},
  {Sasao}, {Shibata}, {Song}, {Tamura}, {Wi}, {Yeom}, \& {Yun}}]{Niinuma2014}
{Niinuma}, K., {Lee}, S.-S., {Kino}, M., {et~al.} 2014, \pasj, 66, 103,
  \dodoi{10.1093/pasj/psu104}

\bibitem[{{Oh} {et~al.}(2015){Oh}, {Trippe}, {Kang}, {Kim}, {Park}, {Lee},
  {Kim}, {Kino}, {Lee}, \& {Sohn}}]{Oh2015}
{Oh}, J., {Trippe}, S., {Kang}, S., {et~al.} 2015, Journal of Korean
  Astronomical Society, 48, 299, \dodoi{10.5303/JKAS.2015.48.5.299}

\bibitem[{{Ostrowski}(1998)}]{Ostrowski1998}
{Ostrowski}, M. 1998, \aap, 335, 134.
\newblock \doarXiv{astro-ph/9803299}

\bibitem[{{O'Sullivan} \& {Gabuzda}(2009)}]{OG2009}
{O'Sullivan}, S.~P., \& {Gabuzda}, D.~C. 2009, \mnras, 393, 429,
  \dodoi{10.1111/j.1365-2966.2008.14213.x}

\bibitem[{{Park} {et~al.}(2020){Park}, {Byun}, {Asada}, \& {Yun}}]{Park2020}
{Park}, J., {Byun}, D.-Y., {Asada}, K., \& {Yun}, Y. 2020, arXiv e-prints,
  arXiv:2011.09713.
\newblock \doarXiv{2011.09713}

\bibitem[{{Park} {et~al.}(2019{\natexlab{a}}){Park}, {Hada}, {Kino},
  {Nakamura}, {Ro}, \& {Trippe}}]{Park2019a}
{Park}, J., {Hada}, K., {Kino}, M., {et~al.} 2019{\natexlab{a}}, \apj, 871,
  257, \dodoi{10.3847/1538-4357/aaf9a9}

\bibitem[{{Park} {et~al.}(2019{\natexlab{b}}){Park}, {Hada}, {Kino},
  {Nakamura}, {Ro}, \& {Trippe}}]{Park2019b}
---. 2019{\natexlab{b}}, \apj, 871, 257, \dodoi{10.3847/1538-4357/aaf9a9}

\bibitem[{{Park} \& {Trippe}(2017)}]{PT2017}
{Park}, J., \& {Trippe}, S. 2017, \apj, 834, 157,
  \dodoi{10.3847/1538-4357/834/2/157}

\bibitem[{{Park} {et~al.}(2019{\natexlab{c}}){Park}, {Lee}, {Kim}, {Hodgson},
  {Trippe}, {Kim}, {Algaba}, {Kino}, {Zhao}, {Lee}, \& {Gurwell}}]{Park2019c}
{Park}, J., {Lee}, S.-S., {Kim}, J.-Y., {et~al.} 2019{\natexlab{c}}, \apj, 877,
  106, \dodoi{10.3847/1538-4357/ab1b27}

\bibitem[{{Park} \& {Trippe}(2014)}]{PT2014}
{Park}, J.-H., \& {Trippe}, S. 2014, \apj, 785, 76,
  \dodoi{10.1088/0004-637X/785/1/76}

\bibitem[{{Peebles}(1972)}]{Peebles1972}
{Peebles}, P.~J.~E. 1972, \apj, 178, 371, \dodoi{10.1086/151797}

\bibitem[{{Penna} {et~al.}(2013){Penna}, {Narayan}, \& {Sadowski}}]{Penna2013}
{Penna}, R.~F., {Narayan}, R., \& {Sadowski}, A.~e. 2013, \mnras, 436, 3741,
  \dodoi{10.1093/mnras/stt1860}

\bibitem[{{Planck Collaboration} {et~al.}(2020){Planck Collaboration},
  {Aghanim}, {Akrami}, {Ashdown}, {Aumont}, {Baccigalupi}, {Ballardini},
  {Banday}, {Barreiro}, {Bartolo}, {Basak}, {Battye}, {Benabed}, {Bernard},
  {Bersanelli}, {Bielewicz}, {Bock}, {Bond}, {Borrill}, {Bouchet}, {Boulanger},
  {Bucher}, {Burigana}, {Butler}, {Calabrese}, {Cardoso}, {Carron},
  {Challinor}, {Chiang}, {Chluba}, {Colombo}, {Combet}, {Contreras}, {Crill},
  {Cuttaia}, {de Bernardis}, {de Zotti}, {Delabrouille}, {Delouis}, {Di
  Valentino}, {Diego}, {Dor{\'e}}, {Douspis}, {Ducout}, {Dupac}, {Dusini},
  {Efstathiou}, {Elsner}, {En{\ss}lin}, {Eriksen}, {Fantaye}, {Farhang},
  {Fergusson}, {Fernandez-Cobos}, {Finelli}, {Forastieri}, {Frailis},
  {Fraisse}, {Franceschi}, {Frolov}, {Galeotta}, {Galli}, {Ganga},
  {G{\'e}nova-Santos}, {Gerbino}, {Ghosh}, {Gonz{\'a}lez-Nuevo}, {G{\'o}rski},
  {Gratton}, {Gruppuso}, {Gudmundsson}, {Hamann}, {Handley}, {Hansen},
  {Herranz}, {Hildebrandt}, {Hivon}, {Huang}, {Jaffe}, {Jones}, {Karakci},
  {Keih{\"a}nen}, {Keskitalo}, {Kiiveri}, {Kim}, {Kisner}, {Knox},
  {Krachmalnicoff}, {Kunz}, {Kurki-Suonio}, {Lagache}, {Lamarre}, {Lasenby},
  {Lattanzi}, {Lawrence}, {Le Jeune}, {Lemos}, {Lesgourgues}, {Levrier},
  {Lewis}, {Liguori}, {Lilje}, {Lilley}, {Lindholm}, {L{\'o}pez-Caniego},
  {Lubin}, {Ma}, {Mac{\'\i}as-P{\'e}rez}, {Maggio}, {Maino}, {Mandolesi},
  {Mangilli}, {Marcos-Caballero}, {Maris}, {Martin}, {Martinelli},
  {Mart{\'\i}nez-Gonz{\'a}lez}, {Matarrese}, {Mauri}, {McEwen}, {Meinhold},
  {Melchiorri}, {Mennella}, {Migliaccio}, {Millea}, {Mitra},
  {Miville-Desch{\^e}nes}, {Molinari}, {Montier}, {Morgante}, {Moss}, {Natoli},
  {N{\o}rgaard-Nielsen}, {Pagano}, {Paoletti}, {Partridge}, {Patanchon},
  {Peiris}, {Perrotta}, {Pettorino}, {Piacentini}, {Polastri}, {Polenta},
  {Puget}, {Rachen}, {Reinecke}, {Remazeilles}, {Renzi}, {Rocha}, {Rosset},
  {Roudier}, {Rubi{\~n}o-Mart{\'\i}n}, {Ruiz-Granados}, {Salvati}, {Sandri},
  {Savelainen}, {Scott}, {Shellard}, {Sirignano}, {Sirri}, {Spencer},
  {Sunyaev}, {Suur-Uski}, {Tauber}, {Tavagnacco}, {Tenti}, {Toffolatti},
  {Tomasi}, {Trombetti}, {Valenziano}, {Valiviita}, {Van Tent}, {Vibert},
  {Vielva}, {Villa}, {Vittorio}, {Wand elt}, {Wehus}, {White}, {White},
  {Zacchei}, \& {Zonca}}]{Planck2020}
{Planck Collaboration}, {Aghanim}, N., {Akrami}, Y., {et~al.} 2020, \aap, 641,
  A6, \dodoi{10.1051/0004-6361/201833910}

\bibitem[{{Pu} {et~al.}(2015){Pu}, {Nakamura}, {Hirotani}, {Mizuno}, {Wu}, \&
  {Asada}}]{Pu2015}
{Pu}, H.-Y., {Nakamura}, M., {Hirotani}, K., {et~al.} 2015, \apj, 801, 56,
  \dodoi{10.1088/0004-637X/801/1/56}

\bibitem[{{Pu} \& {Takahashi}(2020)}]{PT2020}
{Pu}, H.-Y., \& {Takahashi}, M. 2020, \apj, 892, 37,
  \dodoi{10.3847/1538-4357/ab77ab}

\bibitem[{{Pushkarev} {et~al.}(2012){Pushkarev}, {Hovatta}, {Kovalev},
  {Lister}, {Lobanov}, {Savolainen}, \& {Zensus}}]{Pushkarev2012}
{Pushkarev}, A.~B., {Hovatta}, T., {Kovalev}, Y.~Y., {et~al.} 2012, \aap, 545,
  A113, \dodoi{10.1051/0004-6361/201219173}

\bibitem[{{Russell} {et~al.}(2015){Russell}, {Fabian}, {McNamara}, \&
  {Broderick}}]{Russell2015}
{Russell}, H.~R., {Fabian}, A.~C., {McNamara}, B.~R., \& {Broderick}, A.~E.
  2015, \mnras, 451, 588, \dodoi{10.1093/mnras/stv954}

\bibitem[{{Russell} {et~al.}(2018){Russell}, {Fabian}, {McNamara}, {Miller},
  {Nulsen}, {Piotrowska}, \& {Reynolds}}]{Russell2018}
{Russell}, H.~R., {Fabian}, A.~C., {McNamara}, B.~R., {et~al.} 2018, \mnras,
  477, 3583, \dodoi{10.1093/mnras/sty835}

\bibitem[{{Rybicki} \& {Lightman}(1979)}]{RL1979}
{Rybicki}, G.~B., \& {Lightman}, A.~P. 1979, {Radiative processes in
  astrophysics}

\bibitem[{{Sadowski} {et~al.}(2013){Sadowski}, {Narayan}, {Penna}, \&
  {Zhu}}]{Sadowski2013}
{Sadowski}, A., {Narayan}, R., {Penna}, R., \& {Zhu}, Y. 2013, \mnras, 436,
  3856, \dodoi{10.1093/mnras/stt1881}

\bibitem[{{Schwab} \& {Cotton}(1983)}]{SC1983}
{Schwab}, F.~R., \& {Cotton}, W.~D. 1983, \aj, 88, 688, \dodoi{10.1086/113360}

\bibitem[{{Sexton} {et~al.}(2019){Sexton}, {Canalizo}, {Hiner}, {Komossa},
  {Woo}, {Treister}, \& {Hiner Dimassimo}}]{Sexton2019}
{Sexton}, R.~O., {Canalizo}, G., {Hiner}, K.~D., {et~al.} 2019, \apj, 878, 101,
  \dodoi{10.3847/1538-4357/ab21d5}

\bibitem[{{Shepherd}(1997)}]{Shepherd1997}
{Shepherd}, M.~C. 1997, in Astronomical Society of the Pacific Conference
  Series, Vol. 125, Astronomical Data Analysis Software and Systems VI, ed.
  G.~{Hunt} \& H.~{Payne}, 77

\bibitem[{{Sokoloff} {et~al.}(1998){Sokoloff}, {Bykov}, {Shukurov},
  {Berkhuijsen}, {Beck}, \& {Poezd}}]{Sokoloff1998}
{Sokoloff}, D.~D., {Bykov}, A.~A., {Shukurov}, A., {et~al.} 1998, \mnras, 299,
  189, \dodoi{10.1046/j.1365-8711.1998.01782.x}

\bibitem[{{Stawarz} {et~al.}(2006){Stawarz}, {Aharonian}, {Kataoka},
  {Ostrowski}, {Siemiginowska}, \& {Sikora}}]{Stawarz2006}
{Stawarz}, {\L}., {Aharonian}, F., {Kataoka}, J., {et~al.} 2006, \mnras, 370,
  981, \dodoi{10.1111/j.1365-2966.2006.10525.x}

\bibitem[{{Stawarz} \& {Ostrowski}(2002)}]{SO2002}
{Stawarz}, {\L}., \& {Ostrowski}, M. 2002, \apj, 578, 763,
  \dodoi{10.1086/342649}

\bibitem[{{Sudou} {et~al.}(2000){Sudou}, {Taniguchi}, {Ohyama}, {Kameno},
  {Sawada-Satoh}, {Inoue}, {Kaburaki}, \& {Sasao}}]{Sudou2000}
{Sudou}, H., {Taniguchi}, Y., {Ohyama}, Y., {et~al.} 2000, \pasj, 52, 989,
  \dodoi{10.1093/pasj/52.6.989}

\bibitem[{{Sun}(2009)}]{Sun2009}
{Sun}, M. 2009, \apj, 704, 1586, \dodoi{10.1088/0004-637X/704/2/1586}

\bibitem[{{Tavecchio} \& {Ghisellini}(2008)}]{TG2008}
{Tavecchio}, F., \& {Ghisellini}, G. 2008, \mnras, 385, L98,
  \dodoi{10.1111/j.1745-3933.2008.00441.x}

\bibitem[{{Tavecchio} \& {Ghisellini}(2014)}]{TG2014}
---. 2014, \mnras, 443, 1224, \dodoi{10.1093/mnras/stu1196}

\bibitem[{{Tchekhovskoy} {et~al.}(2008){Tchekhovskoy}, {McKinney}, \&
  {Narayan}}]{Tchekhovskoy2008}
{Tchekhovskoy}, A., {McKinney}, J.~C., \& {Narayan}, R. 2008, \mnras, 388, 551,
  \dodoi{10.1111/j.1365-2966.2008.13425.x}

\bibitem[{{Tchekhovskoy} {et~al.}(2009){Tchekhovskoy}, {McKinney}, \&
  {Narayan}}]{Tchekhovskoy2009}
---. 2009, \apj, 699, 1789, \dodoi{10.1088/0004-637X/699/2/1789}

\bibitem[{{Tchekhovskoy} {et~al.}(2011){Tchekhovskoy}, {Narayan}, \&
  {McKinney}}]{Tchekhovskoy2011}
{Tchekhovskoy}, A., {Narayan}, R., \& {McKinney}, J.~C. 2011, \mnras, 418, L79,
  \dodoi{10.1111/j.1745-3933.2011.01147.x}

\bibitem[{{Toma} \& {Takahara}(2013)}]{TT2013}
{Toma}, K., \& {Takahara}, F. 2013, Progress of Theoretical and Experimental
  Physics, 2013, 083E02, \dodoi{10.1093/ptep/ptt058}

\bibitem[{{Trager} {et~al.}(2000){Trager}, {Faber}, {Worthey}, \&
  {Gonz{\'a}lez}}]{Trager2000}
{Trager}, S.~C., {Faber}, S.~M., {Worthey}, G., \& {Gonz{\'a}lez}, J.~J. 2000,
  \aj, 119, 1645, \dodoi{10.1086/301299}

\bibitem[{{Tseng} {et~al.}(2016){Tseng}, {Asada}, {Nakamura}, {Pu}, {Algaba},
  \& {Lo}}]{Tseng2016}
{Tseng}, C.-Y., {Asada}, K., {Nakamura}, M., {et~al.} 2016, \apj, 833, 288,
  \dodoi{10.3847/1538-4357/833/2/288}

\bibitem[{{van der Walt} {et~al.}(2011){van der Walt}, {Colbert}, \&
  {Varoquaux}}]{Numpy2011}
{van der Walt}, S., {Colbert}, S.~C., \& {Varoquaux}, G. 2011, Computing in
  Science Engineering, 13, 22

\bibitem[{{Venturi} {et~al.}(1993){Venturi}, {Giovannini}, {Feretti},
  {Comoretto}, \& {Wehrle}}]{Venturi1993}
{Venturi}, T., {Giovannini}, G., {Feretti}, L., {Comoretto}, G., \& {Wehrle},
  A.~E. 1993, \apj, 408, 81, \dodoi{10.1086/172571}

\bibitem[{{Virtanen} {et~al.}(2020){Virtanen}, {Gommers}, {Oliphant},
  {Haberland}, {Reddy}, {Cournapeau}, {Burovski}, {Peterson}, {Weckesser},
  {Bright}, {van der Walt}, {Brett}, {Wilson}, {Jarrod Millman}, {Mayorov},
  {Nelson}, {Jones}, {Kern}, {Larson}, {Carey}, {Polat}, {Feng}, {Moore}, {Vand
  erPlas}, {Laxalde}, {Perktold}, {Cimrman}, {Henriksen}, {Quintero}, {Harris},
  {Archibald}, {Ribeiro}, {Pedregosa}, {van Mulbregt}, \&
  {Contributors}}]{Scipy2020}
{Virtanen}, P., {Gommers}, R., {Oliphant}, T.~E., {et~al.} 2020, Nature
  Methods, 17, 261, \dodoi{https://doi.org/10.1038/s41592-019-0686-2}

\bibitem[{{Vlahakis}(2004)}]{Vlahakis2004}
{Vlahakis}, N. 2004, \apj, 600, 324, \dodoi{10.1086/379701}

\bibitem[{{Vlahakis}(2015)}]{Vlahakis2015}
---. 2015, Astrophysics and Space Science Library, Vol. 414, {Theory of
  Relativistic Jets}, ed. I.~{Contopoulos}, D.~{Gabuzda}, \& N.~{Kylafis}, 177,
  \dodoi{10.1007/978-3-319-10356-3_7}

\bibitem[{{Vlahakis} \& {K{\"o}nigl}(2004)}]{VK2004}
{Vlahakis}, N., \& {K{\"o}nigl}, A. 2004, \apj, 605, 656,
  \dodoi{10.1086/382670}

\bibitem[{{Wajima} {et~al.}(2016){Wajima}, {Hagiwara}, {An}, {Baan},
  {Fujisawa}, {Hao}, {Jiang}, {Jung}, {Kawaguchi}, {Kim}, {Kobayashi}, {Oh},
  {Roh}, {Wang}, {Xia}, \& {Zhang}}]{Wajima2016}
{Wajima}, K., {Hagiwara}, Y., {An}, T., {et~al.} 2016, in Astronomical Society
  of the Pacific Conference Series, Vol. 502, Frontiers in Radio Astronomy and
  FAST Early Sciences Symposium 2015, ed. L.~{Qain} \& D.~{Li}, 81.
\newblock \doarXiv{1512.03550}

\bibitem[{{Walker} {et~al.}(2000){Walker}, {Dhawan}, {Romney}, {Kellermann}, \&
  {Vermeulen}}]{Walker2000}
{Walker}, R.~C., {Dhawan}, V., {Romney}, J.~D., {Kellermann}, K.~I., \&
  {Vermeulen}, R.~C. 2000, \apj, 530, 233, \dodoi{10.1086/308372}

\bibitem[{{Walker} {et~al.}(2018){Walker}, {Hardee}, {Davies}, {Ly}, \&
  {Junor}}]{Walker2018}
{Walker}, R.~C., {Hardee}, P.~E., {Davies}, F.~B., {Ly}, C., \& {Junor}, W.
  2018, \apj, 855, 128, \dodoi{10.3847/1538-4357/aaafcc}

\bibitem[{{Werner} {et~al.}(2019){Werner}, {McNamara}, {Churazov}, \&
  {Scannapieco}}]{Werner2019}
{Werner}, N., {McNamara}, B.~R., {Churazov}, E., \& {Scannapieco}, E. 2019,
  \ssr, 215, 5, \dodoi{10.1007/s11214-018-0571-9}

\bibitem[{{W}es {M}c{K}inney(2010)}]{Pandas2010}
{W}es {M}c{K}inney. 2010, in {P}roceedings of the 9th {P}ython in {S}cience
  {C}onference, ed. {S}t\'efan van~der {W}alt \& {J}arrod {M}illman, 56 -- 61,
  \dodoi{10.25080/Majora-92bf1922-00a}

\bibitem[{{Worrall} {et~al.}(2003){Worrall}, {Birkinshaw}, \&
  {Hardcastle}}]{Worrall2003}
{Worrall}, D.~M., {Birkinshaw}, M., \& {Hardcastle}, M.~J. 2003, \mnras, 343,
  L73, \dodoi{10.1046/j.1365-8711.2003.06945.x}

\bibitem[{{Worrall} {et~al.}(2007){Worrall}, {Birkinshaw}, {Laing}, {Cotton},
  \& {Bridle}}]{Worrall2007}
{Worrall}, D.~M., {Birkinshaw}, M., {Laing}, R.~A., {Cotton}, W.~D., \&
  {Bridle}, A.~H. 2007, \mnras, 380, 2,
  \dodoi{10.1111/j.1365-2966.2007.11998.x}

\bibitem[{{Yuan} {et~al.}(2015){Yuan}, {Gan}, {Narayan}, {Sadowski}, {Bu}, \&
  {Bai}}]{Yuan2015}
{Yuan}, F., {Gan}, Z., {Narayan}, R., {et~al.} 2015, \apj, 804, 101,
  \dodoi{10.1088/0004-637X/804/2/101}

\bibitem[{{Yuan} \& {Narayan}(2014)}]{YN2014}
{Yuan}, F., \& {Narayan}, R. 2014, \araa, 52, 529,
  \dodoi{10.1146/annurev-astro-082812-141003}

\bibitem[{{Zakamska} {et~al.}(2008){Zakamska}, {Begelman}, \& {Bland
  ford}}]{Zakamska2008}
{Zakamska}, N.~L., {Begelman}, M.~C., \& {Bland ford}, R.~D. 2008, \apj, 679,
  990, \dodoi{10.1086/587870}

\bibitem[{{Zhao} {et~al.}(2019){Zhao}, {Jung}, {Sohn}, {Kino}, {Honma},
  {Dodson}, {Rioja}, {Han}, {Shibata}, {Byun}, {Akiyama}, {Algaba}, {An},
  {Cheng}, {Cho}, {Cui}, {Hada}, {Hodgson}, {Jiang}, {Lee}, {Lee}, {Niinuma},
  {Park}, {Ro}, {Sawada-Satoh}, {Shen}, {Tazaki}, {Trippe}, {Wajima}, \&
  {Zhang}}]{Zhao2019}
{Zhao}, G.-Y., {Jung}, T., {Sohn}, B.~W., {et~al.} 2019, Journal of Korean
  Astronomical Society, 52, 23, \dodoi{10.5303/JKAS.2019.52.1.23}

\end{thebibliography}
\bibliographystyle{aasjournal}

\end{document}